\documentclass[preprint, showpacs, floatfix, nofootinbib]{revtex4}
\usepackage{graphicx}
\usepackage{amsmath}
\usepackage{afterpage}

\begin{document}

\title{Investigating the collision energy dependence of $\eta$/s in RHIC beam energy scan using Bayesian statistics}

\author{Jussi Auvinen}
\email{jaa49@phy.duke.edu}
\affiliation{Department of Physics, Duke University, Durham, NC 27708, USA}
\author{Iurii Karpenko}
\affiliation{INFN - Sezione di Firenze, I-50019 Sesto Fiorentino, Firenze, Italy}
\author{Jonah E. Bernhard}
\affiliation{Department of Physics, Duke University, Durham, NC 27708, USA}
\author{Steffen A. Bass}
\affiliation{Department of Physics, Duke University, Durham, NC 27708, USA}

\begin{abstract}
We determine the probability distributions of shear viscosity over the entropy density ratio $\eta/s$
in Au+Au collisions at $\sqrt{s_{NN}}=19.6, 39$, and $62.4$ GeV,
using Bayesian inference and Gaussian process emulators for a model-to-data statistical analysis
that probes the full input parameter space of a transport+viscous hydrodynamics hybrid model.
We find the most likely value of $\eta/s$ to be larger at smaller $\sqrt{s_{NN}}$,
although the uncertainties still allow for a constant value between 0.10 and 0.15 for the investigated collision energy range.
\end{abstract}


\maketitle

\section{Introduction}
\label{sec:intro}

After successfully verifying the existence of new state of matter, the quark-gluon plasma (QGP),
experiments at RHIC and LHC are now focused on quantifying the transport properties of the QGP and mapping out the QCD phase diagram.
The beam energy scan (BES) program at RHIC in particular is aimed at probing the phase diagram of QCD matter
at varying values of the baryon chemical potential.
It is known that at LHC and top RHIC collision energies,
the phase transition between QGP and hadronic matter is a cross-over \cite{Aoki:2006we,Endrodi:2011gv}.
However, it is not yet known if there exists a critical point marking the change to first-order phase transition at lower collision energies,
where the QCD matter is exposed to typically lower temperatures but higher net-baryon densities.

The beam energy scan has introduced new challenges to the theoretical modeling of relativistic heavy-ion collisions,
as the initial non-equilibrium evolution of the system leading up to thermalization gains in importance and has a large effect on the outcome of the calculation.
In addition, the assumption of a midrapidity plateau does not hold any longer,
requiring a full (3+1)-D hydrodynamical model to describe the evolution of the medium.
Finally, the equation of state needs to accommodate the non-zero baryon chemical potential $\mu_B$.
To address these challenges, we utilize a hybrid model,
which combines a (3+1)-D viscous hydrodynamics model for the QGP phase of the reaction
with a microscopic hadronic transport model,
that describes the non-equilibrium evolution of the system before QGP formation and after its subsequent decay into final state hadrons.

In recent years, a lot of progress has been made in the field regarding the quantification of the
shear viscosity over entropy ratio $\eta/s$ in QCD medium.
In particular, the temperature dependence of $\eta/s$ has been studied in detail \cite{Csernai:2006zz,Niemi:2015qia,Denicol:2015nhu,Bernhard:2016tnd,Plumari:2016lul}.
However, a recent study \cite{Karpenko:2015xea} has found that $\eta/s$ might depend also on baryon chemical potential $\mu_B$,
as earlier calculations had already found to be the case in the hadron resonance gas \cite{Demir:2008tr,Denicol:2013nua}.

Inspired by the success of Bayesian analysis at LHC energies \cite{Bernhard:2016tnd},
we investigate in this article the possible collision energy dependence of $\eta/s$ 
by performing a similar analysis on the input parameters of the hybrid model used in Ref.~\cite{Karpenko:2015xea}.
We describe the main features and the input parameters of the hybrid model in Section \ref{sec:hybrid}.
A summary of the statistical analysis procedure is provided in Section \ref{sec:stats}
and the details of the simulation setup are found in Section \ref{sec:design}.
Results of the analysis are presented in Section \ref{sec:results}.
We summarize our findings in Section \ref{sec:summary}.

\section{Hybrid model}
\label{sec:hybrid}

The hybrid model utilized in our study is the same as  in Ref.~\cite{Karpenko:2015xea}.
In this model, the initial non-equilibrium evolution of the system is modeled with the UrQMD hadron-string cascade \cite{Bass:1998ca,Bleicher:1999xi}, which is followed by a transport-to-hydro transition (''fluidization'').
The earliest possible starting time for this transition is when the two colliding nuclei have passed through each other:
$\tau_0 \geq \frac{2R_{\text{nucleus}}}{\sqrt{\gamma_{CM}^2-1}}$.
At the transport-to-hydro transition the microscopic particle properties (energy, baryon number)
are mapped to hydrodynamic energy, momentum and charge densities using 3-D Gaussians with ''smearing'' parameters $W_{\text{trans}}$, $W_{\text{long}}$ ($\equiv \sqrt{2}\sigma$).
From the hydrodynamic side, cells with high enough local energy density correspond to regions of matter in the QGP phase. 

The hydrodynamic evolution is modeled with (3+1)D relativistic viscous hydrodynamics \cite{Karpenko:2013wva}.
For simplicity we utilize a constant value of $\eta/s$ throughout the QGP evolution,
which is an input parameter of the model.
In an event-by-event hydrodynamic calculation at the given energy range,
temperature as well as baryon chemical potential vary significantly in space-time.
Therefore an equation of state which covers the large-$\mu_B$ sector is a necessary ingredient for the hydrodynamic stage.
For the present study a chiral model equation of state \cite{Steinheimer:2010ib} is used.
In this equation of state the transition between hadron and quark-gluon phases is a crossover for all $\mu_B$.

The transition from hydrodynamics back to microscopic non-equilibrium transport (''particlization'') is performed
when the local rest frame energy density $\epsilon$ in the hydro cells falls below a specified switching value $\epsilon_{SW}$.
The iso-energy density hypersurface is constructed using the Cornelius routine \cite{Huovinen:2012is}.
At this hypersurface hadrons are sampled via Monte Carlo procedure using Cooper-Frye prescription
and including viscous corrections to their distribution functions.
The generated hadrons are then fed into the UrQMD cascade to simulate final state hadronic rescatterings and resonance decays.

\section{Statistical analysis}
\label{sec:stats}

The basics of Bayesian analysis procedure and model emulation have been described in detail in \cite{Novak:2013bqa,Bernhard:2015hxa}.
In this section we summarize the main points and describe the details specific to the current implementation.

\subsection{Bayes' theorem}

Given a set $X=\{\vec{x}_k\}_{k=1}^{N}$ of points in parameter space (design points),
and a corresponding set $Y=\{\vec{y}_k\}_{k=1}^{N}$ of points in observable space (output points),
the Bayes' theorem states that the conditional probability distribution (aka {\em posterior}) of an input parameter combination $\vec{x}^*$,
given the observation $(X, Y, \vec{y}^{\text{\,exp}})$,  is
\begin{equation}
  P(\vec{x}^*|X, Y, \vec{y}^{\text{\,exp}})\propto P(X, Y, \vec{y}^{\text{\,exp}}|\vec{x}^*)P(\vec{x}^*),
\end{equation}
where $P(\vec{x}^*)$ is the {\em prior} probability distribution of $\vec{x}^*$, and
$P(X, Y, \vec{y}^{\text{\,exp}}|\vec{x}^*)$
is the {\em likelihood} of the observation $(X, Y, \vec{y}^{\text{\,exp}})$ for a given $\vec{x}^*$.

As in Ref.~\cite{Bernhard:2015hxa}, the likelihood in our analysis will be of the form
\begin{equation}
P(X, Y, \vec{y}^{\text{\,exp}}|\vec{x}^*) \propto e^{-\frac{1}{2}(\vec{y}^{\,*}-\vec{y}^{\text{\,exp}})^T\Sigma^{-1}(\vec{y}^{\,*}-\vec{y}^{\text{\,exp}})},
\end{equation}
where $\vec{y}^{\,*}$ is model output for the input parameter point $\vec{x}^*$,
$\vec{y}^{\text{\,exp}}$ is the target value and $\Sigma$ is the covariance matrix.

\subsection{Gaussian emulator}

To calculate the likelihood, one must be able to determine the model output $\vec{y}^*$ for an arbitrary $\vec{x}^*$.
As the simulations typically take several hours to run, it is highly impractical to run the full model during statistical analysis.
Instead, the model is emulated with a {\em Gaussian process}. That is, the model is taken as a mapping of a set $X$ of points in the parameter space to a set $Y_a$ of values of the observable $y_a$,
\begin{equation}
F: X \rightarrow Y_a,
\end{equation}
where $Y_a$ is assumed to have a multivariate normal distribution
\begin{equation}
Y_a \sim \mathcal{N}(\mu(X),\Sigma_{X,X})
\end{equation}
with $\mu(X)=\{\mu(\vec{x}_1),...,\mu(\vec{x}_N)\}$ as the mean and
\begin{equation}
 \Sigma_{X,X}=
 \begin{pmatrix}
 \sigma(\vec{x}_1,\vec{x}_1) & \cdots & \sigma(\vec{x}_1,\vec{x}_N) \\
 \vdots & \ddots & \vdots \\
 \sigma(\vec{x}_N,\vec{x}_1) & \cdots & \sigma(\vec{x}_N,\vec{x}_N)
 \end{pmatrix}
\end{equation}
as the covariance matrix with covariance function $\sigma(\vec{x},\vec{x}')$.
The covariance matrix is further reformulated in terms of a triangular matrix $S$ using the Cholesky decomposition $\Sigma = SS^T$.
A sample from a Gaussian process is then of the form
\begin{equation}
\vec{y}_{GP}(\vec{x})=\mu(\vec{x})+S\vec{u},
\end{equation}
where the components of $\vec{u}$ are normally distributed random variables, $u_i \sim N(0,1)$.

To predict the model output $y_0$ in an arbitrary point $\vec{x_0}$, one writes the joint distribution
\begin{equation}
 \begin{pmatrix}
 y_0 \\
 Y
 \end{pmatrix}
 =
 \mathcal{N}\left(
  \begin{pmatrix}
  \mu(\vec{x_0})\\
  \mu(X)
 \end{pmatrix}
 ,
  \begin{pmatrix}
  \Sigma_{0,0} & \Sigma_{0,X}\\
  \Sigma_{X,0} & \Sigma_{X,X}
 \end{pmatrix}
 \right)
\end{equation}
and calculates the conditional predictive mean
\begin{equation}
 \bar{\mu}(\vec{x_0})=\mu(\vec{x_0})+\Sigma_{0,X}\Sigma_{X,X}^{-1}(Y-\mu(X)).
\end{equation}
Typically we define the mean function $\mu(\vec{x}) \equiv 0$ and the prediction is simply
\begin{equation}
\mu(\vec{x_0})=\Sigma_{0,X}\Sigma_{X,X}^{-1}Y
\end{equation}
with the associated predictive (co)variance
\begin{equation}
\label{eq:predict_cov}
 \bar{\Sigma} = \Sigma_{0,0}-\Sigma_{0,X}\Sigma_{X,X}^{-1}\Sigma_{X,0}.
\end{equation}

Clearly the crucial component of the Gaussian process is the covariance matrix $\Sigma$.
The covariance function $\sigma(\vec{x},\vec{x}')$ can have any form from a simple constant to an exponential or a periodic function,
and the right choice depends on the features of the emulated model.
For this study we have chosen the squared-exponential covariance function with a noise term
\begin{equation}
\sigma(\vec{x},\vec{x}')=\theta_0\exp\left(-\sum\limits_{i=1}^{n}\frac{(x_i-x'_i)^2}{2\theta_i^2}\right)+\theta_{\text{noise}}\delta_{\vec{x}\vec{x}'}
\end{equation}
The {\em hyperparameters} $\vec{\theta}=(\theta_0,\theta_1,...\theta_n,\theta_{\text{noise}})$ are estimated from the provided data $(X,Y)$.
This ``training'' of the emulator is done by maximising the multivariate Gaussian log-likelihood function
\begin{equation}
\begin{split}
\log P(Y|X,\vec{\theta}) =&-\frac{1}{2}Y^T\Sigma_{X,X}^{-1}(\vec{\theta})Y\\
&-\frac{1}{2}\log|\Sigma_{X,X}(\vec{\theta})|-\frac{N}{2}\log(2\pi).
\end{split}
\end{equation}
In principle, using maximum likelihood values for hyperparameters is only an approximative method,
as in the rigorous Bayesian approach the hyperparameters are allowed to change during calibration procedure,
resulting to posterior distributions also for the hyperparameters.
However, in practice 100 training points creates strong enough constraints on the Gaussian processes,
that the precise value of the hyperparameters has little effect on the final results.

\subsection{Principal component analysis}

Generally, the analysis requires one emulator per data point.
However, the number of emulators can be reduced with a principal component analysis,
which is an orthogonal linear transformation of the data onto the directions of maximal variance.

We use our $N$ simulation points and $m$ observables to form a $N$ x $m$ data matrix $Y$.
Data is normalized with the corresponding experimental values to obtain dimensionless quantities,
and centered by subtracting the mean of each observable from all points.
The singular value decomposition of $Y$ is
\begin{equation}
Y = U S V^{T},
\end{equation}
where $S$ is a diagonal matrix containing the singular values,
and $U$ and $V^{T}$ are orthogonal matrices containing the left- and right-singular vectors, respectively.
Eigenvalue decomposition of the covariance matrix $Y^TY$ becomes
\begin{equation}
Y^{T}Y=V S^2V^{T}.
\end{equation}
Singular values in $S$ are square roots of eigenvalues $\lambda_i$,
and right singular vectors $\vec{v_i}$ represented by columns in $V^{T}$ are eigenvectors of $Y^TY$;
these are the principal components (PCs).
The eigenvalues are proportional to the total variance of the data,
such that the eigenvalue of the first principal component explains the largest fraction of the variance.

We can now write our data matrix in principal component space:
\begin{equation}
Z=\sqrt{N}YV.
\end{equation}

Since $\lambda_1 \geq \lambda_2 \geq ... \geq \lambda_m$,
the fraction of the total variance explained by the $q$th principal component,
$\lambda_q / (\sum\limits_{j=1}^m \lambda_j)$,
becomes negligible starting from some index $q < m$.
This allows us to define a lower-rank approximation of the original $Z$ as $Z_q=\sqrt{N}YV_q$.

In addition to dimensional reduction, we perform data whitening,
i.e. divide each principal component with their singular value,
such that they all have unit variance.
The full prescription for transforming a vector $\vec{y}$ from observable space to vector $z$ in whitened principal component space is thus defined as
\begin{equation}
\vec{z} = \vec{y}\,\tilde{V_q} \equiv \vec{y}\sqrt{N}V_qS_q^{-1}.
\end{equation}

To compare an emulator prediction $\vec{z}^{\,*}$ against physical observables, we use inverse transformation
\begin{equation}
\vec{y}^{\,*} = \vec{z}^{\,*} \, \frac{1}{\sqrt{N}} S_q V_q^T.
\end{equation}

\subsection{Likelihood function}

The log-likelihood function used for determining the posterior probability distribution in this analysis has the following form:
\begin{equation}
\begin{split}
\log P(\vec{z}^{\text{\,exp}}|\vec{x}^*) = &-\frac{1}{2} (\vec{z}^{*}\negmedspace-\vec{z}^{\text{\,exp}}) \Sigma_{z}^{-1} (\vec{z}^*\negmedspace-\vec{z}^{\text{\,exp}})^{T}\\
&-\frac{1}{2} \log |2\pi\Sigma_z|,
\end{split}
\end{equation}
where $\vec{z}^*$ is the emulator prediction at the input parameter point $\vec{x}^*$
and $\vec{z}^{\text{\,exp}}$ represents the experimental data transformed to principal component space.

The covariance matrix includes both experimental and emulator uncertainties: $\Sigma_z=\Sigma_z^{\text{\,exp}}+\Sigma_z^{\text{\,GP}}$.
The uncertainty of experimental data $\Sigma_z^{\text{\,exp}}$ is obtained
by transforming the observable space covariance into principal component space:\footnote{
In the previous Bayesian beam energy scan analyses \cite{Auvinen:2016uuv,Auvinen:2016tgi,Auvinen:2017pny},
the experimental uncertainties were treated as simple diagonal contributions in the principal component space:
$\Sigma_z^{\text{\,exp}}=\text{diag}((\sigma'_0)^2,..,(\sigma'_q)^2)$, where $\sigma'=(\sigma_0,...,\sigma_m)\tilde{V}_q$.
However, this neglected the covariances between the principal components
and exaggerated the variance for the first principal component.
The uncertainties were overestimated as a net result.
}
\begin{equation}
\Sigma_z^{\text{\,exp}}=\tilde{V}_q^T\Sigma_y^{\text{\,exp}}\tilde{V}_q=\tilde{V}_q^T\text{diag}(\sigma_0^2,...,\sigma_m^2)\tilde{V}_q.
\end{equation}
For many of the observables, the reported values $\sigma_0,...,\sigma_m$ include both statistical and systematic uncertainties.
In cases where the uncertainties are reported separately, they have been added quadratically:
$\sigma_i^2 = \sigma_{i,\text{stat}}^2+\sigma_{i,\text{sys}}^2$.
In principle, systematic uncertainties introduce also off-diagonal terms into $\Sigma_y^{\text{\,exp}}$.
However, the task of estimating these off-diagonal terms and investigating their effect on the final results
requires detailed knowledge of the detector systems in the experiments and is referred to a future study.

The emulator uncertainty is included as
\[
\Sigma_z^{\text{\,GP}}=\text{diag}(\sigma_{z0}^2,...,\sigma_{zq}^2),
\]
where $\sigma_{zi}^2$ refers to the GP predictive variance (Eq.~\eqref{eq:predict_cov}) for $i$th principal component.
Finally, the term containing the determinant of the total covariance matrix
$|2\pi\Sigma_z|$ is included to provide more value for points which both fit the data and have small uncertainties,
as opposed to points which fit the data but have large uncertainties.

\subsection{Data calibration using random walkers}

The posterior distribution is sampled using the Markov chain Monte Carlo (MCMC) method,
which is a random walk in parameter space, where each step is accepted or rejected based on a relative likelihood.
It converges to posterior distribution at the limit $N_{\text{steps}}\rightarrow \infty$.

An important measure for the quality of the resulting random walk is the {\em acceptance fraction},
which should be between 0.2 and 0.5 \cite{ForemanMackey:2012ig}.
A very low value of the acceptance fraction indicates that the walkers are stuck,
while values close to 1 mean all proposals are accepted and the walk is purely random.

Another measure is the {\em autocorrelation time}, which is an estimate of the number of steps required between independent samples.
This can be used to evaluate if the random walk contained enough steps to produce a proper sample size.

We initialize $\mathcal{O}(1000)$ random walkers in random positions in the input parameter space.
This is followed by the ''burn-in'' phase, which allows the random walkers to converge to the high-likelihood areas.
At the end of the burn-in, the samples produced during this phase are removed and the proper sampling phase is initialized.

Once the MCMC is complete, we verify that we had sufficient number of steps in the burn-in phase to cover a couple autocorrelation lengths,
and enough steps in the sampling phase to cover $\sim \mathcal{O}(10)$ autocorrelations.
Having thousands of walkers ensures a sufficiently large sample of independent observations for generating posterior distributions.

\section{Simulation setup}
\label{sec:design}

\subsection{Input parameters}

The parameters of the computational model encode the physical properties of the system that we wish to extract from our analysis.
These are the specific shear viscosity of the QGP, $\eta/s$, the thermalization time of the system $\tau_0$,
the initial condition smearing parameters $W_{\text{trans}}$ and $W_{\text{long}}$
as well as the hydro-to-transport transition energy density that can provide information on the temperature scale
at which non-equilibrium effects in the hadronic evolution become important.
The following parameter ranges were used for the design points:
\begin{enumerate}
 \item \mbox{Shear viscosity over entropy density $\eta/s$:} \mbox{0.001 -- 0.3} (extended to 0.4 for \mbox{$\sqrt{s_{NN}}=19.6$ GeV})
 \item \mbox{Transport-to-hydro transition time $\tau_0$:} \mbox{0.4 -- 3.1 fm/$c$}
 \item \mbox{Transverse Gaussian smearing of particles $W_{\text{trans}}$:} \mbox{0.2 -- 2.2 fm}
 \item \mbox{Longitudinal Gaussian smearing of particles $W_{\text{long}}$:} \mbox{0.2 -- 2.2 fm}
 \item \mbox{Hydro-to-transport transition energy density $\epsilon_{SW}$:} \mbox{0.15 -- 0.75 GeV/fm$^3$}
\end{enumerate}

As described in Section \ref{sec:hybrid}, $\tau_0$ is constrained by the condition that the two nuclei must have passed each other,
and thus the lower limit depends on collision energy:

\begin{equation}
 \label{eq:taumin}
 \tau_0 \geq 2R / \sqrt{s_{NN}/(2m_N)^2-1}
\end{equation}

The cells in the hydro grid have a transverse length $\Delta x = \Delta y = 0.25$ fm and a longitudinal length $\Delta \eta = 0.15$,
which sets a soft lower limit for the smearing parameters,
as the limited resolution will lead to the ''over-smearing'' of the Gaussians much narrower than the cell size.

As in Ref.~\cite{Bernhard:2015hxa}, the input parameter combinations were sampled using maximin Latin hypercube method,
which attempts to optimize the sample by maximizing the minimum distance between the design points.

\subsection{Normality of the data}

Principal component analysis assumes that variance provides a reasonable measure of the spread of the simulation output values,
i.~e.~the simulation data is approximately normally distributed.
However, this is not generally true for the distributions of observable values calculated from the model output.
We aim to normalize the distributions by scaling the output with the corresponding experimental data
and applying the Box-Cox transformation \cite{Box:1964} on these dimensionless values $\tilde{y}=y / y^{\text{exp}}$:
\begin{equation}
y^{(\lambda)}= \left\{
\begin{array}{ll}
(\tilde{y}^{\lambda}-1)/\lambda & : \lambda \neq 0\\
\log \tilde{y} & : \lambda = 0
\end{array}
\right.
\end{equation}

We check the normality of the resulting distribution with a probability plot, which is a method for comparing
the quantiles of the sample data against the respective quantiles of the normal distribution.
We perform a least-squares fit on the points in probability plot:
the goodness of the fit can be estimated by the coefficient of determination $R^2$, which should be close to 1.
We use $R^2_c=0.91$ as a (very lenient) critical test value for too large deviations from normal distribution.
Observables with $R^2 < R^2_c$ are excluded from the analysis.

\subsection{Event selection}

Centrality bins were determined based on the number of participants $N_{\text{part}}$,
such that the average number of participants $\langle N_{\text{part}} \rangle$ in each bin matches the respective experimental estimate
\cite{Back:2002wb,Alver:2010ck,Adamczyk:2017iwn,Das:2015phd,Abelev:2008ab}.

To save computation time, the centrality selection was done by running the full simulation for the 50\% of the initial UrQMD events with highest number of participants $N_{\text{part}}$,
instead of running the full simulation for all initial states and sorting the events by final charged particle multiplicity $N_{\text{ch}}$.
In addition, each switching energy density hypersurface has been used 50 times for particle sampling (''oversampling'') to increase statistics.
As the number of independent hypersurfaces in any centrality bin is at least an order of magnitude greater,
this method is unlikely to produce any notable bias in the final results.

\section{Results}
\label{sec:results}

\subsection{Emulator verification}

\begin{figure}[htb]
\includegraphics[width=8cm]{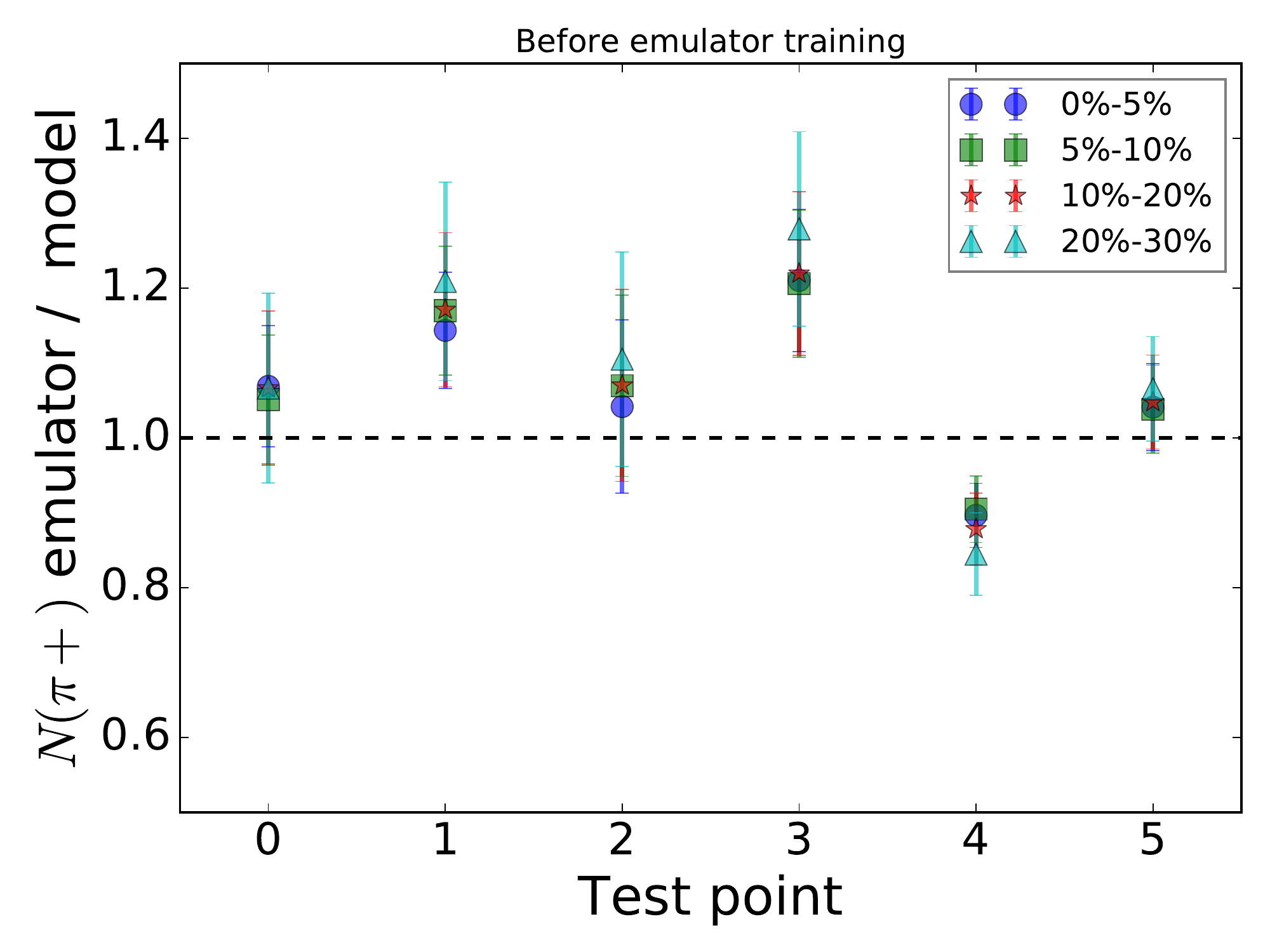}
\includegraphics[width=8cm]{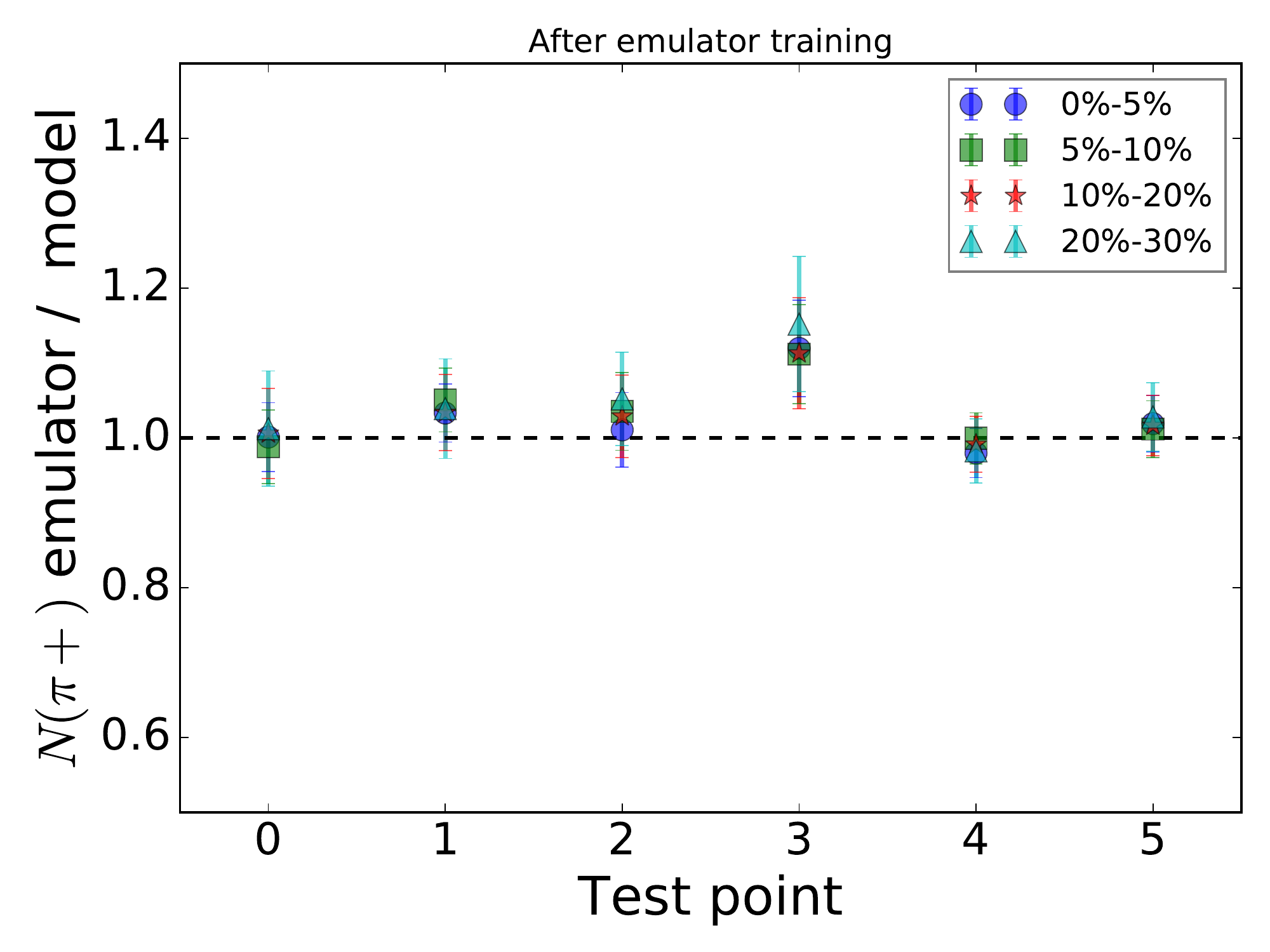}
\caption{(Color online) The ratio of the Gaussian process prediction over the model result for positively charged pion yields,
for the 6 test points at $\sqrt{s_{NN}}=19.6$ GeV.
Top panel: Untrained emulator. Bottom panel: After emulator training.}
\label{fig:verificationN19}
\end{figure}

The quality of the GP emulator was verified against 6 randomly selected test points from the simulation output data,
which were not included in the analysis.
As an example, Fig.~\ref{fig:verificationN19} shows the verification results on $\pi^+$ yields at $\sqrt{s_{NN}}=19.6$ GeV.
The untrained emulator is typically able to make reasonably accurate predictions already,
which indicates that the GPs are already considerably constrained by the density of design points in the input parameter space.
After training the agreement with the test point values is very good.
This assures us that the principal components of the model variance have been correctly identified
and the chosen kernel for Gaussian processes provides a good description of the model behavior
with respect to the principal components.

\subsection{Presentation of analysis results}

The Bayesian posterior probability distributions are illustrated in Figs.~\ref{fig:corner19}, \ref{fig:corner39}, and \ref{fig:corner62}.
We visualize these 5-dimensional probability distributions using corner plots.
The histograms on the diagonal panels represent the probability distributions of each input parameter
integrated over the possible values of the other four parameters,
while the off-diagonal panels show pairwise correlations between the parameters.

Model results after the calibration are compared to experimental data in the following way:
To illustrate how the uncertainty in parameter values reflects to the observables,
100 random parameter combinations were drawn from the posterior
and the model outputs for these combinations were predicted by the Gaussian process emulator.
These emulator predictions are drawn as box-and-whisker plots.
The single red horizontal line within each box represents the median prediction;
that is, half of the predictions are smaller and half are larger than this value.
Boxes represent the values between first and third quartile $Q_1$ and $Q_3$;
this means 25\% of the predictions were smaller than the values in the box and 25\% were larger.
Finally, the whiskers represent either the complete range of predictions,
or 1.5 times the interquartile range $Q_3-Q_1$, whichever is smaller.
In the latter case, the predictions outside the whisker range (''outliers'') are represented by individual points.
Thus, the box-and-whisker representation illustrates the shape of the distribution of emulator predictions.

To verify the quality of the calibration,
full model simulations were run with the median values of the one-dimensional projections of each posterior probability distribution,
represented as dashed red lines and red triangles in figures \ref{fig:corner19}, \ref{fig:corner39}, and \ref{fig:corner62}.

\subsection{$\sqrt{s_{NN}}=19.6$ GeV}

\begin{figure*}[htb]
\includegraphics[width=16cm]{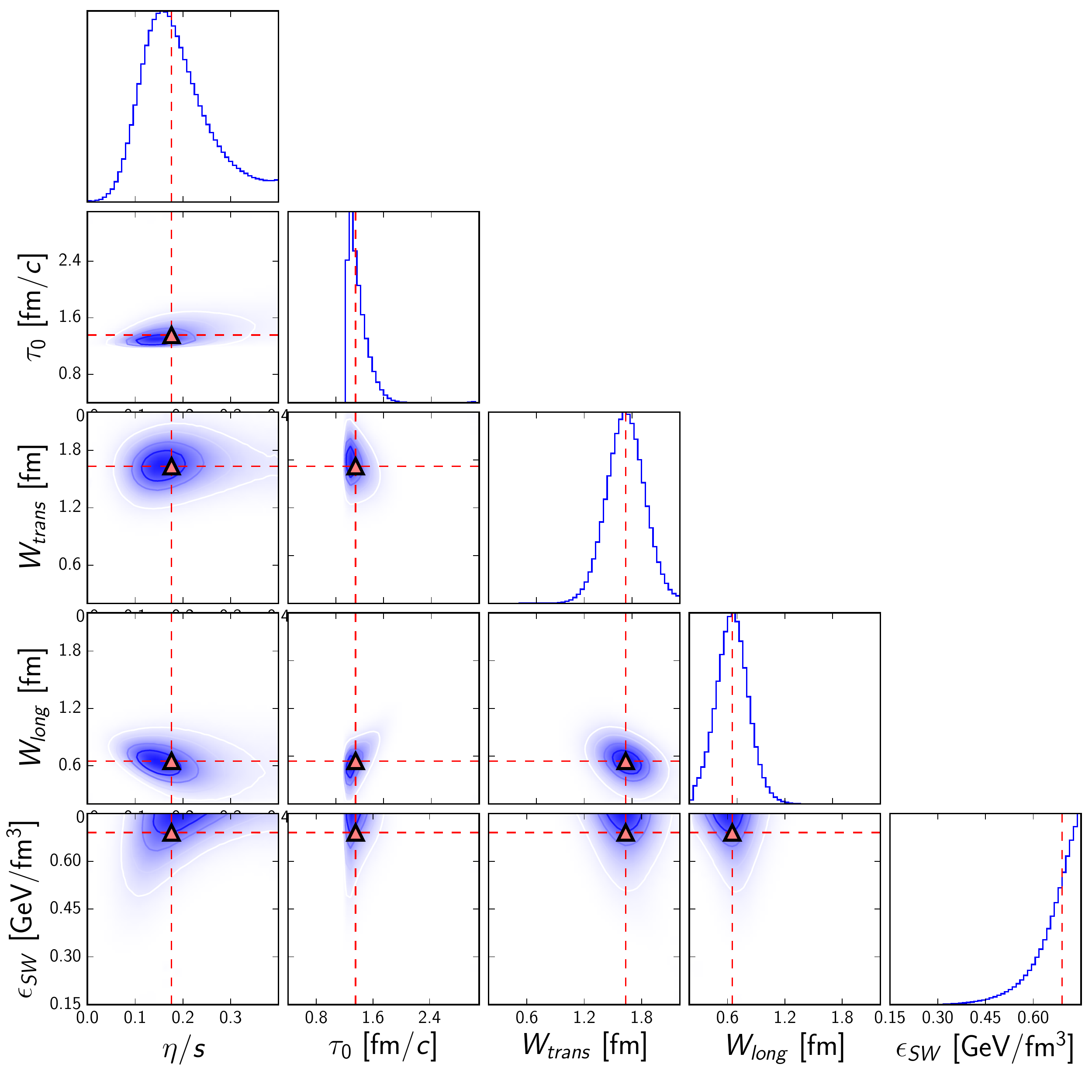}
\caption{(Color online) Representation of the posterior probability distribution of the input parameters at $\sqrt{s_{NN}}=19.6$ GeV.
Panels on a diagonal show one-dimensional projections of the posterior for each parameter axis,
while the off-diagonal panels display projections on the different 2-D planes of the input parameter space.
Dashed red lines indicate median values of the one-dimensional distributions.}
\label{fig:corner19}
\end{figure*}

\begin{figure*}[htb]
\includegraphics[width=16cm]{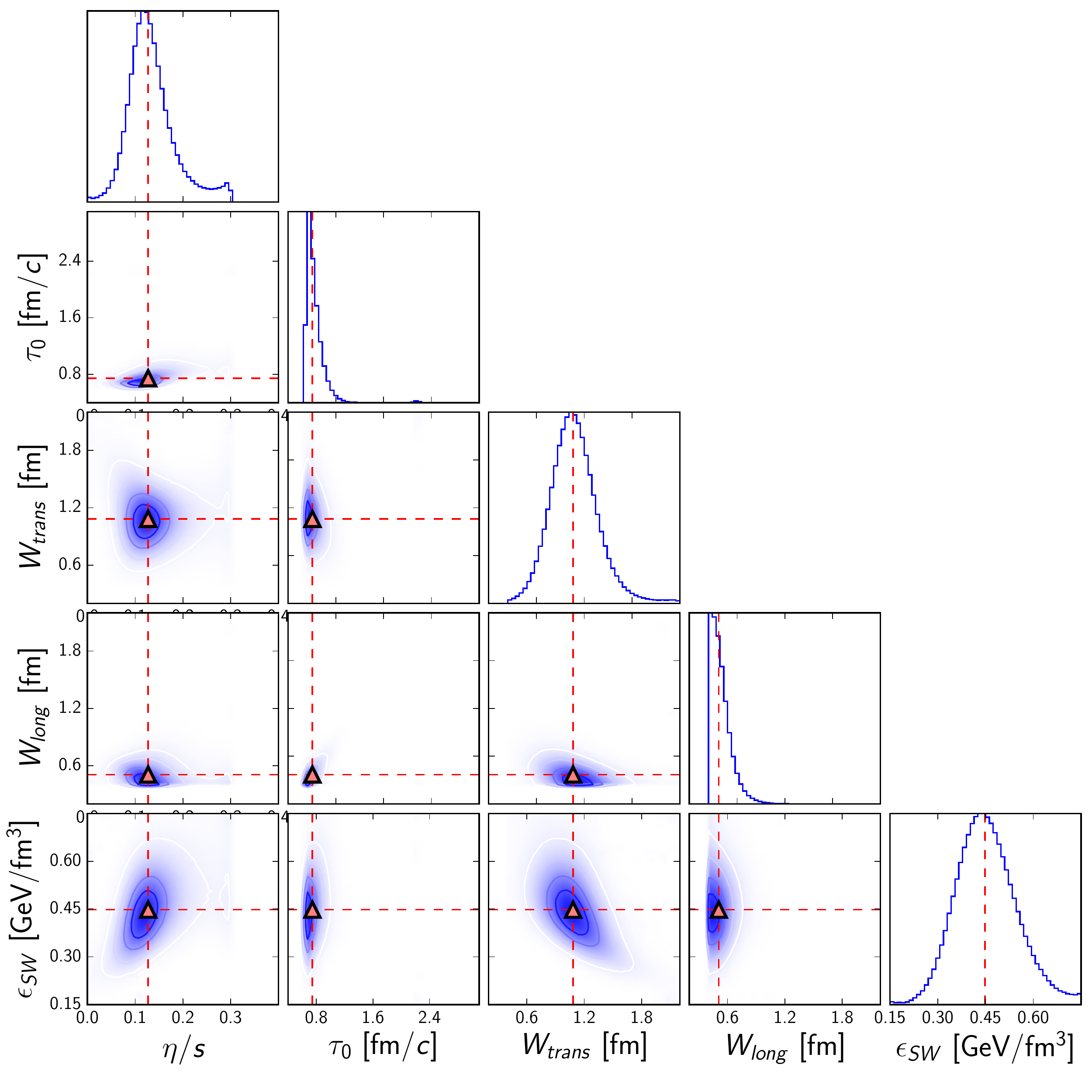}
\caption{(Color online) Representation of the posterior probability distribution of the input parameters at $\sqrt{s_{NN}}=39$ GeV.
See caption of Fig.~\ref{fig:corner19} for details.}
\label{fig:corner39}
\end{figure*}

The posterior distribution for $\sqrt{s_{NN}}=19.6$ GeV is presented in Fig.~\ref{fig:corner19}.
While the data supports the earliest possible starting time and longitudinally narrow Gaussians as density representations of the initial particles,
the transverse plane of the initial density profile is strongly smeared.
An early transition from hydro to hadron transport is preferred,
which is in contrast to the earlier studies where $\Omega$ baryon was given more weight in the analysis \cite{Auvinen:2016uuv,Auvinen:2016tgi,Auvinen:2017pny}.

\begin{figure}[htb]
\includegraphics[width=8cm]{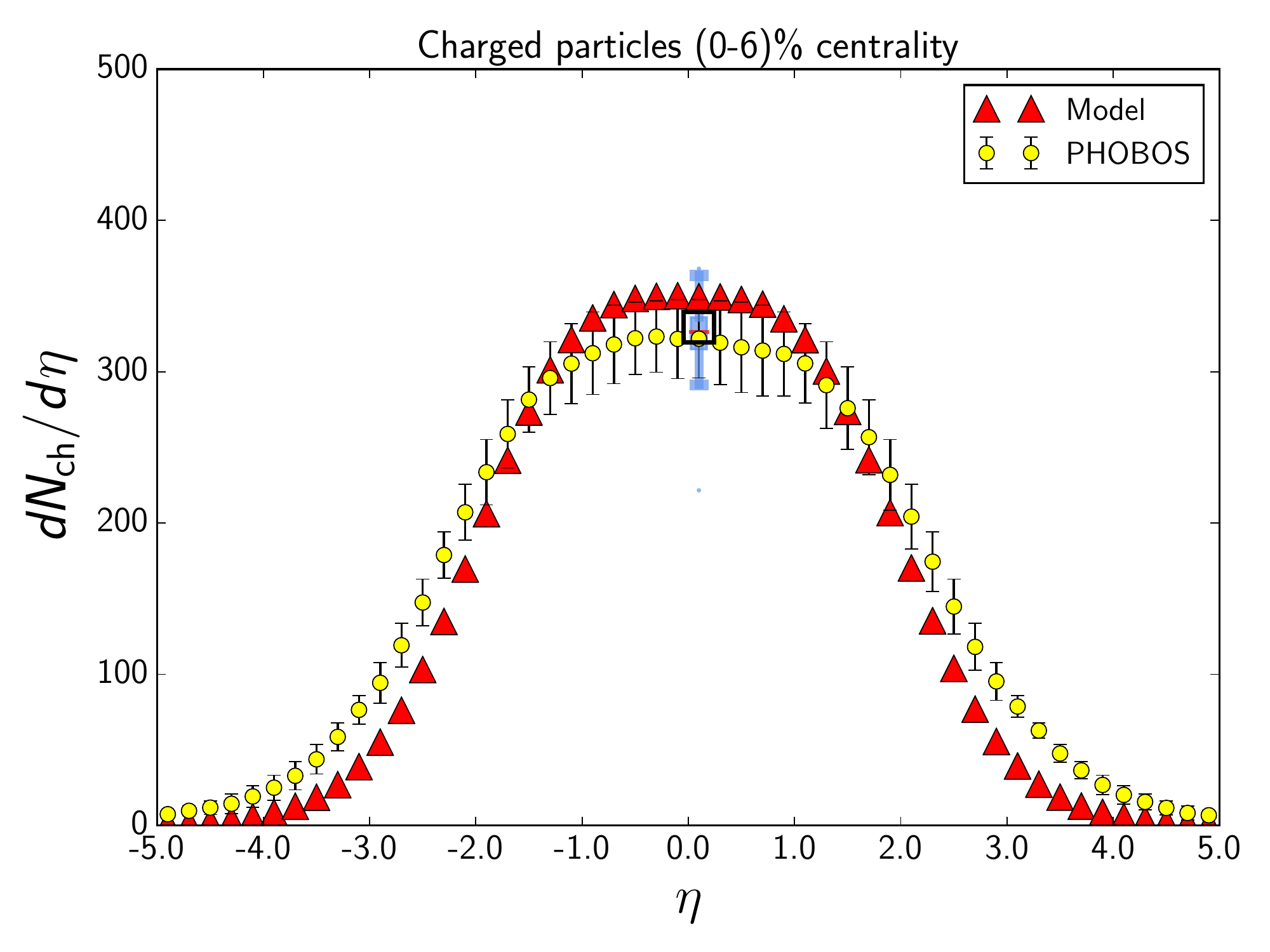}
\includegraphics[width=8cm]{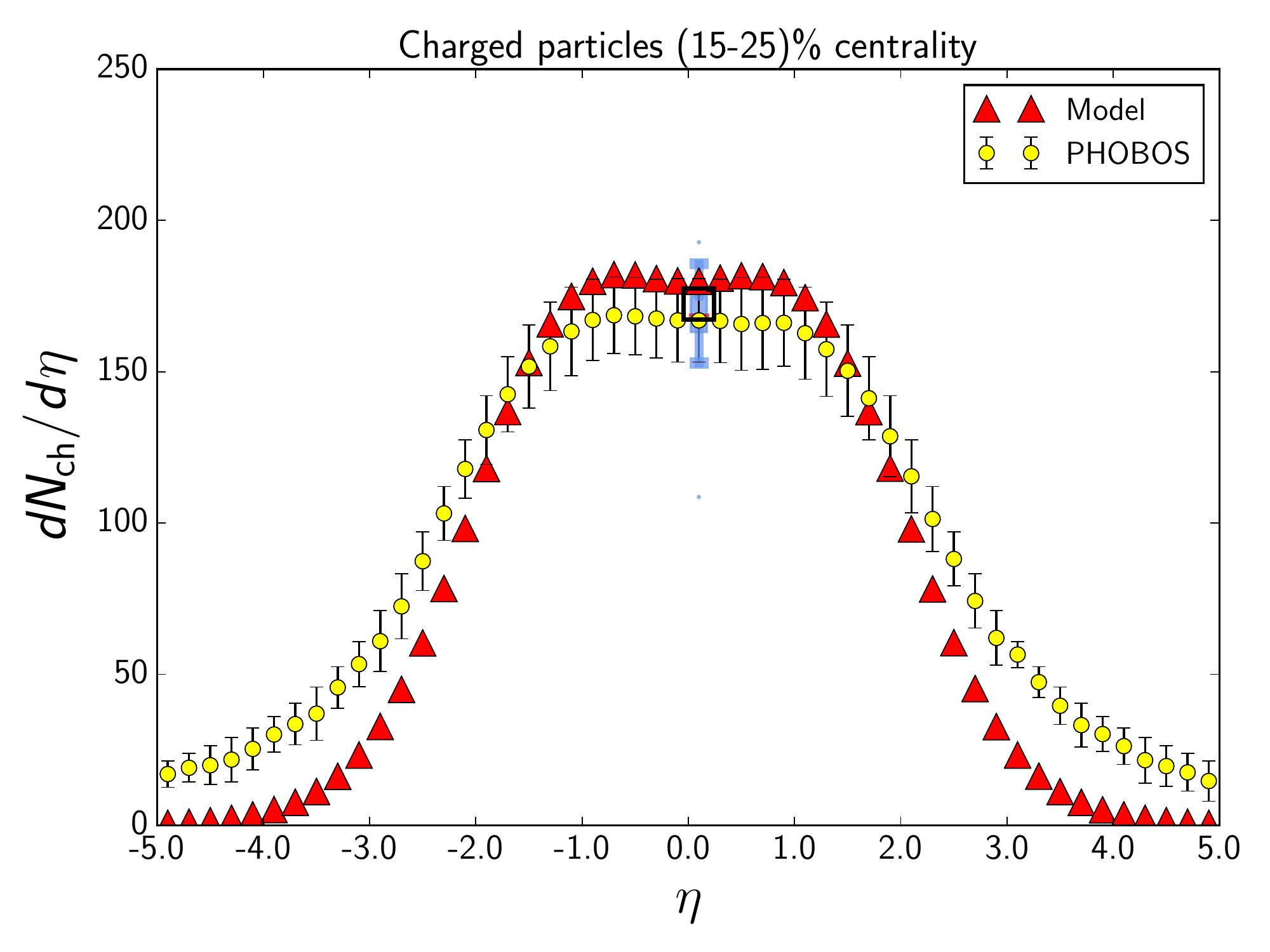}
\caption{(Color online) Charged particle pseudorapidity distribution at $\sqrt{s_{NN}}=19.6$ GeV for (0-6)\% centrality (top)
and (15-25)\% centrality (bottom) from the simulations using the posterior peak values for model parameters.
Blue boxes with whiskers represent the range of Gaussian process emulator predictions for 100 samples from the posterior.
Black open squares are emulator predictions for the model output at distribution median values (red triangles).
PHOBOS data from \cite{Alver:2010ck}.}
\label{fig:dndeta19}
\end{figure}

\begin{figure}[htb]
\includegraphics[width=8cm]{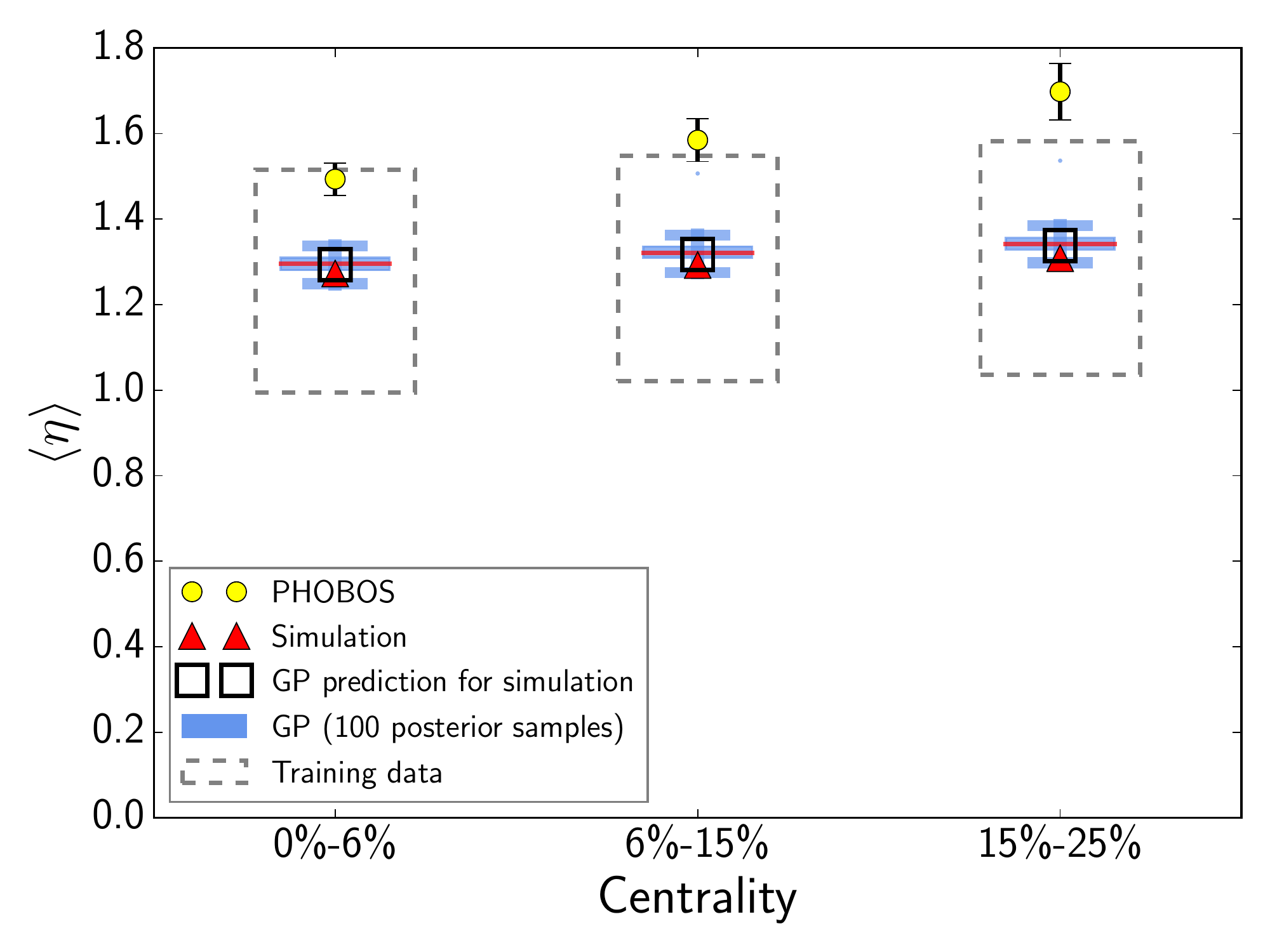}
\caption{(Color online) Centrality dependence of average charged particle pseudorapidity $\langle \eta \rangle$ for $\eta > 0.0$ at $\sqrt{s_{NN}}=19.6$ GeV.
Blue box-whisker plots are GP estimates for samples from the posterior distribution.
Black open squares are emulator predictions for the model output at distribution median values (red triangles).
Dashed grey boxes represent the range of values from the training data.
Experimental values are estimated from PHOBOS data \cite{Alver:2010ck}.}
\label{fig:meaneta19}
\end{figure}

We can investigate the quality of statistical analysis in more detail by looking the results for each observable in turn.
In Figs.~\ref{fig:dndeta19} and \ref{fig:meaneta19},
results from simulations using the posterior median values are shown for the charged particle pseudorapidity distributions
$dN_{\text{ch}}/d\eta$ for (0-6)\% and (15-25)\% centrality at $\sqrt{s_{NN}}=19.6$ GeV.
For the purposes of the statistical analysis, the pseudorapidity distribution was reduced to two descriptive numbers;
the value of $dN/d\eta$ at $\eta=0.1$ bin, and the average $\langle \eta \rangle$ over the positive part of the distribution.

We can observe from Fig.~\ref{fig:dndeta19} that the midrapidity model-to-data agreement is very good when using the posterior median.
Emulator predictions (black open boxes) appear to underestimate the real simulation result (red triangles).
As will become clear, the pseudorapidity distribution is one of the rare observables where the emulator prediction is systematically off,
although still reasonably close to the true value.

At larger pseudorapidities, the longitudinal expansion is systematically underestimated, as Fig.~\ref{fig:meaneta19} clearly illustrates.
The range of values over all training points are represented as dashed boxes in the figure,
which lies below the $\langle \eta \rangle$ value calculated from experimental data;
thus, it would not be possible to fully match the PHOBOS data with the currently used input parameter priors.

The recently published STAR data on identified particle yields and mean transverse momenta \cite{Adamczyk:2017iwn}
finally makes it possible to rigorously test the bulk production capabilities of relativistic heavy ion collision models
over the whole beam energy scan range.
However, as the proton and antiproton data suffers from additional uncertainties related to feed-down corrections,
we limit ourselves to pion and kaon data in this analysis.

\begin{figure}[htb]
\includegraphics[width=8cm]{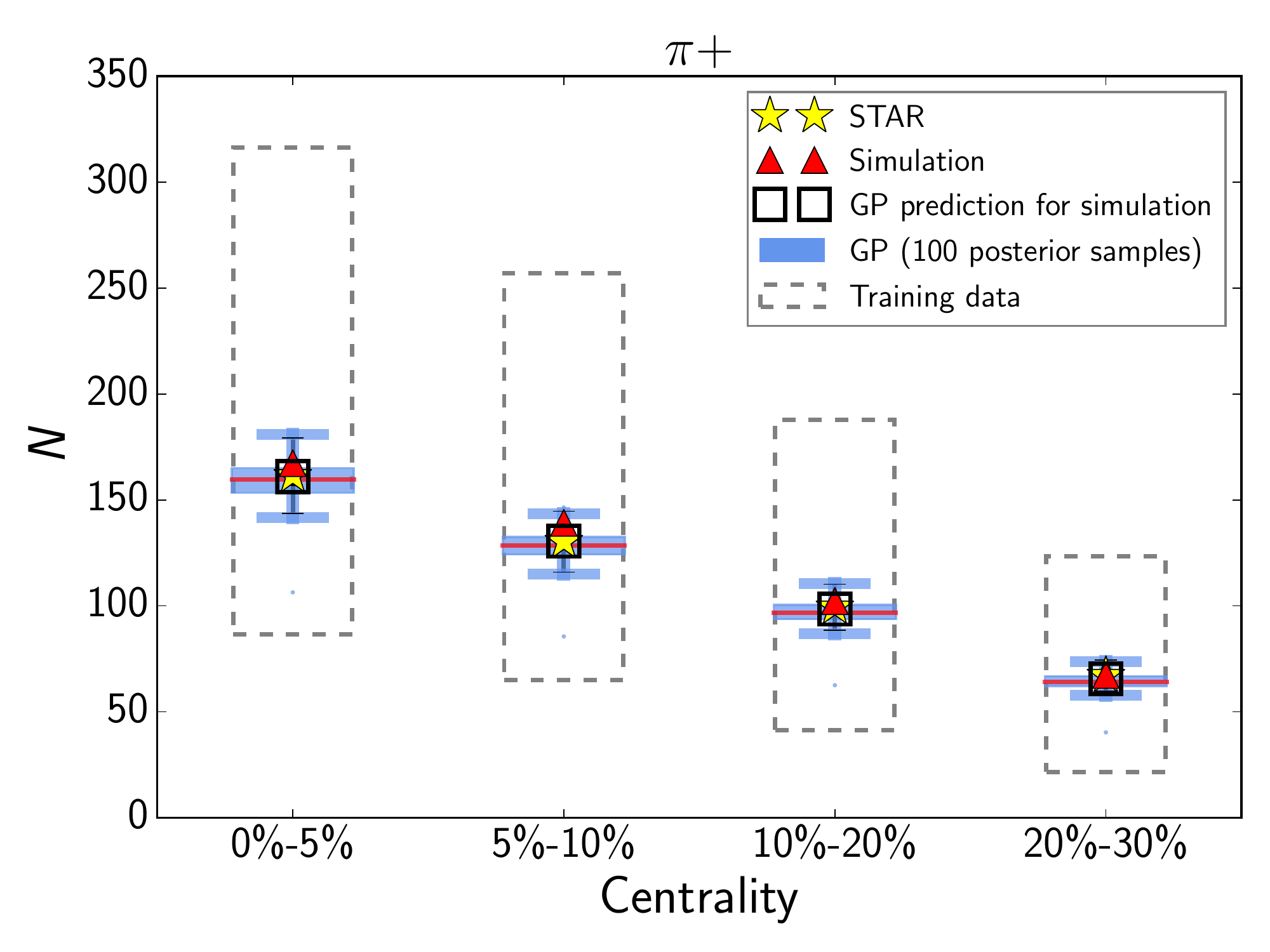}
\includegraphics[width=8cm]{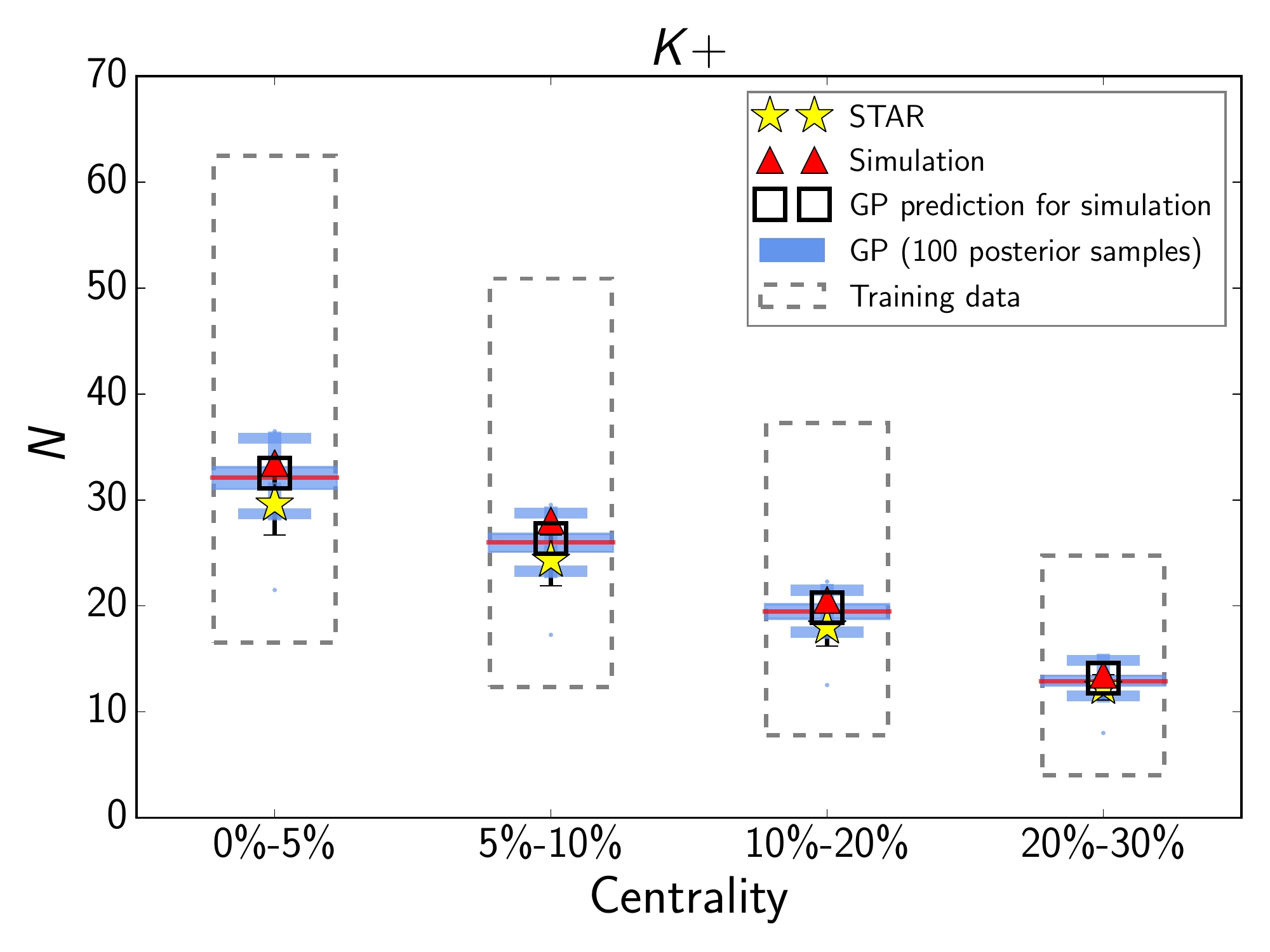}
\caption{(Color online) Multiplicities of $\pi^+$ and $K^+$ at midrapidity as a function of centrality at $\sqrt{s_{NN}}=19.6$ GeV.
Experimental data from \cite{Adamczyk:2017iwn}.}
\label{fig:nid19}
\end{figure}

The results for positive-charge pion and kaon multiplicities are shown in Fig.~\ref{fig:nid19}.
Pion yields are well reproduced for the investigated centrality range.
There is a tendency to overestimate kaon yields, especially at the more central collisions.
The median value result is still within statistical uncertainties, however.

\begin{figure}[htb]
\includegraphics[width=8cm]{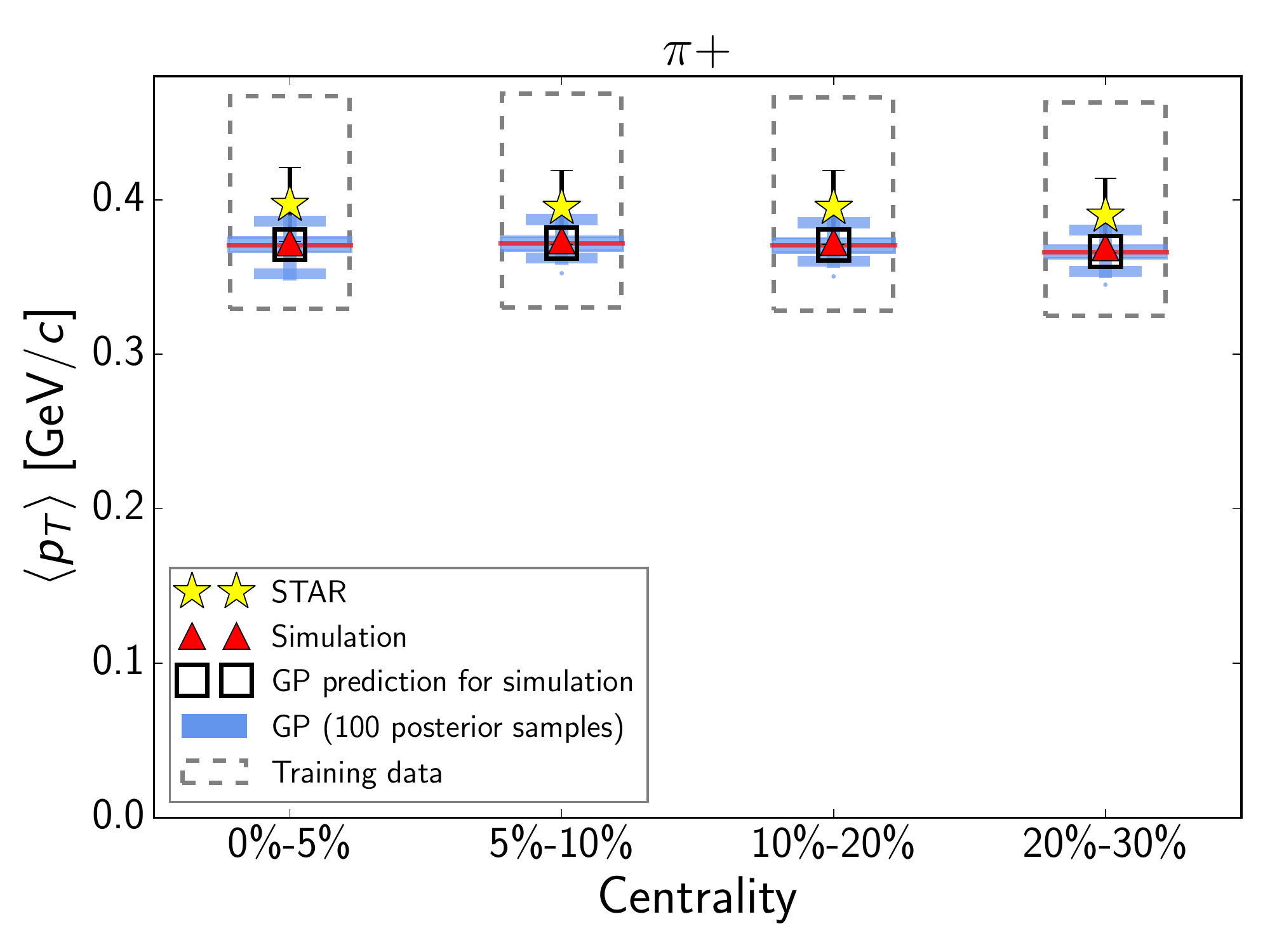}
\includegraphics[width=8cm]{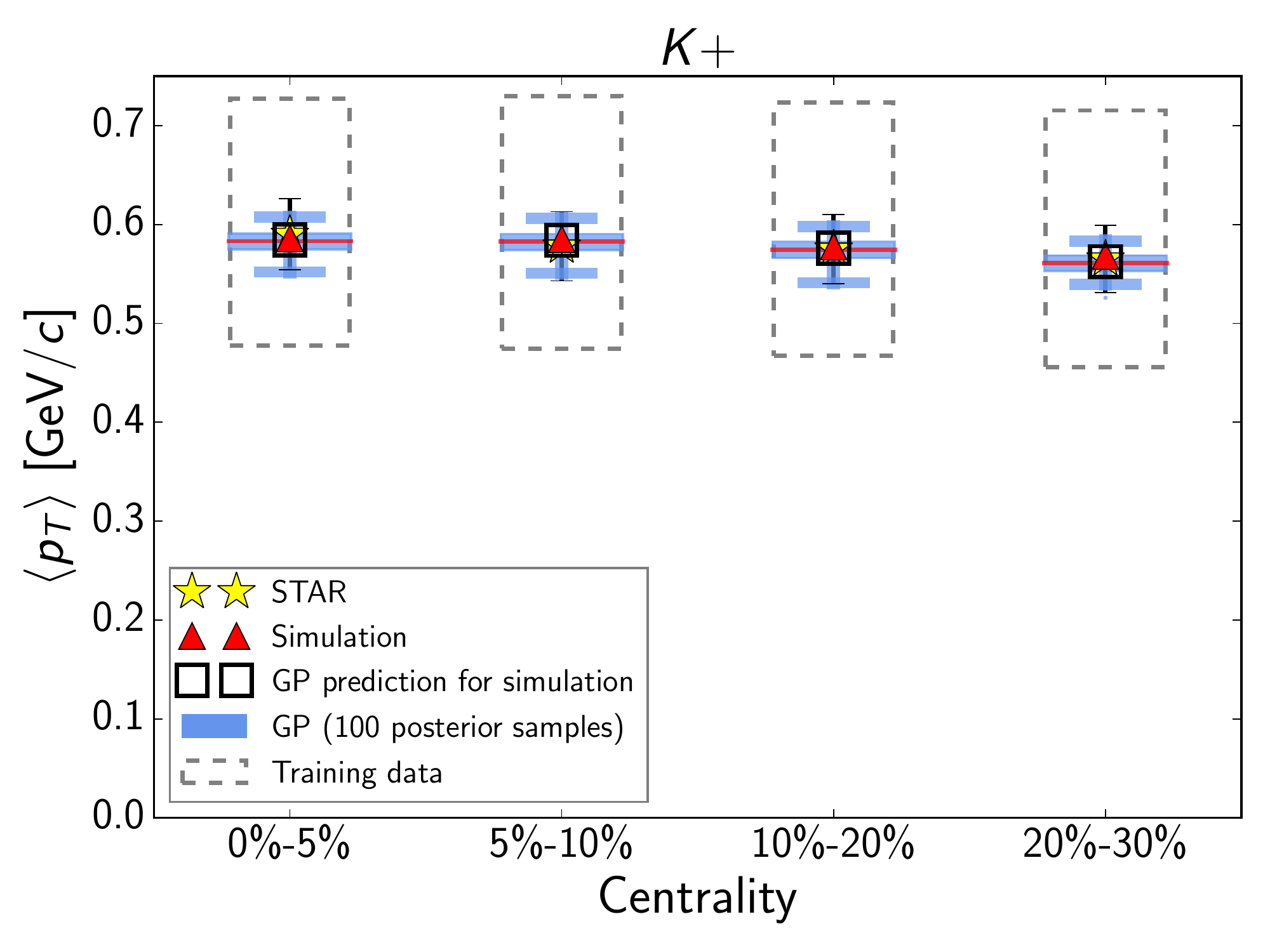}
\caption{(Color online) Mean transverse momentum of $\pi^+$ and $K^+$ as a function of centrality at $\sqrt{s_{NN}}=19.6$ GeV.
Experimental data from \cite{Adamczyk:2017iwn}.}
\label{fig:meanpt19}
\end{figure}

\begin{figure}[htb]
\includegraphics[width=8cm]{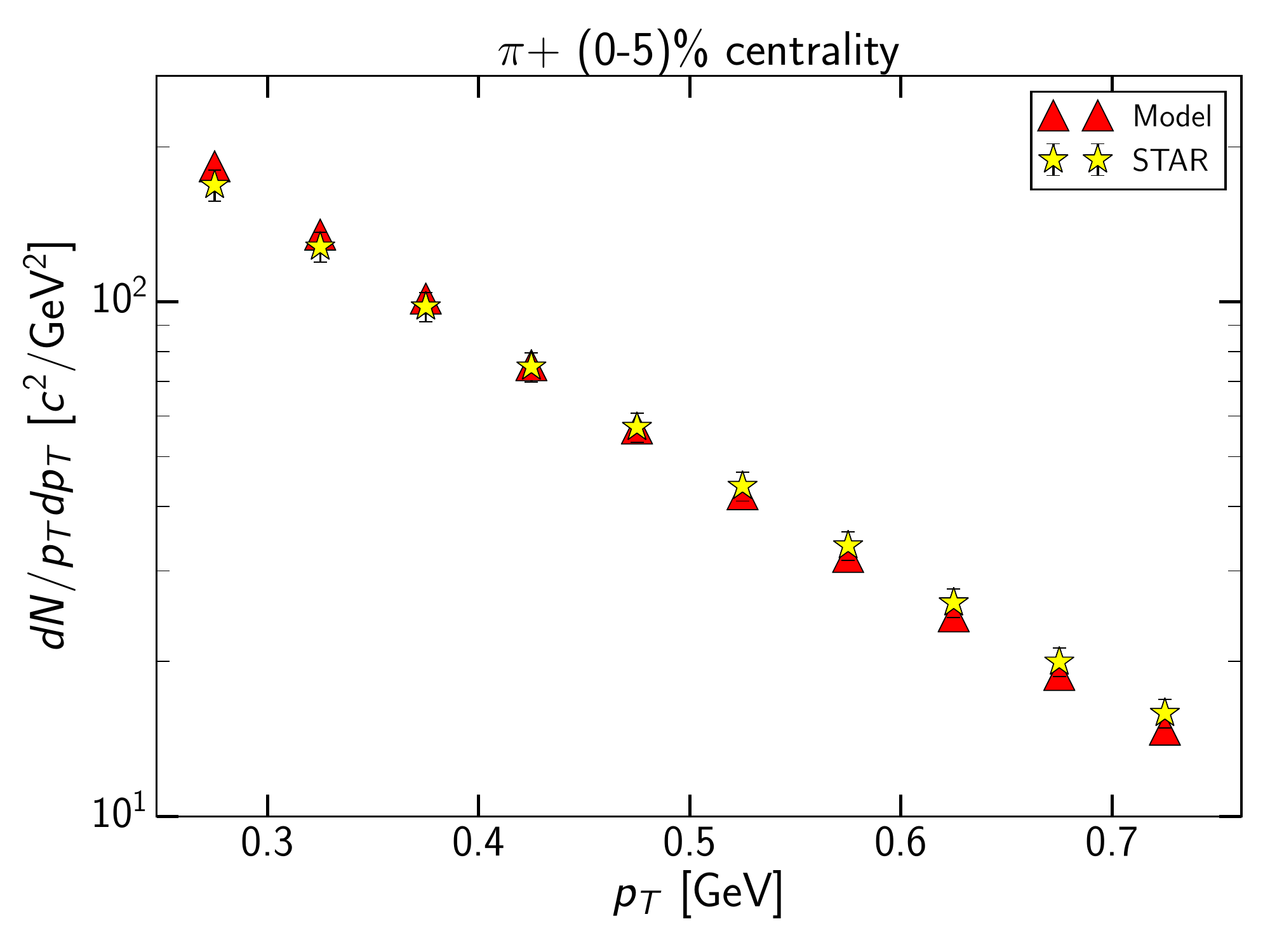}
\includegraphics[width=8cm]{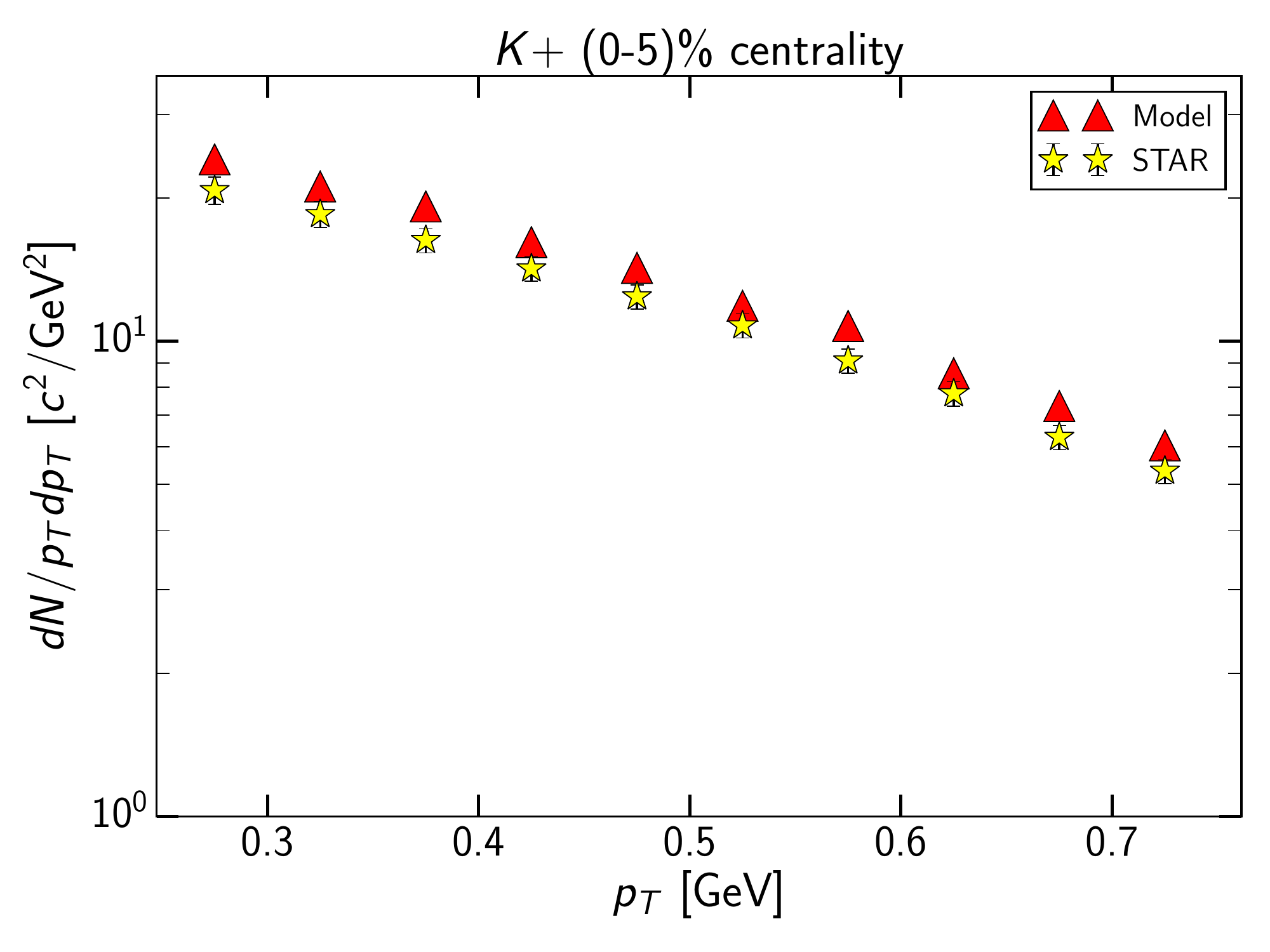}
\caption{(Color online) Transverse momentum spectra of $\pi^+$ and $K^+$ for (0-5)\% centrality at $\sqrt{s_{NN}}=19.6$ GeV.
Experimental data from \cite{Adamczyk:2017iwn}.}
\label{fig:dndptid19}
\end{figure}

To convince us that the model has captured the correct dynamical evolution of the system,
it is important to check not only particle yields but also the transverse momentum spectra $dN/dp_T$.
It was postulated in Ref.~\cite{Novak:2013bqa} that the combination of integrated yield and mean $p_T$ is sufficient to describe the $p_T$ spectrum.
Results for mean transverse momenta of $\pi^+$ and $K^+$ are shown in Fig.~\ref{fig:meanpt19}.
The pion transverse momenta is slightly below reported value for all centralities,
but still within error bars, whereas the agreement with kaon data is very good.
Overall pion and kaon spectra are well reproduced, as shown in Fig.~\ref{fig:dndptid19},
although the kaon excess in the most central collisions is evident also on this figure.

\begin{figure}[htb]
\includegraphics[width=6cm]{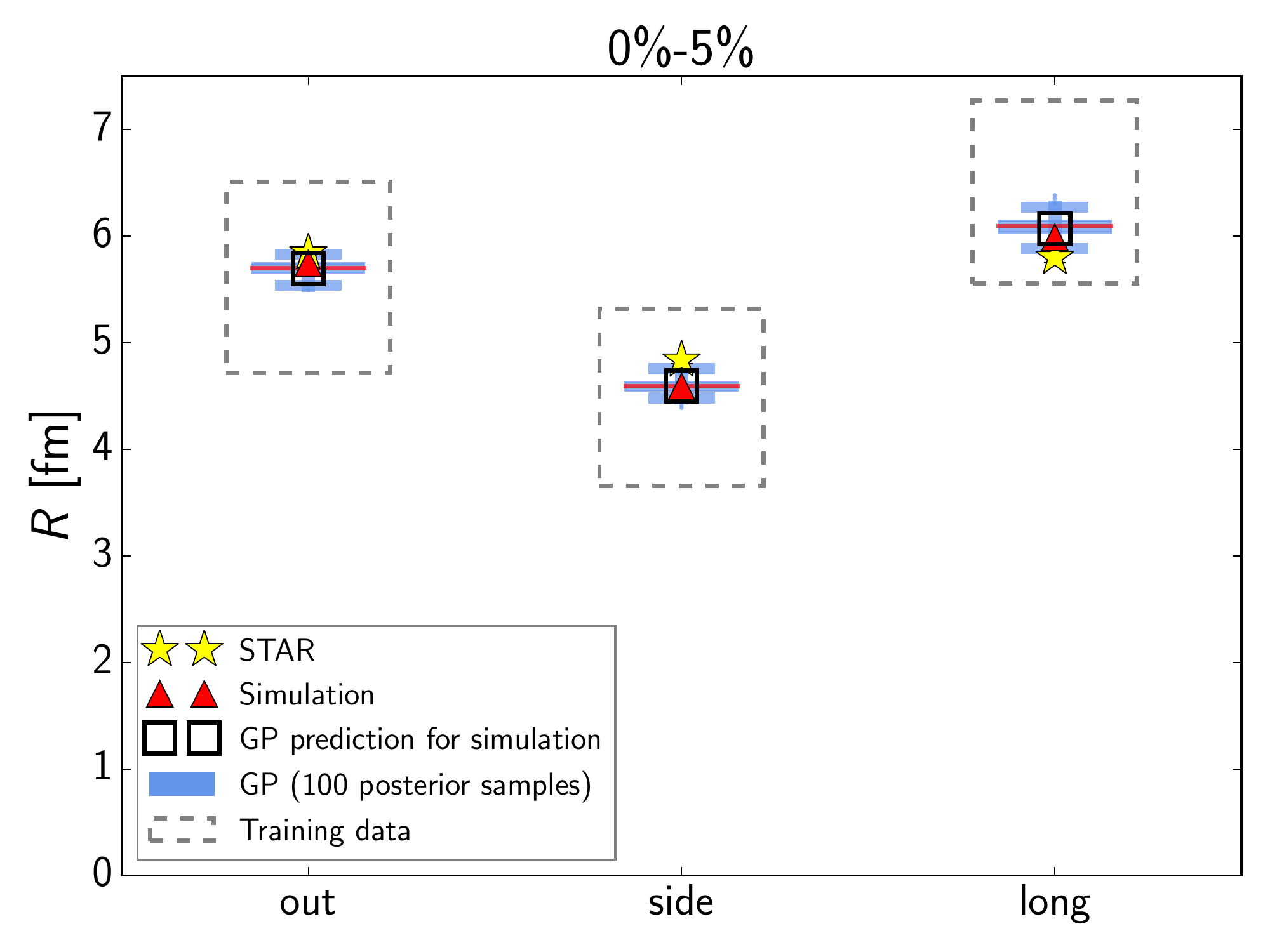}
\includegraphics[width=6cm]{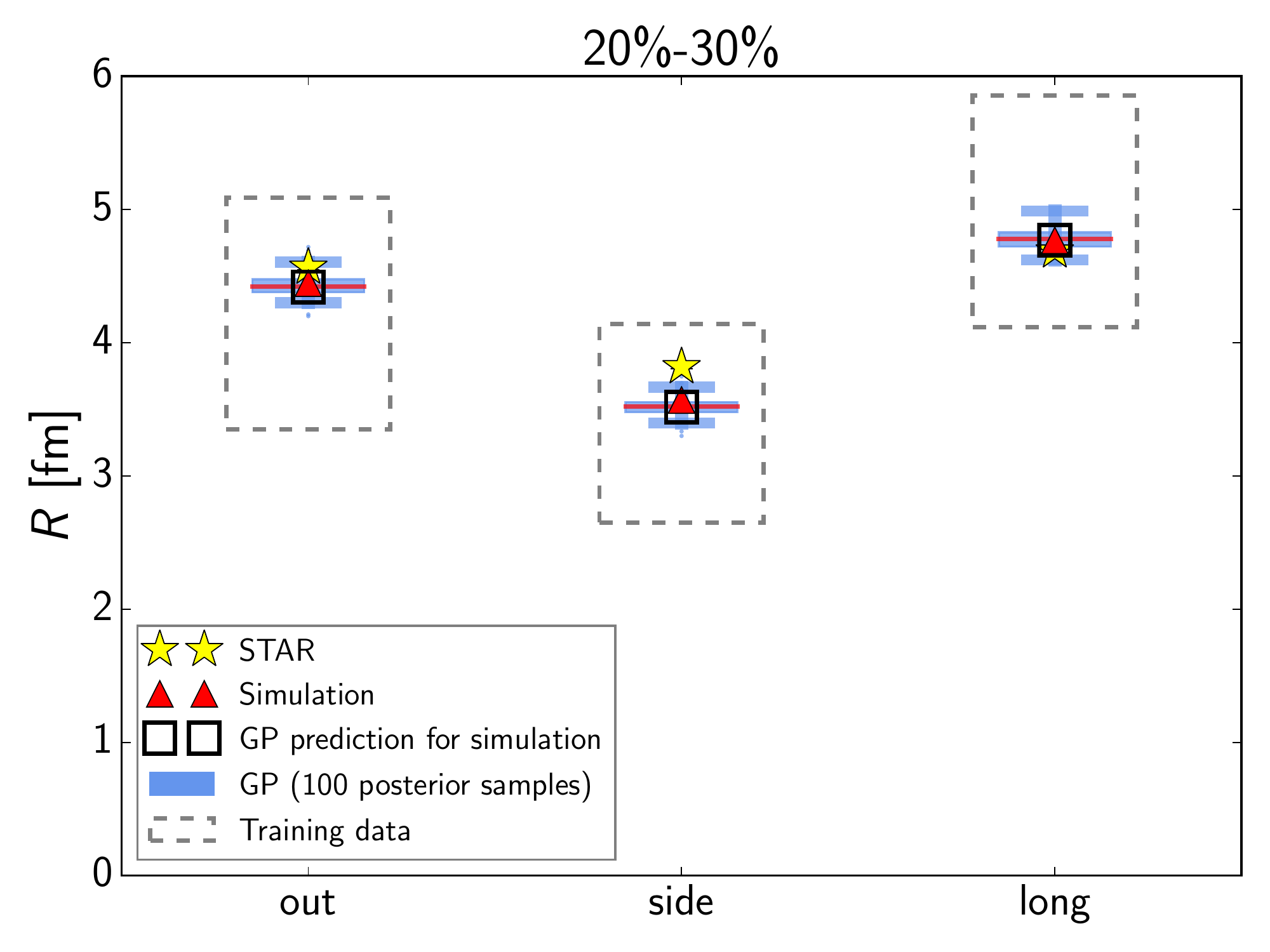}
\caption{(Color online) $R_{\text{out}}$, $R_{\text{side}}$, $R_{\text{long}}$ of charged $\pi$ \mbox{at $\langle k_T \rangle \approx 0.22$ GeV/$c$}
at (0-5)\% centrality and (20-30)\% centrality at $\sqrt{s_{NN}}=19.6$ GeV.
Experimental data from \cite{Adamczyk:2014mxp}.}
\label{fig:hbt_onebin_19}
\end{figure}

\begin{figure}[htb]
\includegraphics[width=6cm]{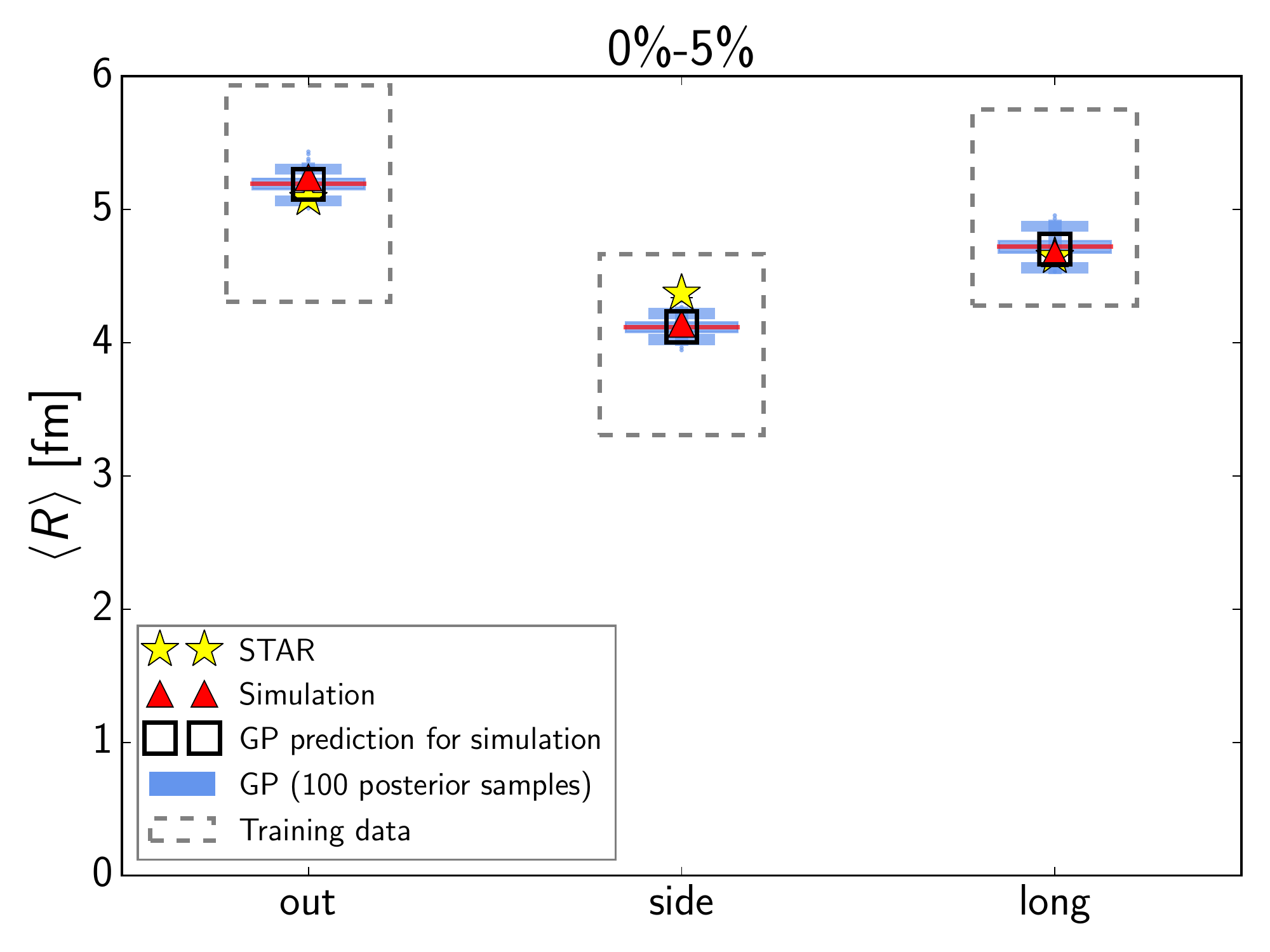}
\includegraphics[width=6cm]{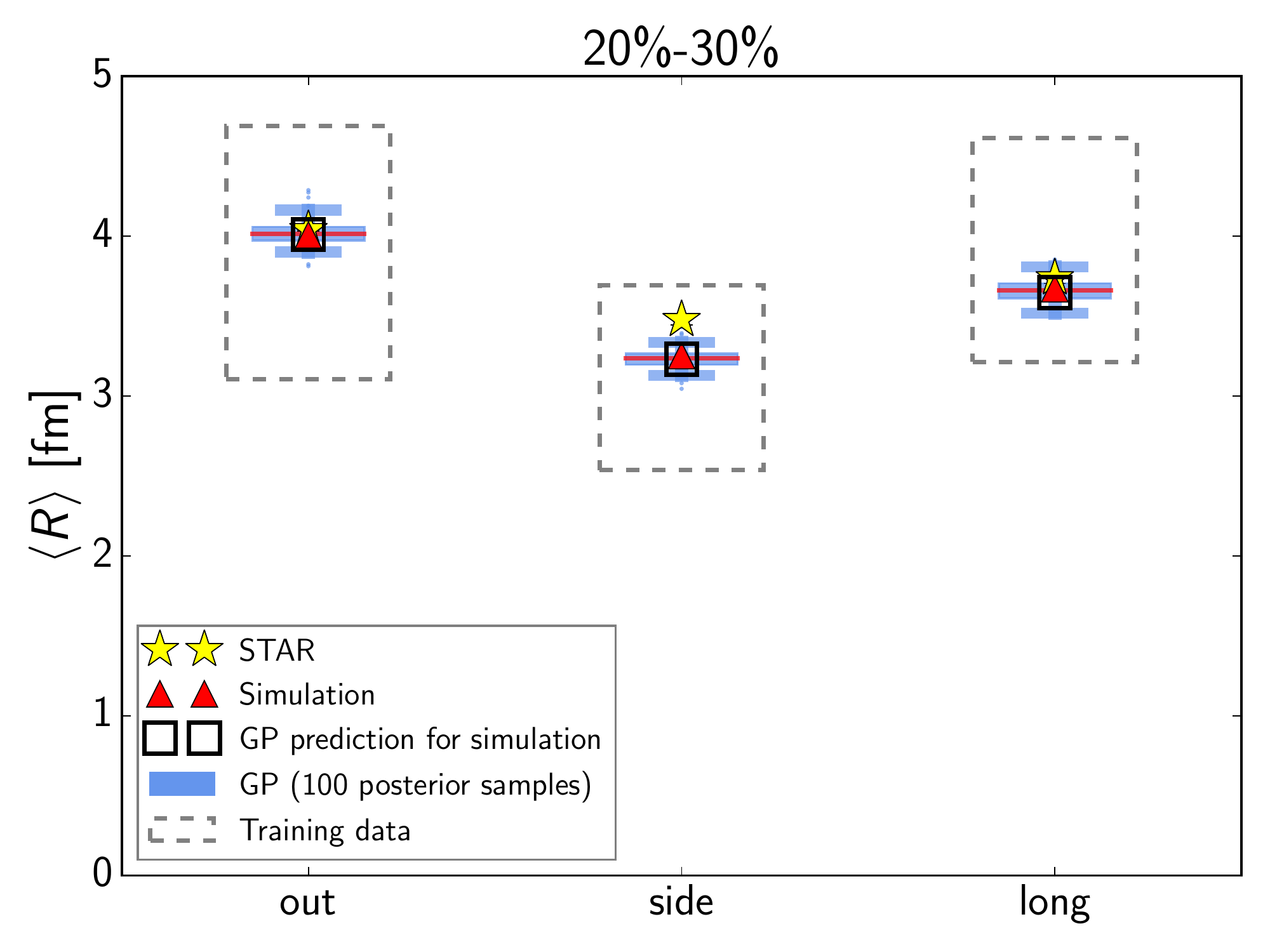}
\caption{(Color online) $R_{\text{out}}$, $R_{\text{side}}$, $R_{\text{long}}$ of charged $\pi$ averaged over 0.15 GeV/$c$ $<k_T<$ 0.6 GeV/$c$
at (0-5)\% centrality and (20-30)\% centrality at $\sqrt{s_{NN}}=19.6$ GeV.
Experimental data from \cite{Adamczyk:2014mxp}.}
\label{fig:hbt_mean_19}
\end{figure}

Two-pion interferometry, commonly referred as ``Hanbury-Brown-Twiss analysis'',
is considered as a probe of the spacetime geometry of the kinetic freezeout region \cite{Lisa:2005dd}.
Results for the HBT radii are summarized in Fig.~\ref{fig:hbt_onebin_19} for the lowest $\langle k_T \rangle$ bin,
which corresponds to the observable used for the comparison over several different experiments in Ref.~\cite{Adamczyk:2014mxp}.
Again following the example of Ref.~\cite{Novak:2013bqa},
we also investigate mean radii averaged over several $k_T$ bins in Fig.~\ref{fig:hbt_mean_19}.

Using the median values for input parameters, the model is able to match $R_{\text{out}}$ and $R_{\text{long}}$ quite well,
but $R_{\text{side}}$ is systematically underestimated.
The same is true for the averaged radii.
Emphasizing the importance of HBT radii has thus potential to impose strong constraints on the input parameters.

\begin{figure}[htb]
\includegraphics[width=8cm]{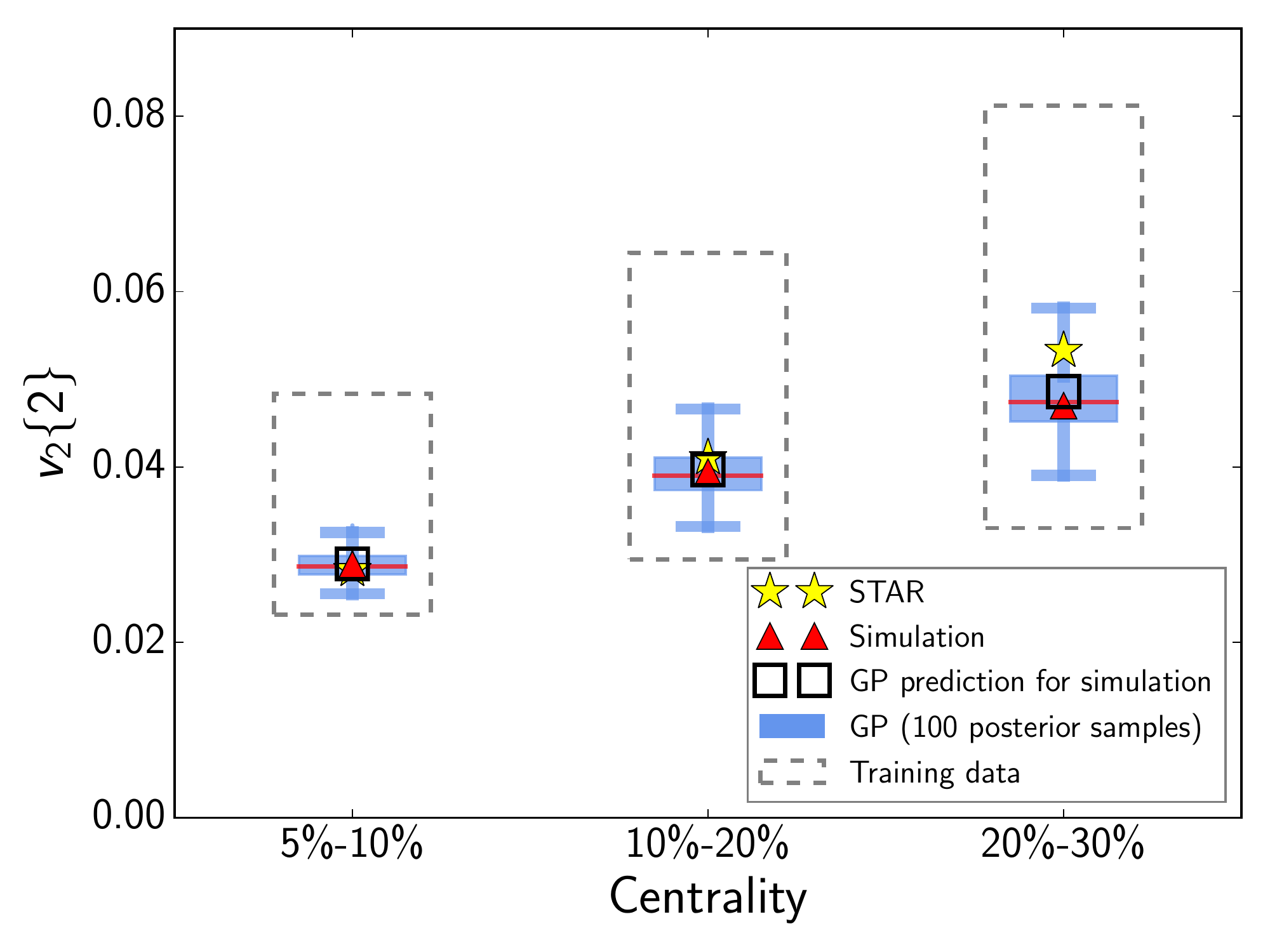}
\caption{(Color online) Centrality dependence of charged particle  $v_2\{2\}$ at $\sqrt{s_{NN}}=19.6$ GeV.
Experimental data from \cite{Adamczyk:2012ku}.}
\label{fig:v2219}
\end{figure}

The centrality dependence of $v_2\{2\}$ for charged particles is presented in Fig.~\ref{fig:v2219}.
Agreement with the STAR data is very good for the (5-10)\% and (10-20)\% centralities;
in the (20-30)\% centrality bin, the median input parameters produce slightly less elliptic flow compared to observations,
which may indicate an issue using the same value of $\tau_0$ for all centralities.
The lower limit of $\tau_0$, the time for the two nuclei to pass each other, becomes smaller at higher centralities,
in principle enabling earlier hydro starting times.
This might also effect $W_{\text{long}}$ which tends to have strong correlation with $\tau_0$.
We leave the investigation of the effect of centrality-dependent $\tau_0$ for future work.

\begin{figure}[htb]
\includegraphics[width=8cm]{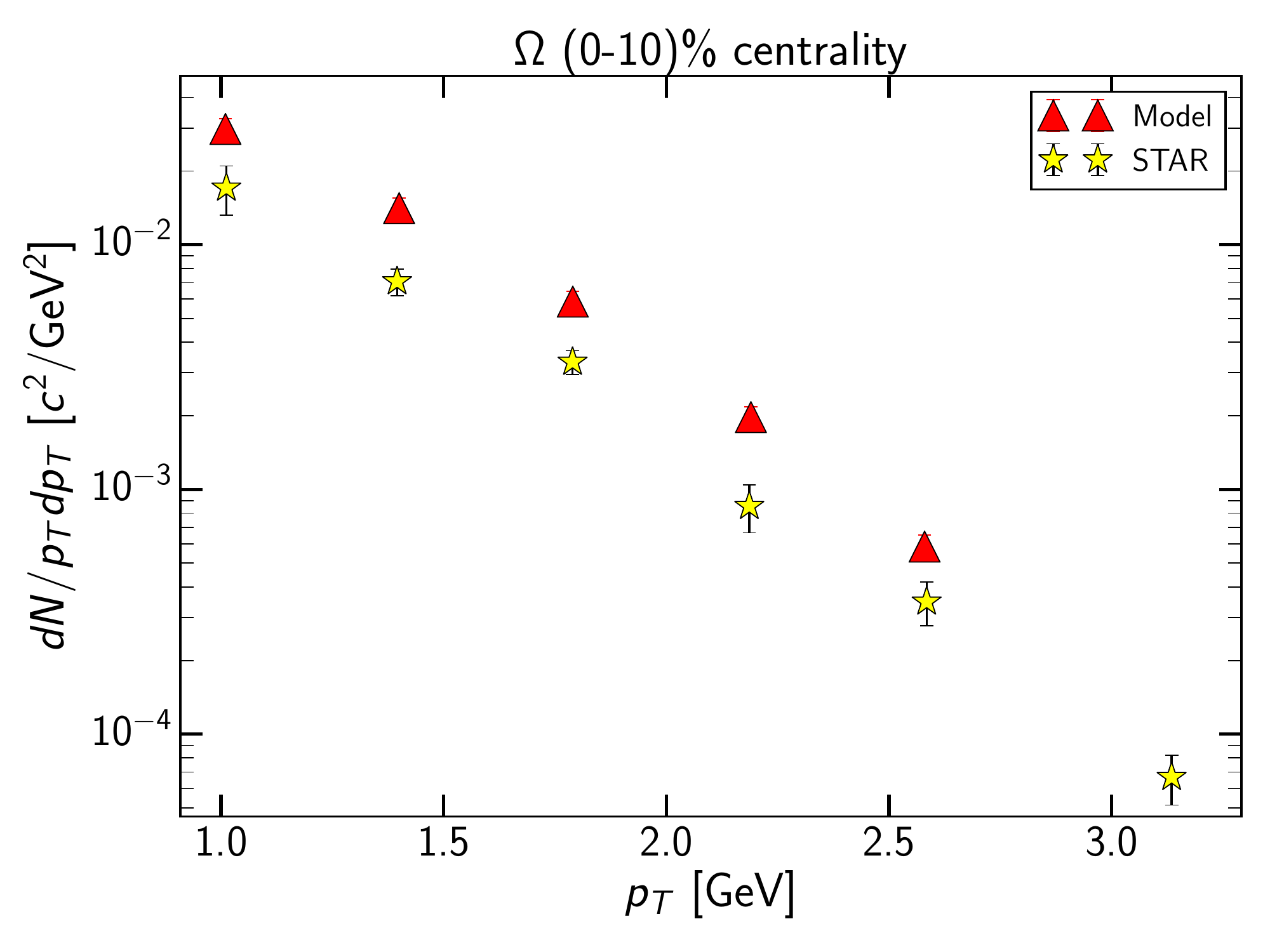}
\caption{(Color online) Transverse momentum spectrum of $\Omega$ at $\sqrt{s_{NN}}=19.6$ GeV.
Experimental data from \cite{Adamczyk:2015lvo}.}
\label{fig:dndptomega_19}
\end{figure}

Finally, we compare the transverse momentum spectrum of $\Omega$ against the STAR data in Fig.~\ref{fig:dndptomega_19}.
It is evident from the figure that $\Omega$ yield is overestimated when using the median values of the posterior distribution.
However, the fit here and at 39 GeV was done using the $p_T=1.01$ GeV/$c$ data point
and the estimated mean transverse momentum for the $p_T$ range 1-5 GeV/$c$,
which was calculated from a Levy fit on the $p_T$ spectra \cite{Auvinen:2016uuv}.
Based on results at 62.4 GeV (Fig.~\ref{fig:dndptomega_62}),
using the fully $p_T$-integrated yield and $\langle p_T \rangle$ is expected to produce better results.
Once such data becomes available, its effect on the posterior distribution needs be investigated.

\clearpage

\subsection{$\sqrt{s_{NN}}=39$ GeV}

The experimental data available at $\sqrt{s_{NN}}=39$ GeV is a subset of that at 19.6 GeV;
information about the pseudorapidity dependence of charged particle yields is not available.
As the earliest possible starting time remains favored, the $\tau_0$ distribution shifts below 1.0 fm,
and the other distributions also peak at lower values, in particular the switching energy density $\epsilon_{SW}$.

\begin{figure}[htb]
\includegraphics[width=8cm]{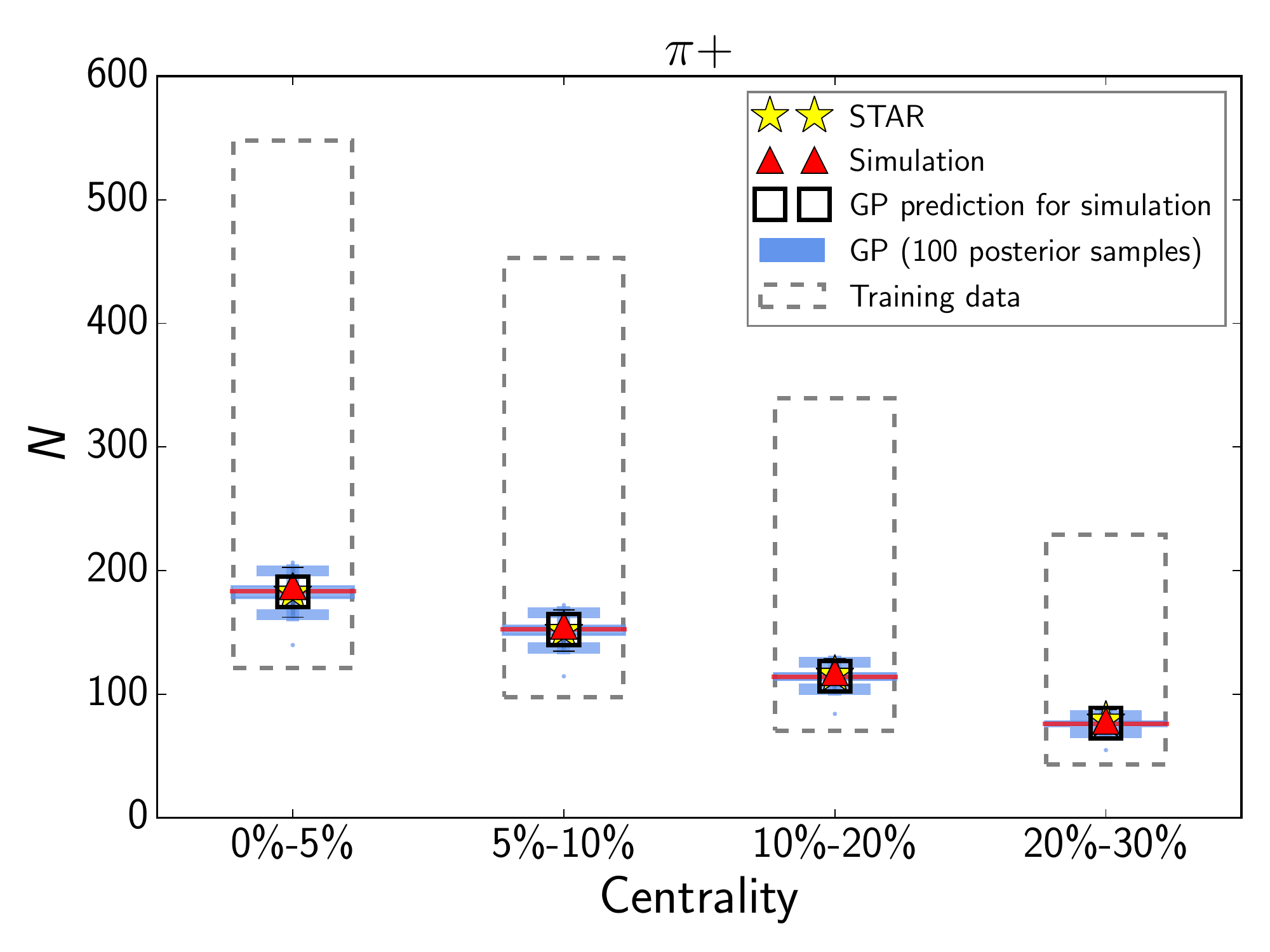}
\includegraphics[width=8cm]{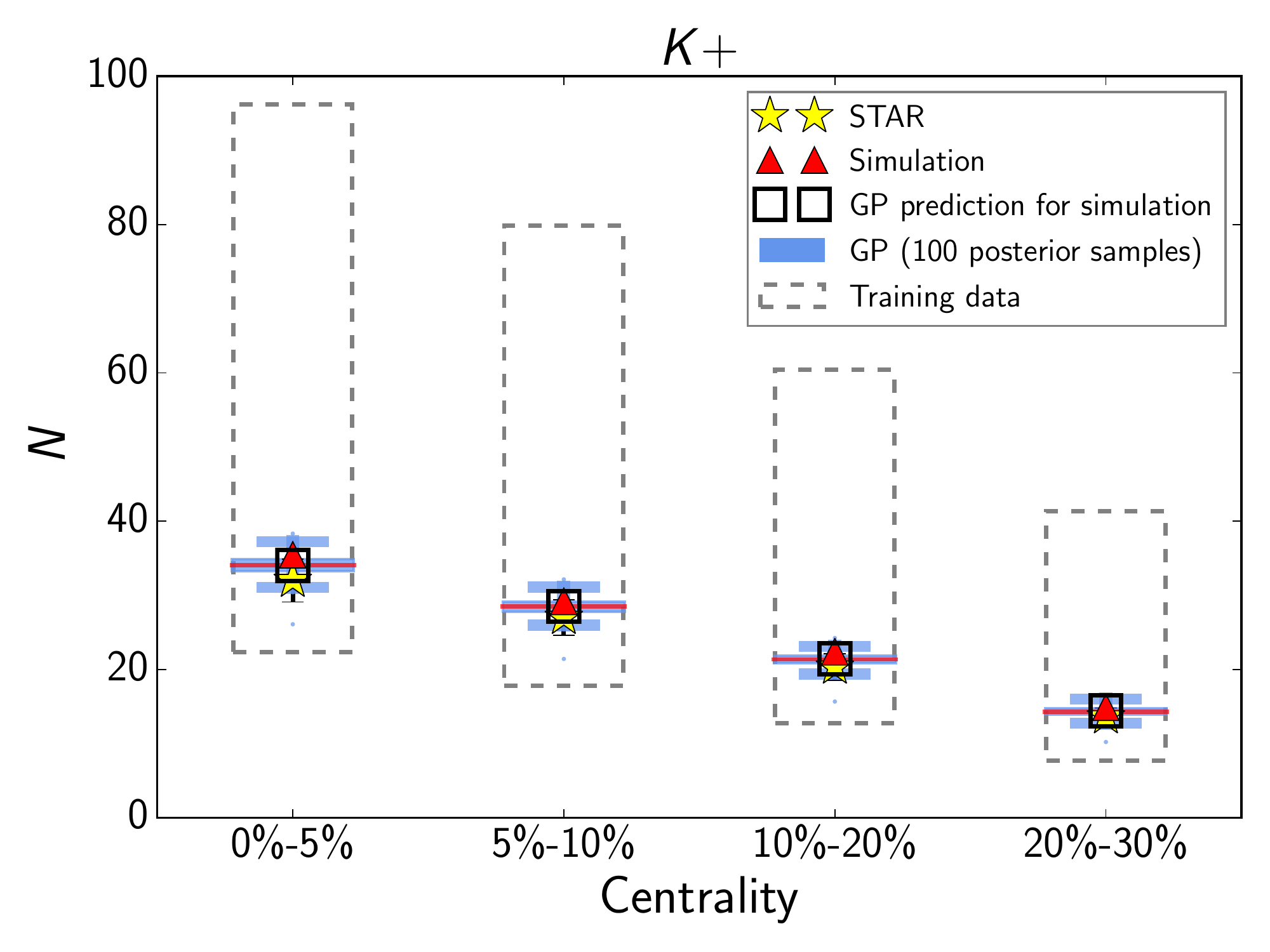}
\caption{(Color online) Centrality dependence of positive pion and kaon yields at midrapidity at $\sqrt{s_{NN}}=39$ GeV.
Experimental data from \cite{Adamczyk:2017iwn}.}
\label{fig:nid39}
\end{figure}

\begin{figure}[htb]
\includegraphics[width=8cm]{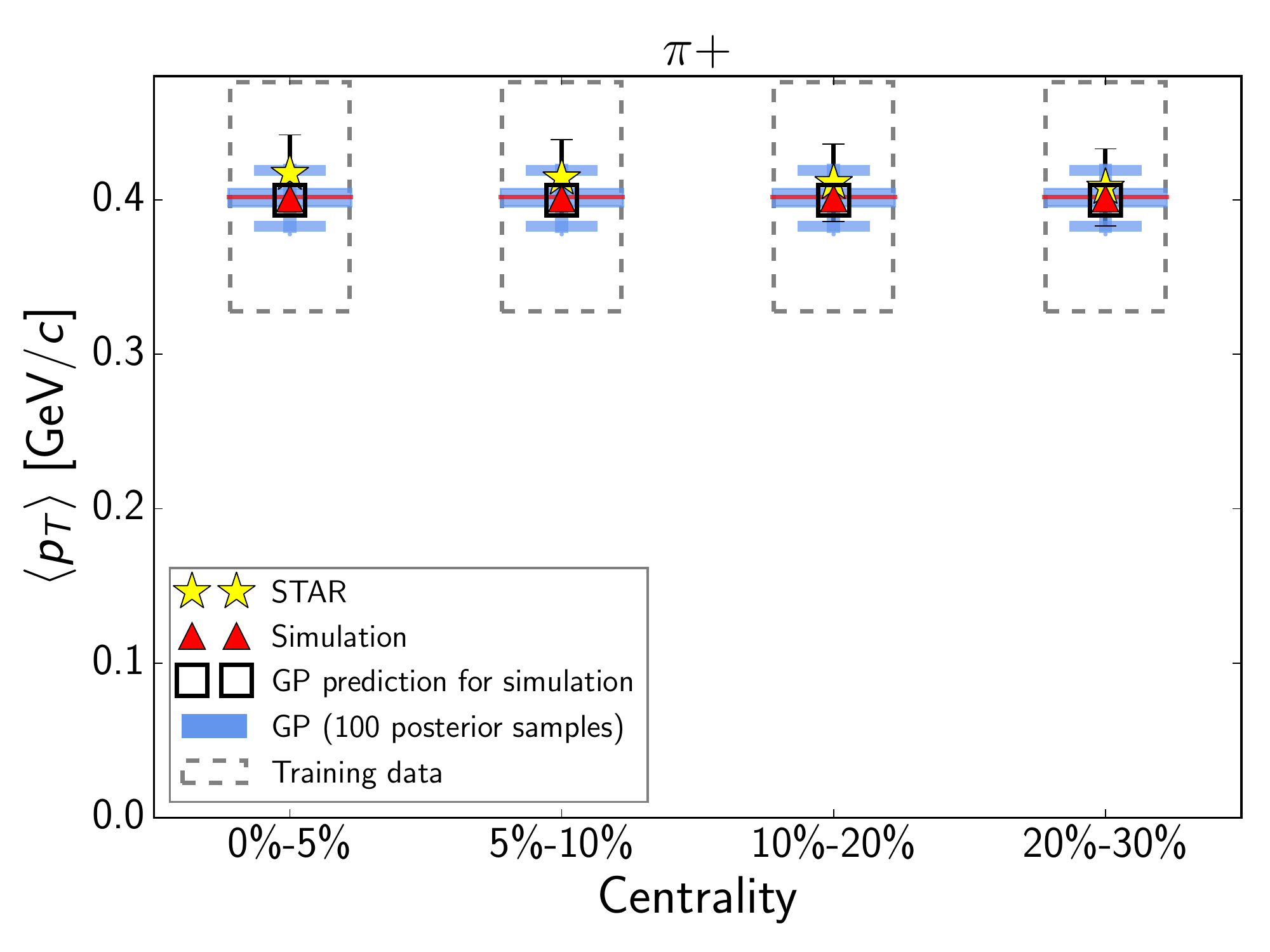}
\includegraphics[width=8cm]{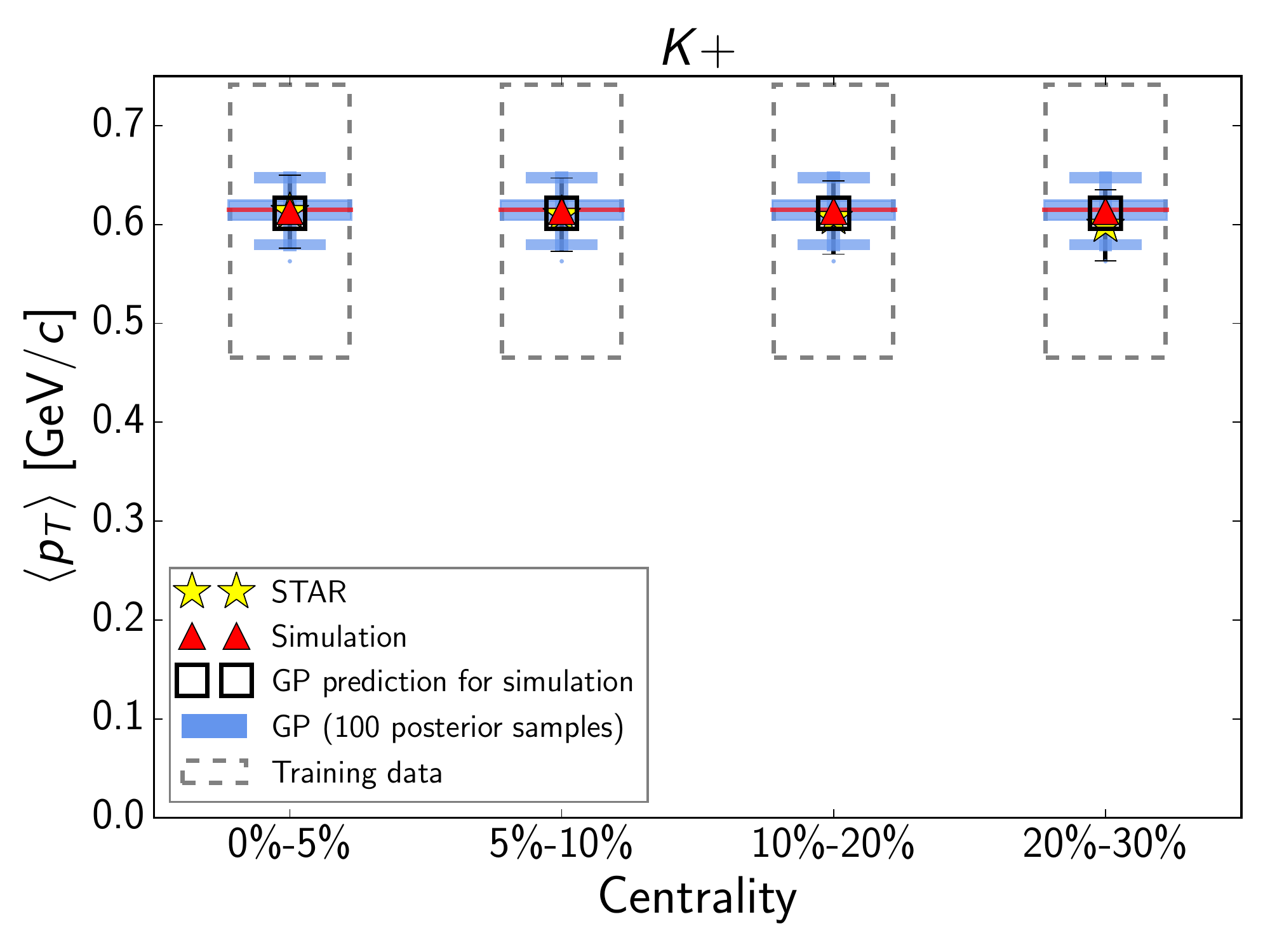}
\caption{(Color online) Centrality dependence of mean transverse momentum of $\pi^+$ and $K^+$ at $\sqrt{s_{NN}}=39$ GeV.
Experimental data from \cite{Adamczyk:2017iwn}.}
\label{fig:meanpt39}
\end{figure}

\begin{figure}[htb]
\includegraphics[width=8cm]{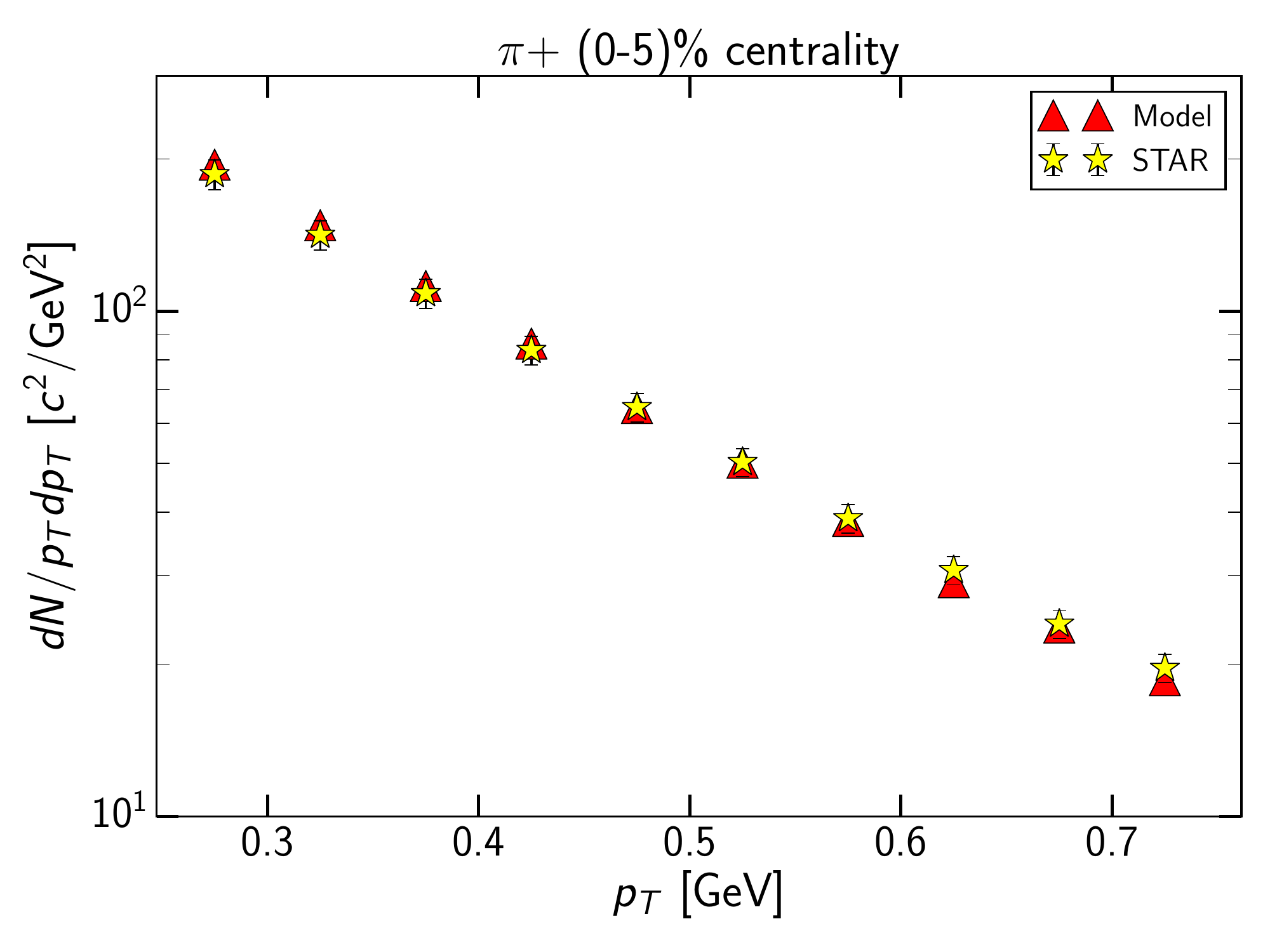}
\includegraphics[width=8cm]{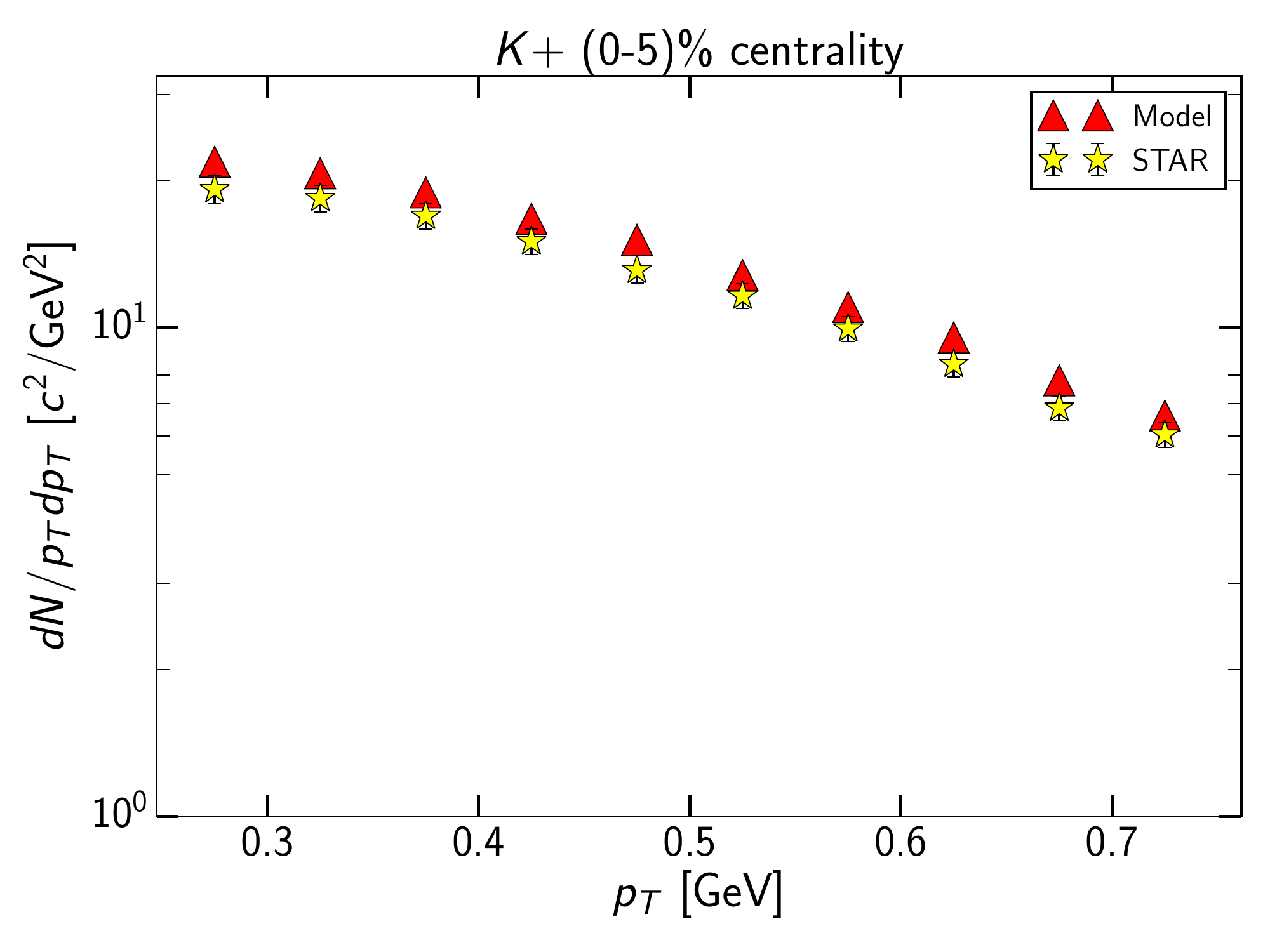}
\caption{(Color online) Transverse momentum of $\pi^+$ and $K^+$ for (0-5)\% centrality at $\sqrt{s_{NN}}=39$ GeV.
Experimental data from \cite{Adamczyk:2017iwn}.}
\label{fig:dndptid39}
\end{figure}

The results for the actual observables are very similar to $\sqrt{s_{NN}}=19.6$ GeV.
Pion and kaon yields are again well reproduced (Fig. \ref{fig:nid39}),
as is their mean $p_T$ (Fig.~\ref{fig:meanpt39}),
resulting in a good agreement with the measured transverse momentum spectra (Fig.~\ref{fig:dndptid39}).

\begin{figure}[htb]
\includegraphics[width=6cm]{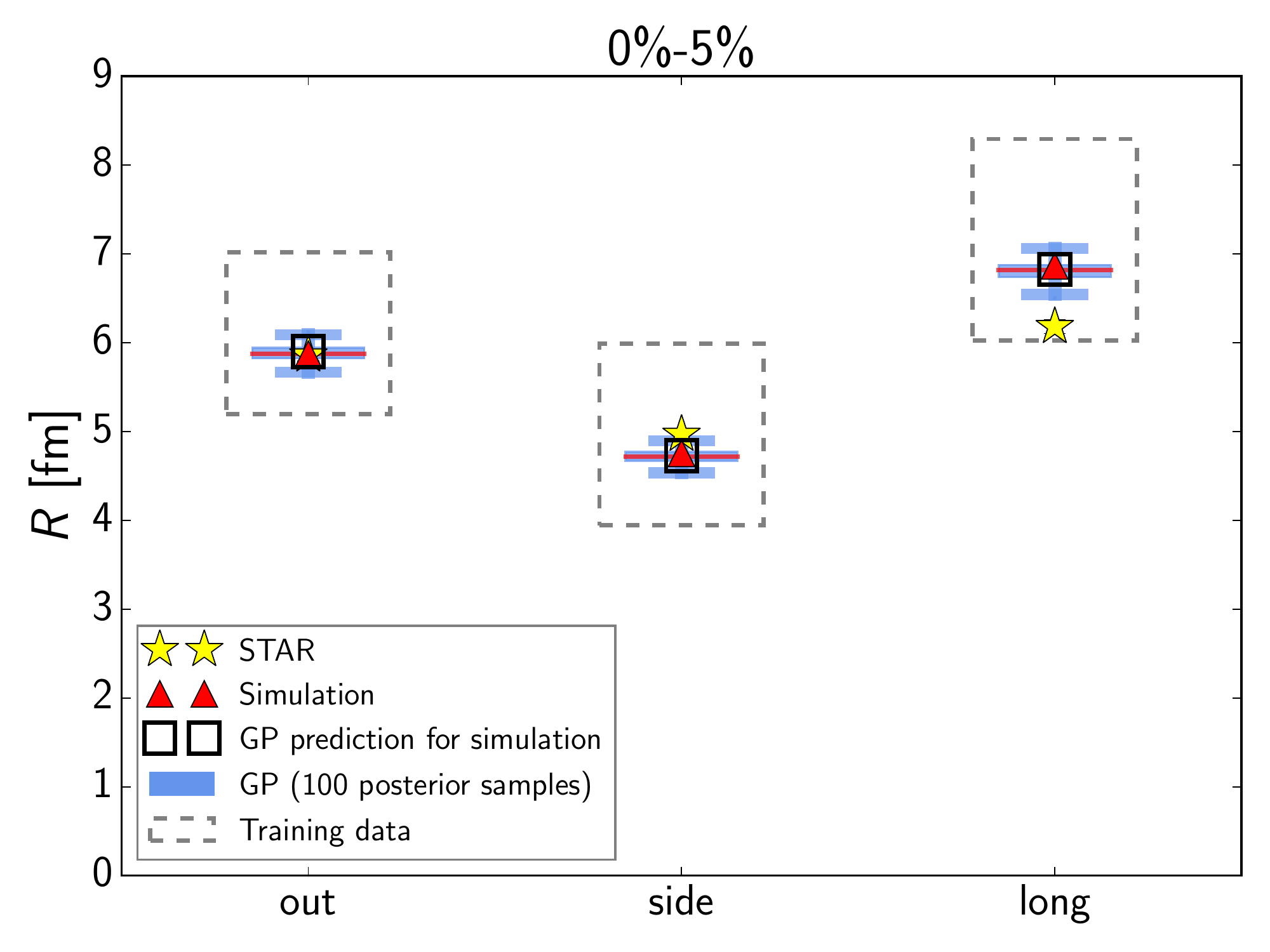}
\includegraphics[width=6cm]{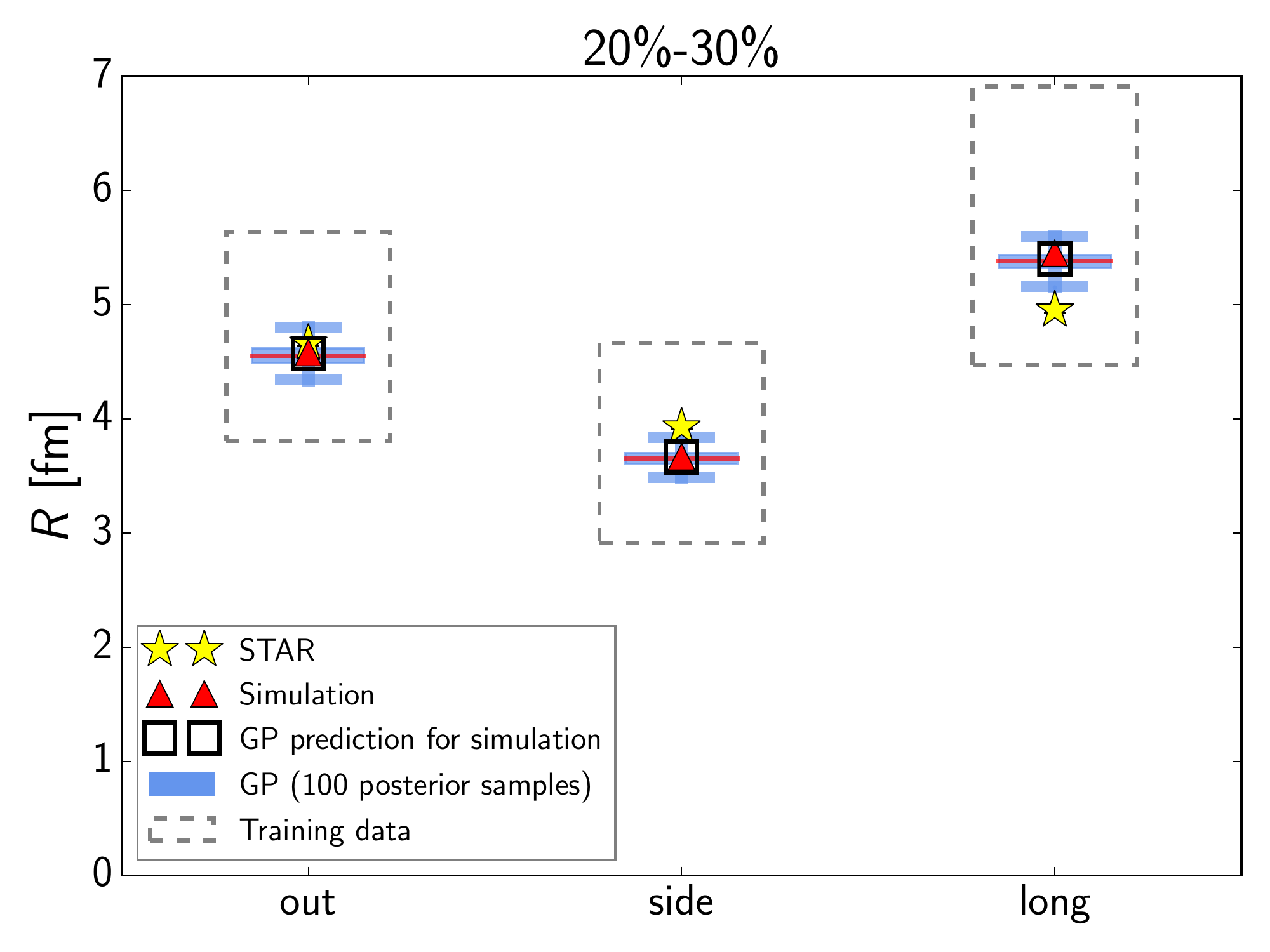}
\caption{(Color online) $R_{\text{out}}$, $R_{\text{side}}$, $R_{\text{long}}$ of charged $\pi$ \mbox{at $\langle k_T \rangle \approx 0.22$ GeV/$c$}
at (0-5)\% centrality (top) and (20-30)\% centrality (bottom) at $\sqrt{s_{NN}}=39$ GeV.
Experimental data from \cite{Adamczyk:2014mxp}.}
\label{fig:hbt_onebin_39}
\end{figure}

\begin{figure}[htb]
\includegraphics[width=6cm]{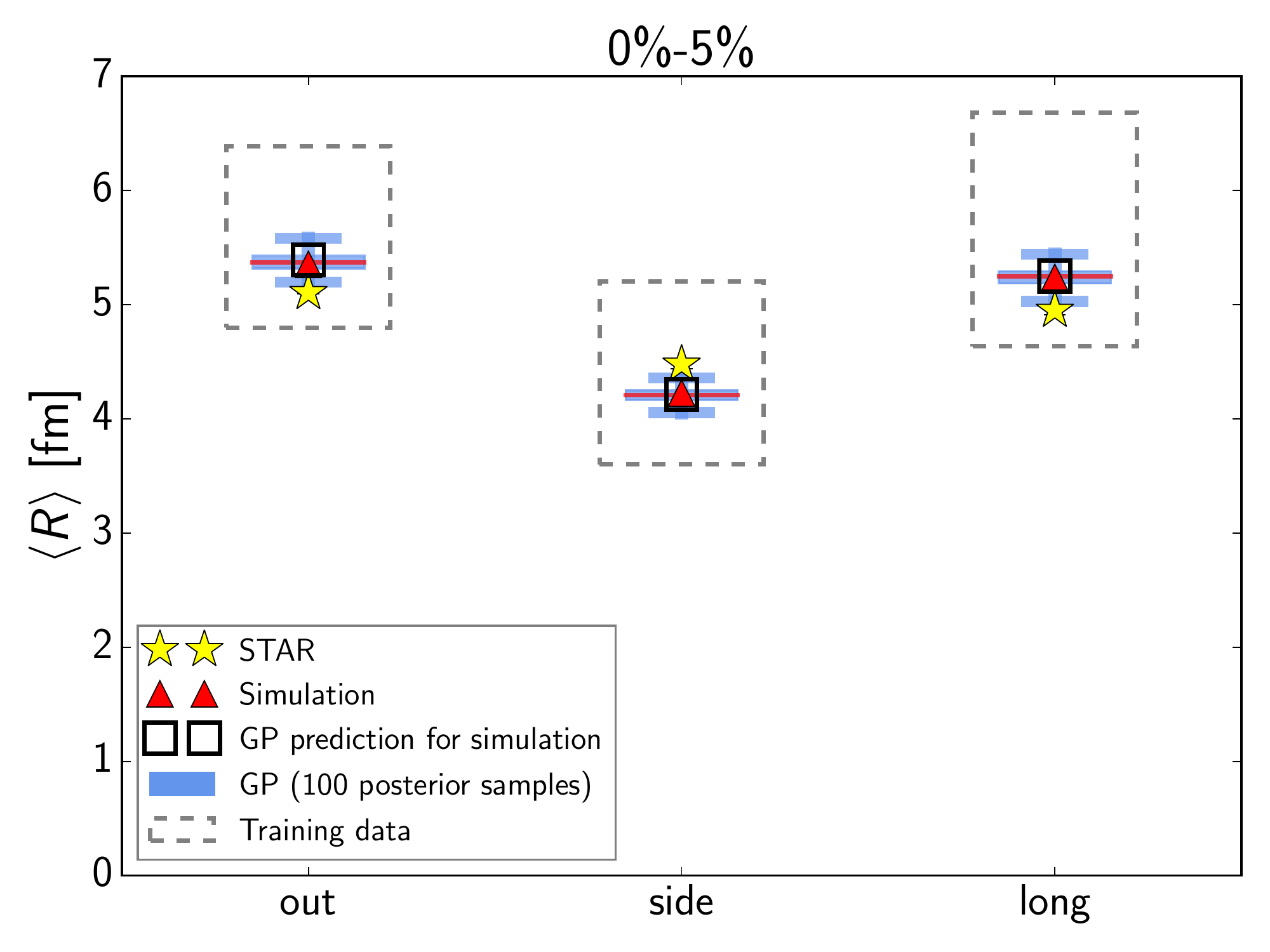}
\includegraphics[width=6cm]{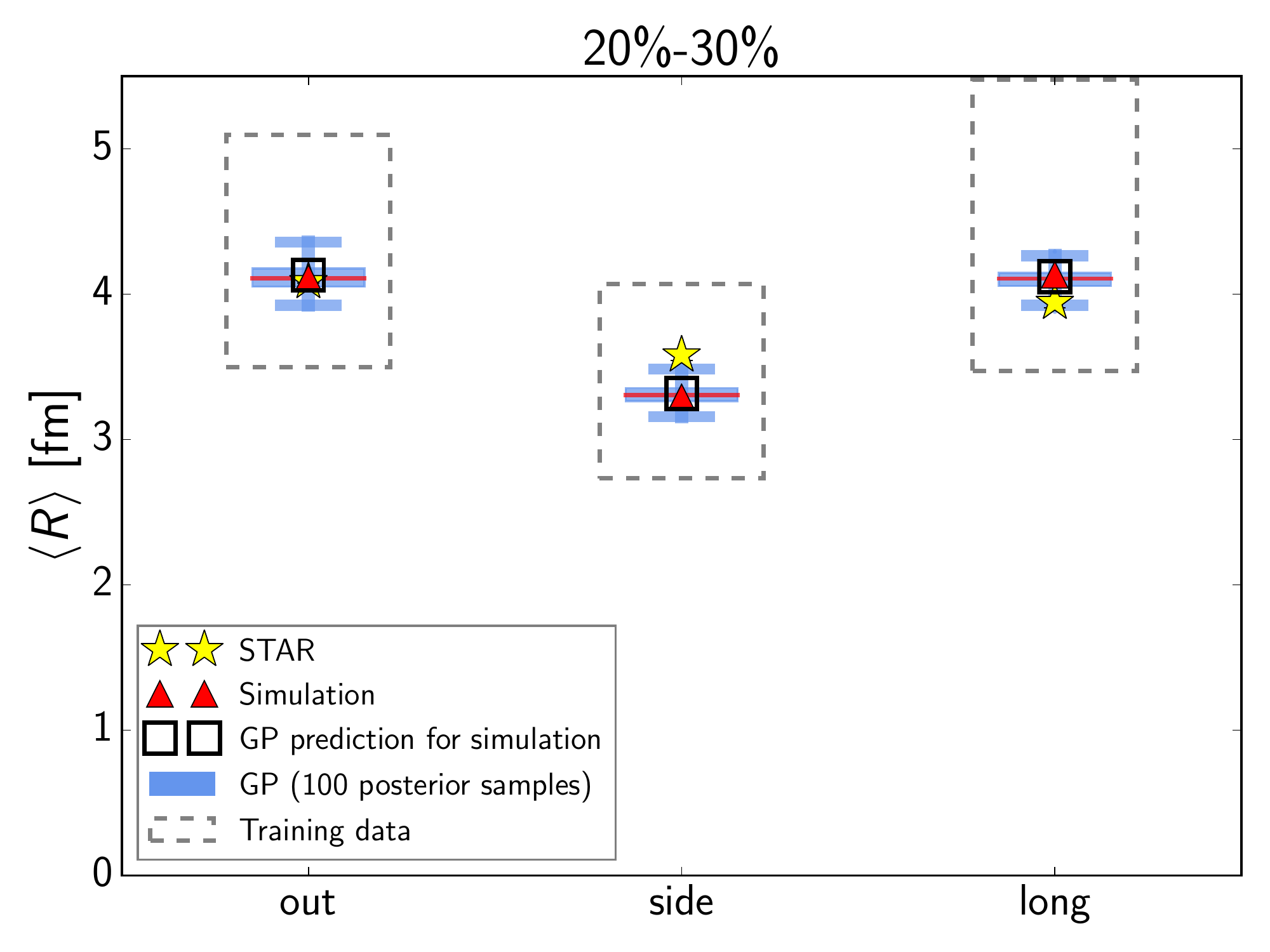}
\caption{(Color online) $R_{\text{out}}$, $R_{\text{side}}$, $R_{\text{long}}$ of charged $\pi$ averaged over 0.15 GeV/$c$ $<k_T<$ 0.6 GeV/$c$
at (0-5)\% centrality (top) and (20-30)\% centrality (bottom) at $\sqrt{s_{NN}}=39$ GeV.
Experimental data from \cite{Adamczyk:2014mxp}.}
\label{fig:hbt_mean_39}
\end{figure}

Regarding the HBT radii (Figs.~\ref{fig:hbt_onebin_39} and \ref{fig:hbt_mean_39}),
$R_{\text{out}}$ is still well described with peak parameter values.
However, in addition to the underestimation of $R_{\text{side}}$,
we find $R_{\text{long}}$ to be overestimated at this energy.
As $R_{\text{long}}$ has been associated with the freezeout time and and temperature of the system \cite{Herrmann:1994rr},
and should also be sensitive to shear viscosity \cite{Teaney:2003kp},
increasing the weight of HBT observables at this energy might have an effect on the most probable values of $\eta/s$ and $\epsilon_{SW}$.

\begin{figure}[htb]
\includegraphics[width=8cm]{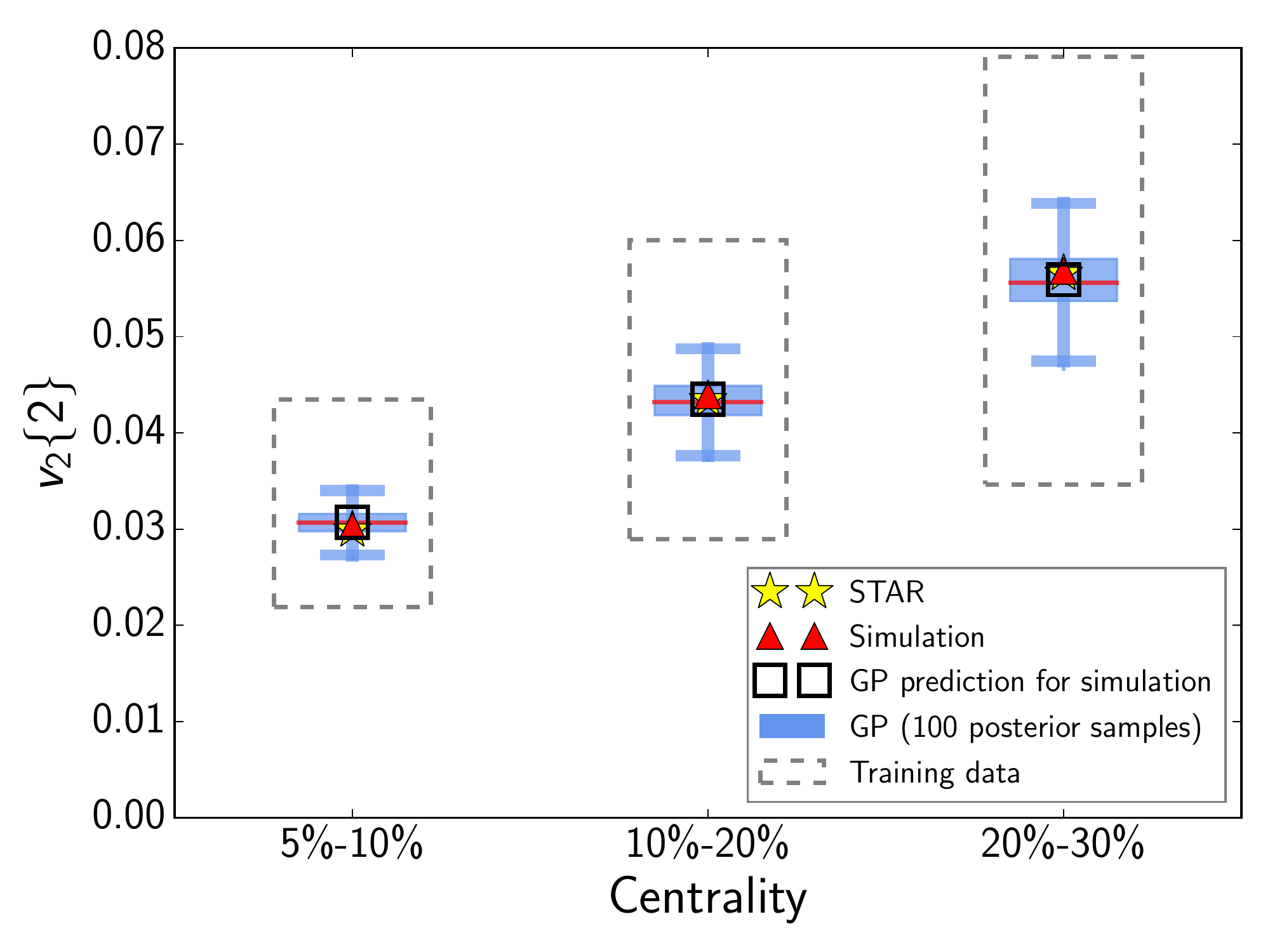}
\caption{(Color online) Centrality dependence of charged particle  $v_2\{2\}$ at $\sqrt{s_{NN}}=39$ GeV.
Experimental data from \cite{Adamczyk:2012ku}.}
\label{fig:v2239}
\end{figure}

\begin{figure}[htb]
\includegraphics[width=8cm]{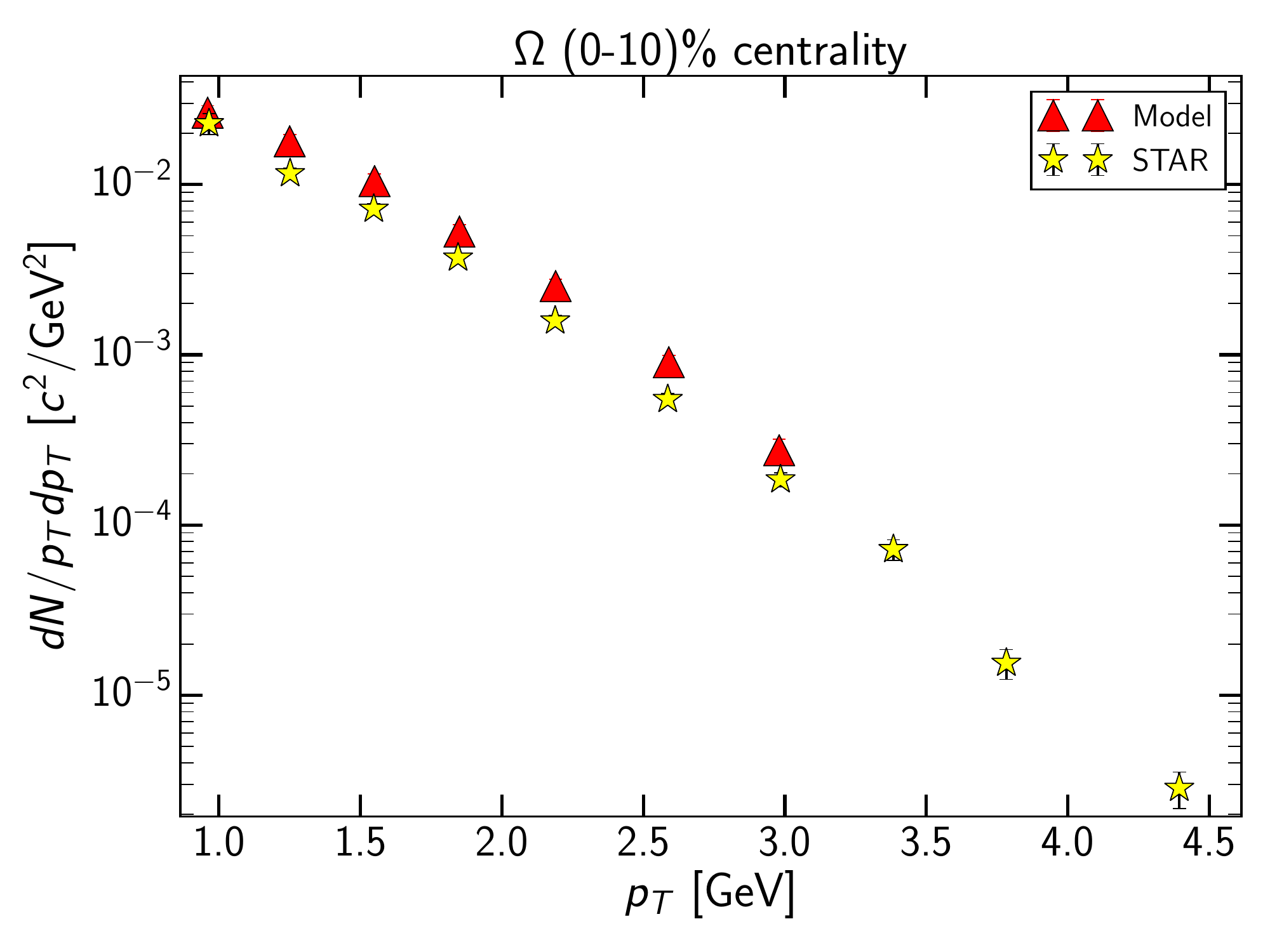}
\caption{(Color online) Transverse momentum spectrum of $\Omega$ at $\sqrt{s_{NN}}=39$ GeV.
Experimental data from \cite{Adamczyk:2015lvo}.}
\label{fig:dndptomega_39}
\end{figure}

Also the results for $v_2\{2\}$ (Fig.~\ref{fig:v2239}) and $\Omega$ (Fig.~\ref{fig:dndptomega_39})
support the conclusions drawn from $\sqrt{s_{NN}}=19.6$ GeV results.
Here the agreement with experimental data is good also for (20-30)\% centrality,
as the initial allowed hydro starting time is lower and likely appropriate for wider range of centralities.

The transverse momentum spectrum of $\Omega$ is in somewhat better agreement with the STAR data at $\sqrt{s_{NN}}=39$ GeV compared to $\sqrt{s_{NN}}=19.6$ GeV.
The total yield is still overestimated, however.

\subsection{$\sqrt{s_{NN}}=62.4$ GeV}

The highest collision energy investigated in this analysis has also the most comprehensive set of observables,
including information about the $p_T$ spectra of identified particles for several centralities.
Instead of calibrating the model on the yield of pions and kaons separately, the $K^+/\pi^+$ ratio was used instead.
As the majority of charged particles consists of pions,
calibrating the model on the centrality dependence of $N_{\text{ch}}$ was deemed sufficient to constrain the total number of pions
when combined with the kaon/pion ratio.

\begin{figure*}[htb]
\includegraphics[width=16cm]{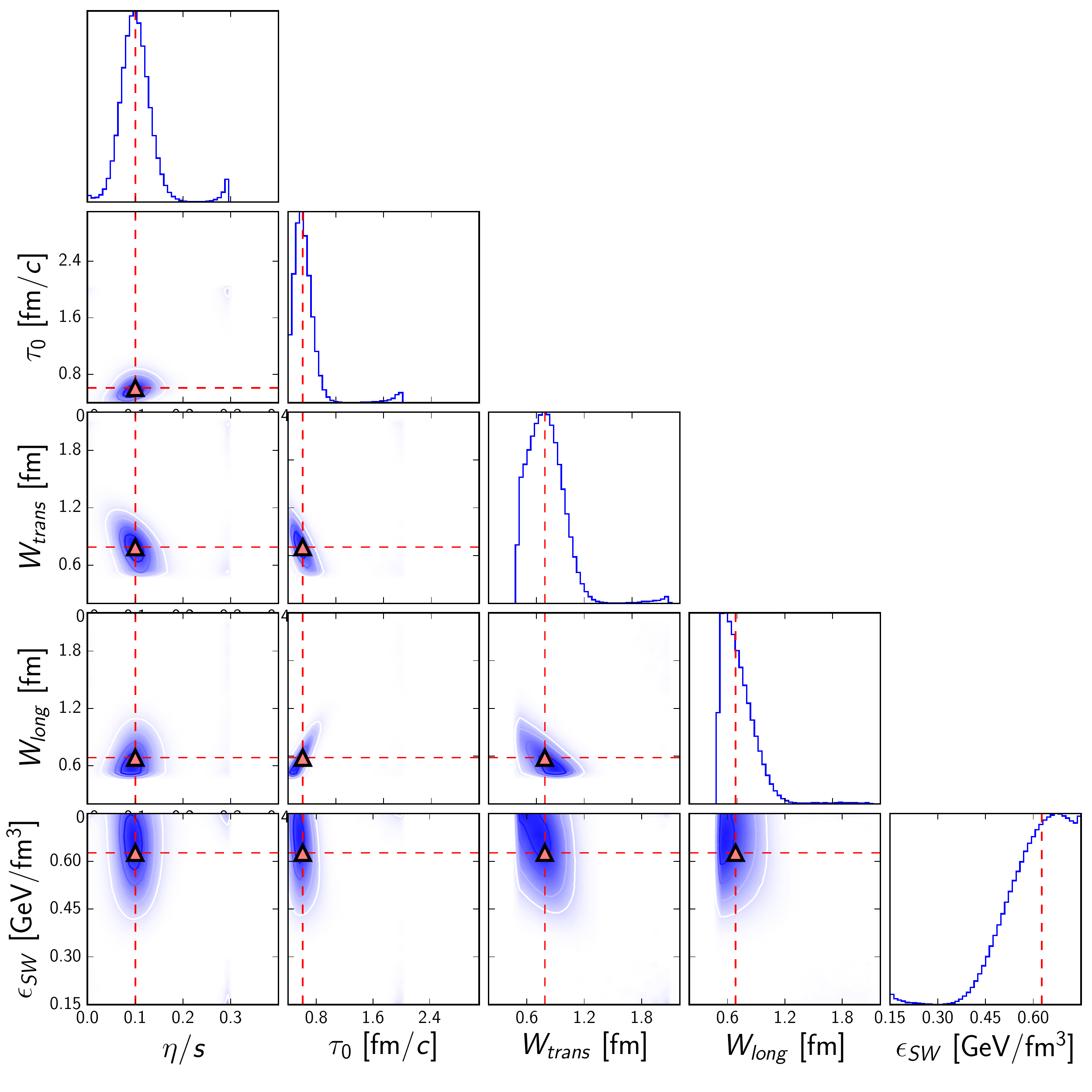}
\caption{(Color online) Representation of the posterior probability distribution of the input parameters at $\sqrt{s_{NN}}=62.4$ GeV.
See caption of Fig.~\ref{fig:corner19} for details.}
\label{fig:corner62}
\end{figure*}

The posterior probability distribution for 62.4 GeV, shown in Fig.~\ref{fig:corner62},
shows similar trends as the distributions at the lower energies.
Early starting time for the hydrodynamic evolution is preferred,
together with a longitudinally narrow Gaussian smoothing for the initial density.
The transverse width of the smearing functions has narrowed further,
leading to $W_{\text{trans}}$ and $W_{\text{long}}$ being roughly equivalent for this energy.
The shear viscosity over entropy density ratio has a clear peak at 0.1;
overall the posterior probability distribution of $\eta/s$ does not change much over the investigated beam energies.
Like at 19.6 GeV, and unlike at 39 GeV, a high value of the switching energy density is preferred.
As the pseudorapidity distribution is the one observable which is included in the data calibration at 19.6 GeV and 62.4 GeV,
but not included in the analysis at 39 GeV,
it seems that the early transition from hydrodynamics to hadron transport is mainly driven by the longitudinal expansion.

\begin{figure}[htb]
\includegraphics[width=8cm]{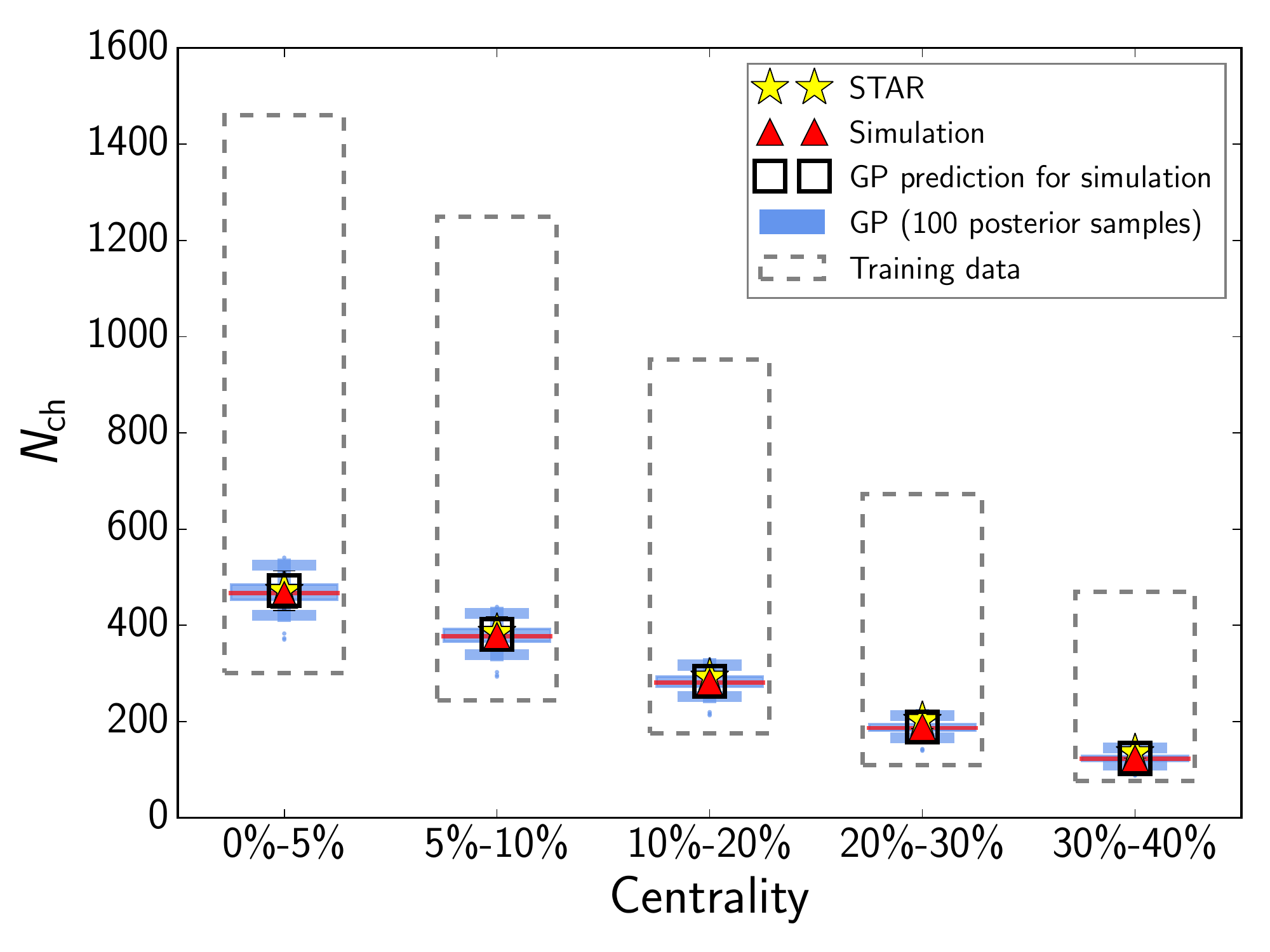}
\caption{(Color online) Centrality dependence of charged particle multiplicity in $|\eta|<0.5$ at $\sqrt{s_{NN}}=62.4$ GeV.
Experimental data from STAR \cite{Abelev:2008ab}.}
\label{fig:nch62}
\end{figure}

As illustrated in Fig.~\ref{fig:nch62},
the simulation using the peak parameter values gives a good description of the centrality dependence of charged particle multiplicity at midrapidity,
especially considering the wide range of $N_{\text{ch}}$ in the prior range.

\begin{figure}[htb]
\includegraphics[width=8cm]{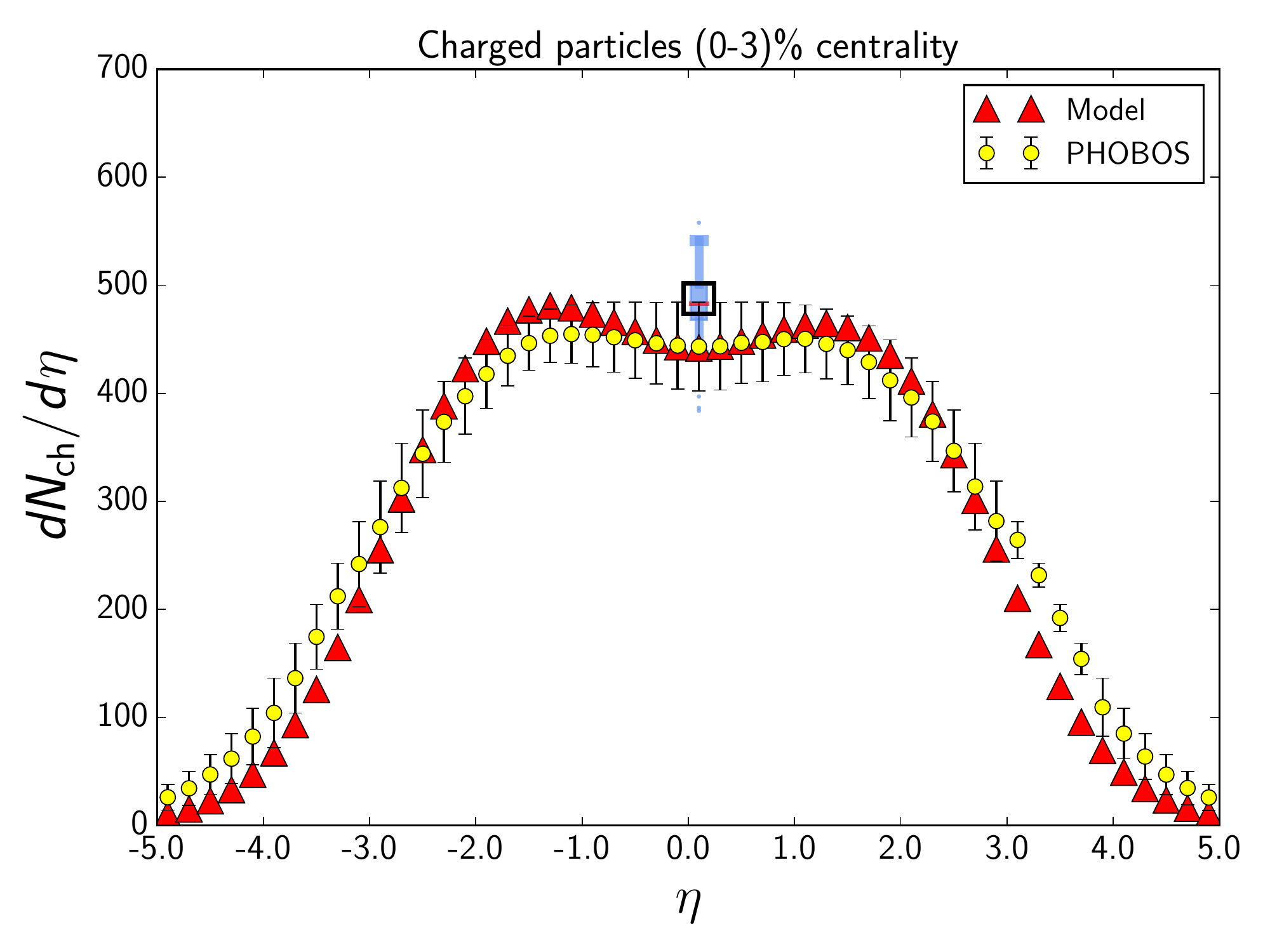}
\includegraphics[width=8cm]{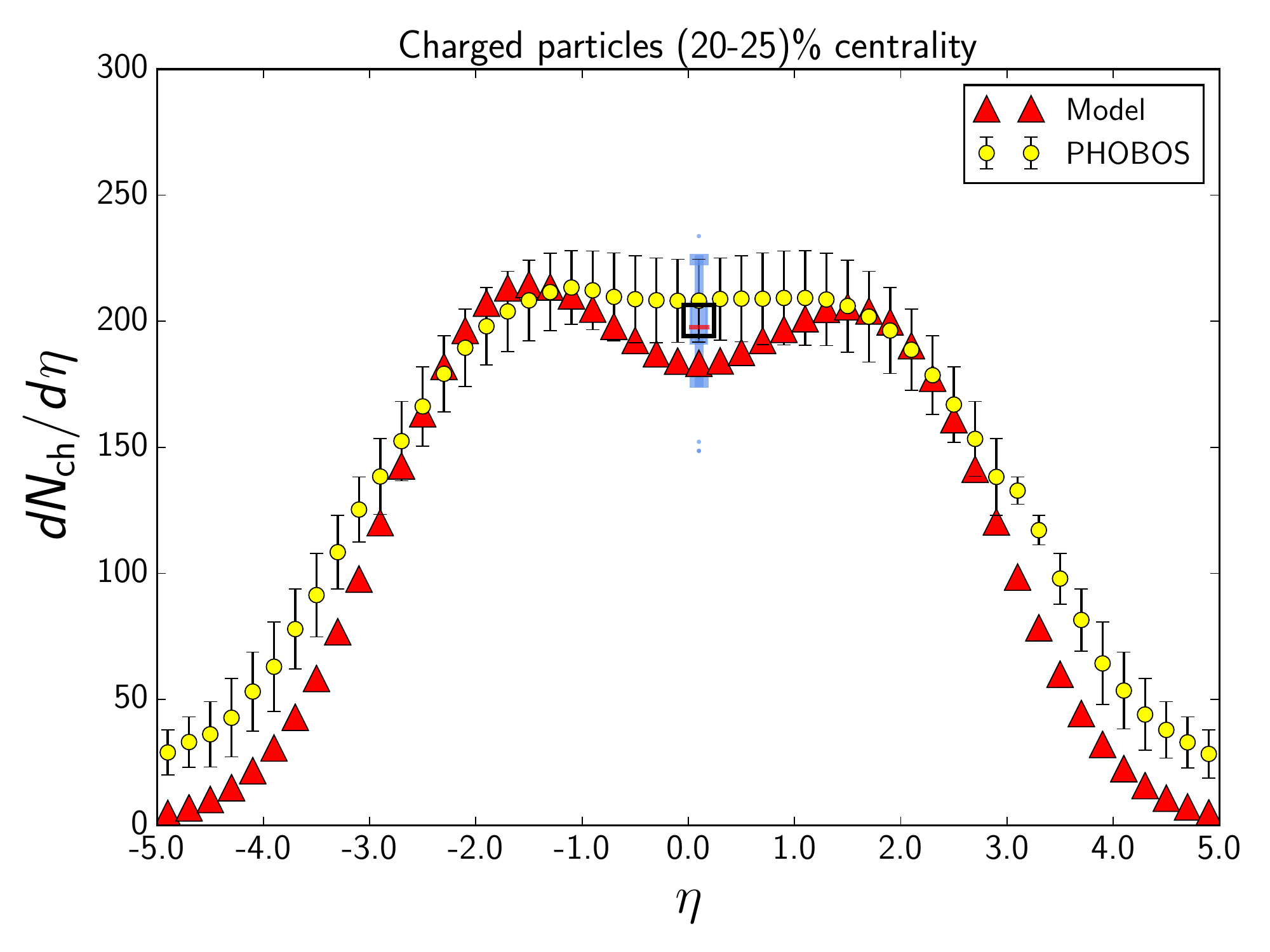}
\caption{(Color online) Charged particle pseudorapidity distribution at $\sqrt{s_{NN}}=62.4$ GeV for (0-3)\% centrality (top)
and (20-25)\% centrality (bottom). PHOBOS data from \cite{Alver:2010ck}.}
\label{fig:dndeta62}
\end{figure}

The full charged particle pseudorapidity distributions for (0-3)\% and (20-25)\% centrality at $\sqrt{s_{NN}}=62.4$ GeV are shown in Fig.~\ref{fig:dndeta62}.
The systematic difference between emulator prediction and actual simulation,
observed at $\sqrt{s_{NN}}=19.6$ GeV, is present also at this higher collision energy.
However, here the emulator prediction is systematically overestimating the midrapidity yield,
in contrast to the systematic underestimation seen at 19.6 GeV.

The simulations run with median values agree with PHOBOS data overall quite well at (0-3)\% centrality.
At (20-25)\% centrality, the charged particle yields are somewhat below the data both at $\eta \approx 0$
and again at $|\eta|>2.0$.

\begin{figure}[htb]
\includegraphics[width=8cm]{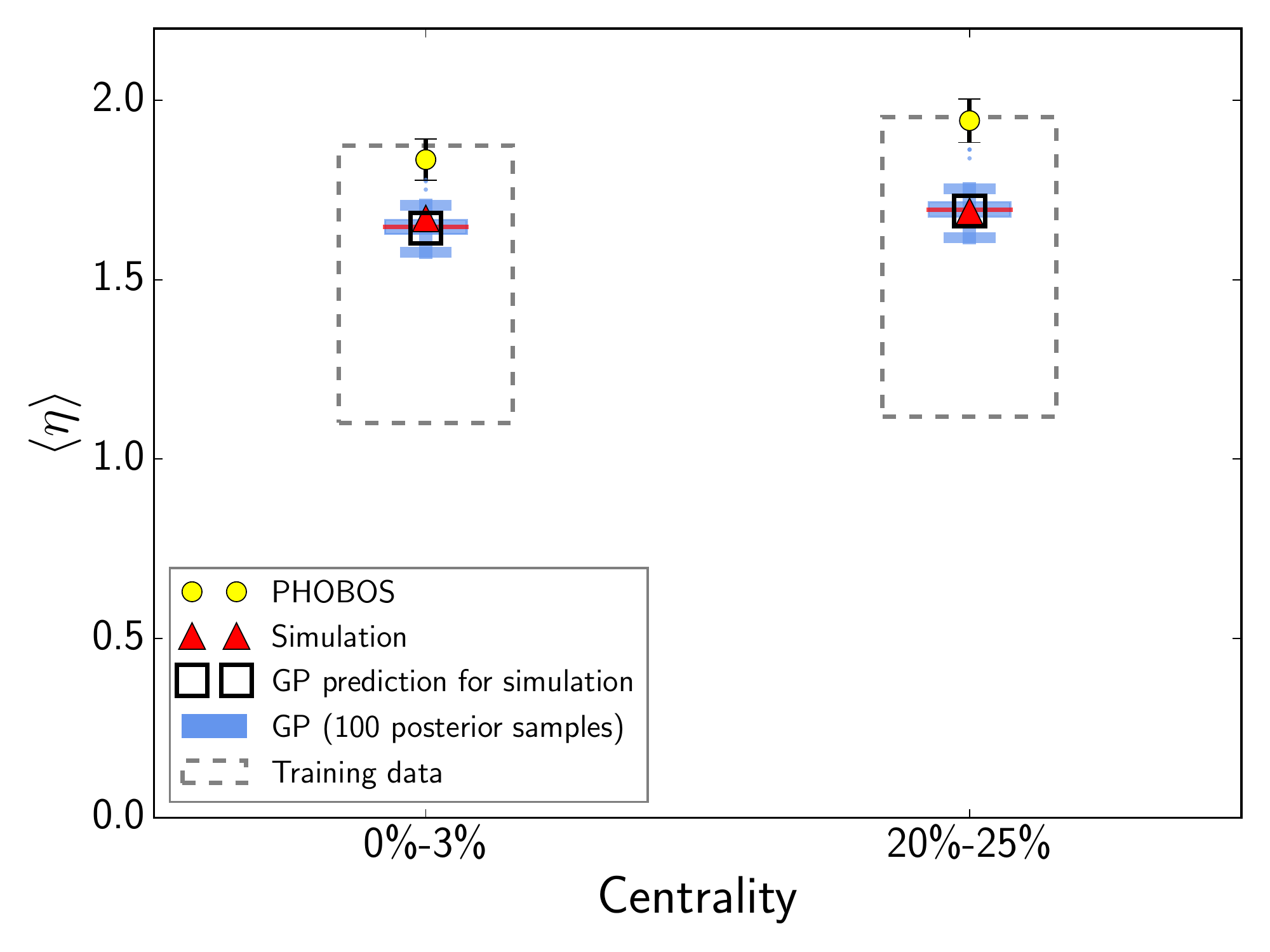}
\caption{(Color online) Centrality dependence of average charged particle pseudorapidity $\langle \eta \rangle$ for $\eta > 0.0$ at $\sqrt{s_{NN}}=62.4$ GeV.
Experimental values are estimated from PHOBOS data \cite{Alver:2010ck}.}
\label{fig:meaneta62}
\end{figure}

The mean pseudorapidities remain below the measurement, as shown in Fig.~\ref{fig:meaneta62}.
At this energy, it appears possible to match PHOBOS result within the parameter prior.
Based on the shape of $dN_{\text{ch}}/d\eta$ from Fig.~\ref{fig:dndeta62},
matching $\langle \eta \rangle$ is likely to lead to a notable depletion of charged particles at $\eta \approx 0$, however.

\begin{figure}[htb]
\includegraphics[width=8cm]{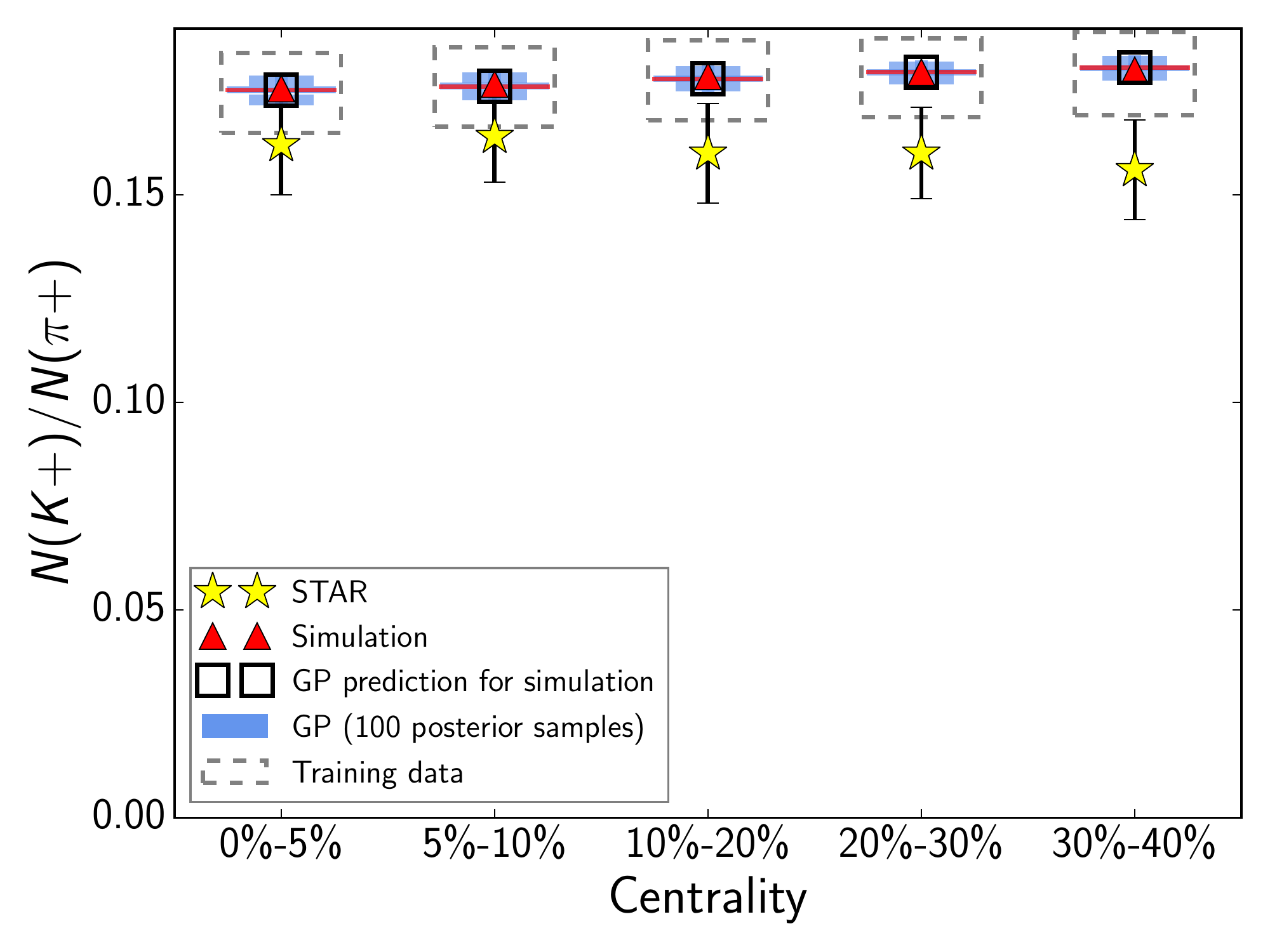}
\caption{(Color online) Centrality dependence of kaon/pion ratio at $\sqrt{s_{NN}}=62.4$ GeV.
Experimental data from STAR \cite{Abelev:2008ab}.}
\label{fig:kpi62}
\end{figure}

\begin{figure}[htb]
\includegraphics[width=8cm]{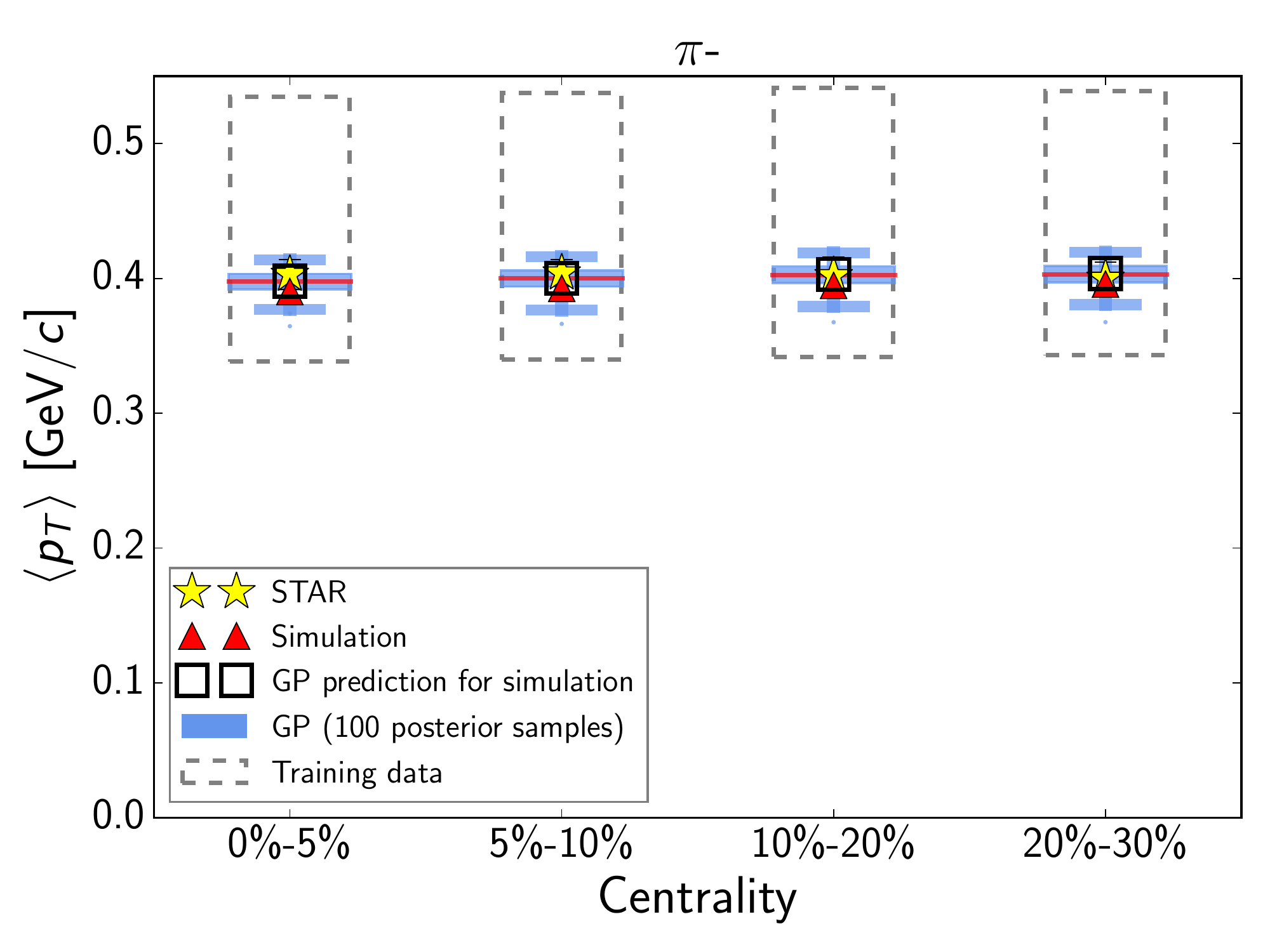}
\includegraphics[width=8cm]{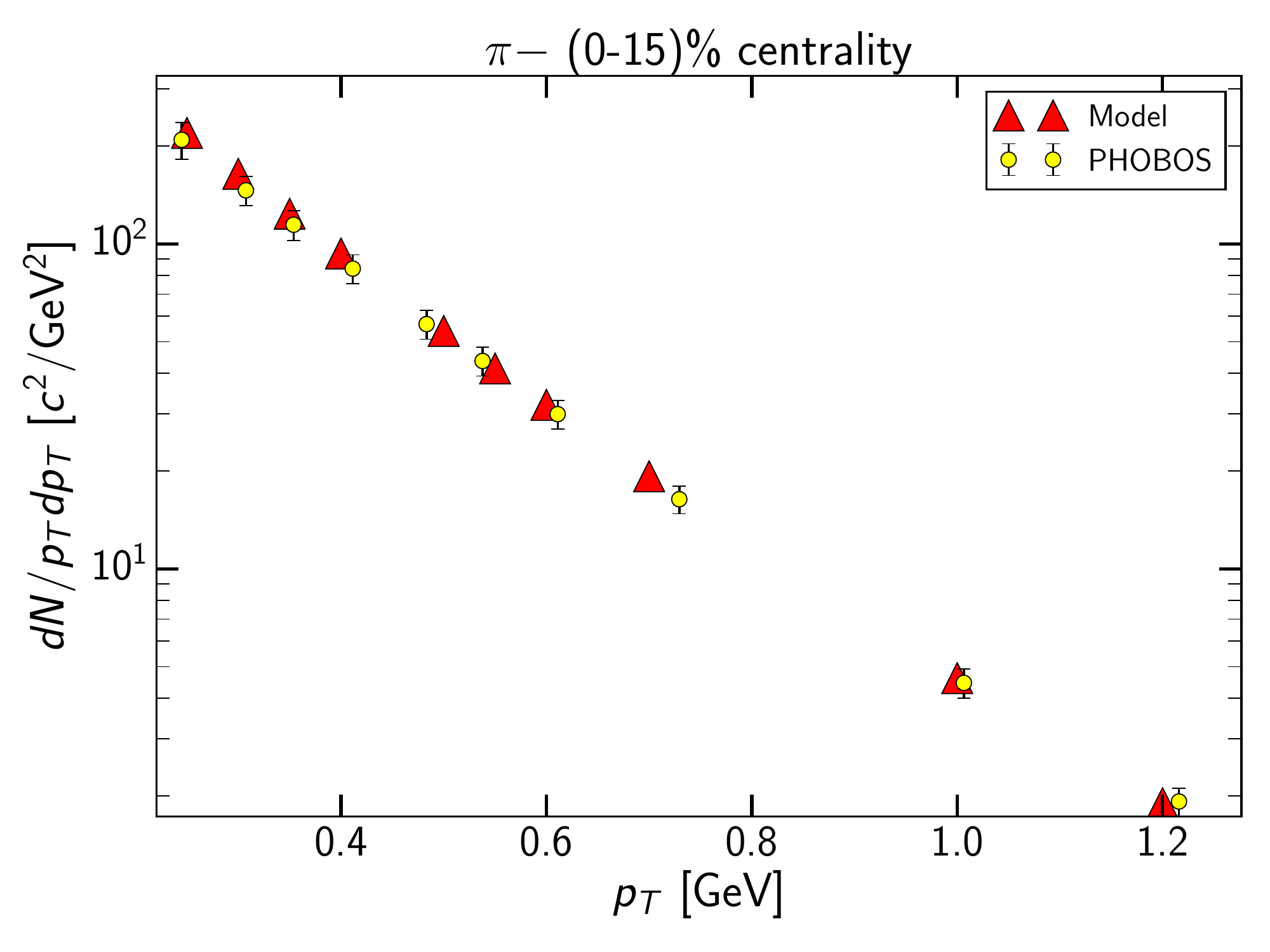}
\caption{(Color online) The centrality dependence of $\pi^-$ mean transverse momentum (top)
and the transverse momentum distribution for (0-15)\% centrality (bottom) at $\sqrt{s_{NN}}=62.4$ GeV.
STAR data from \cite{Adamczyk:2017iwn} and PHOBOS data from \cite{Back:2006tt}.}
\label{fig:dndptpiminus62}
\end{figure}

\begin{figure}[htb]
\includegraphics[width=8cm]{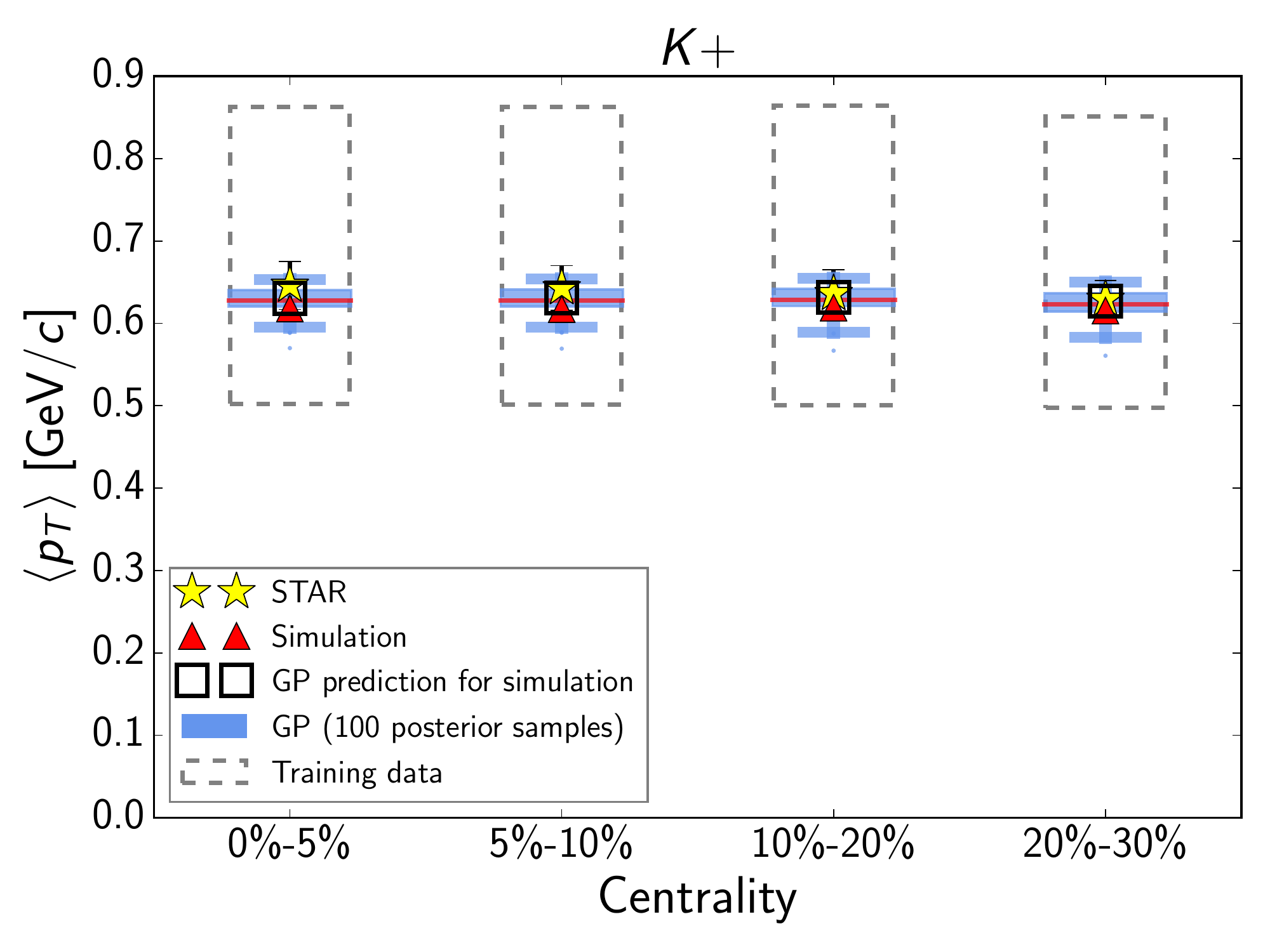}
\includegraphics[width=8cm]{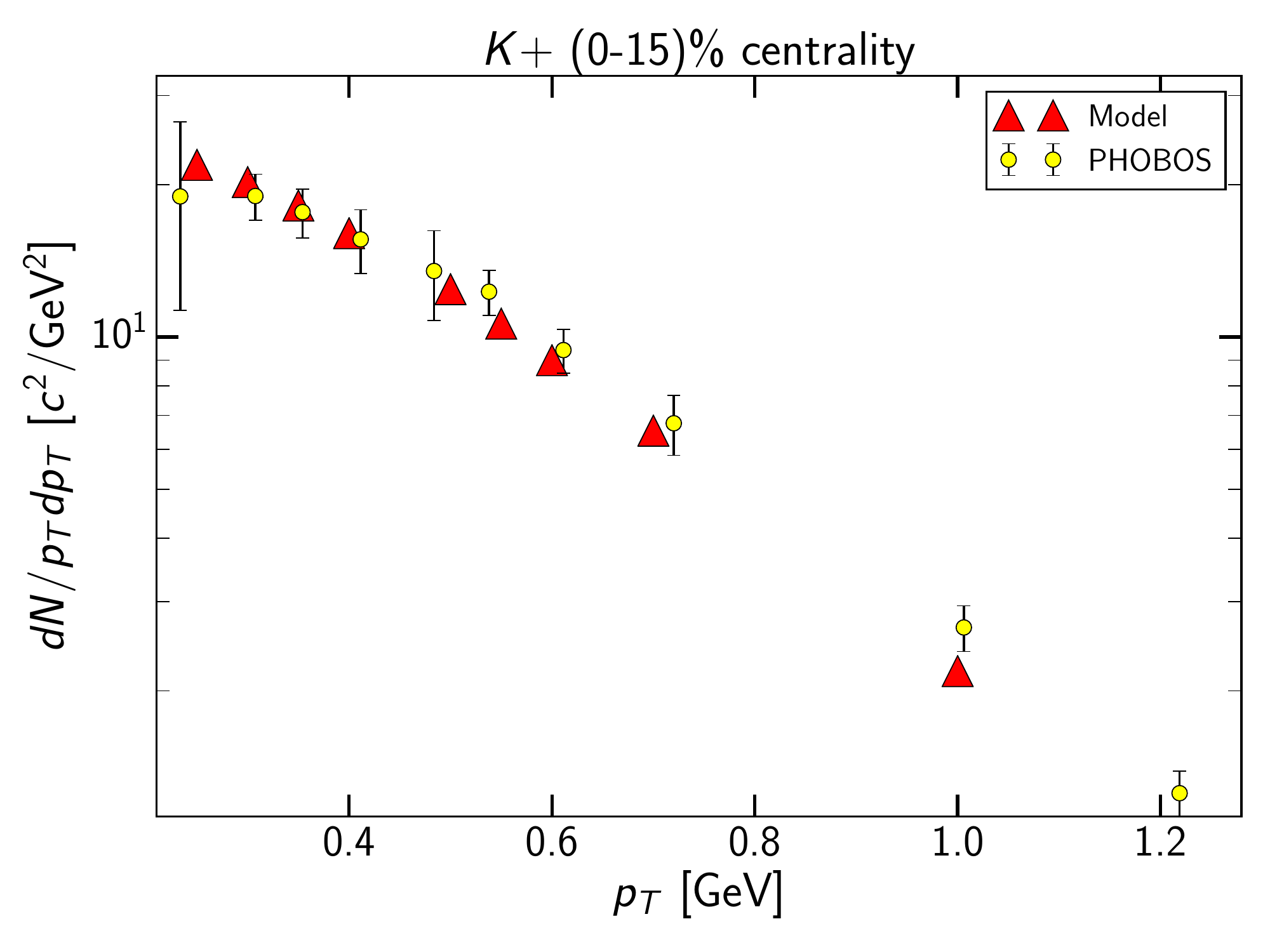}
\caption{(Color online) The centrality dependence of $K^+$ mean transverse momentum (top)
and the transverse momentum distribution for (0-15)\% centrality (bottom) at $\sqrt{s_{NN}}=62.4$ GeV.
STAR data from \cite{Adamczyk:2017iwn} and PHOBOS data from \cite{Back:2006tt}.}
\label{fig:dndptkplus62}
\end{figure}

Figure \ref{fig:kpi62} shows that there is some tendency for the model to overestimate the abundance of kaons over pions.
Overall there is very little room variation even in the prior range of the particle ratio,
so the shape of posterior will have little effect on this observable in any case.
Despite the observed slight excess of kaons compared to pions,
the mean transverse momenta of $\pi^-$ and $K^+$ is reproduced very well with the median parameter values
and the $p_T$ spectra for $\pi^-$, $K^+$ agrees well with PHOBOS results,
as shown in Figs.~\ref{fig:dndptpiminus62} and \ref{fig:dndptkplus62}.

\begin{figure}[htb]
\includegraphics[width=6cm]{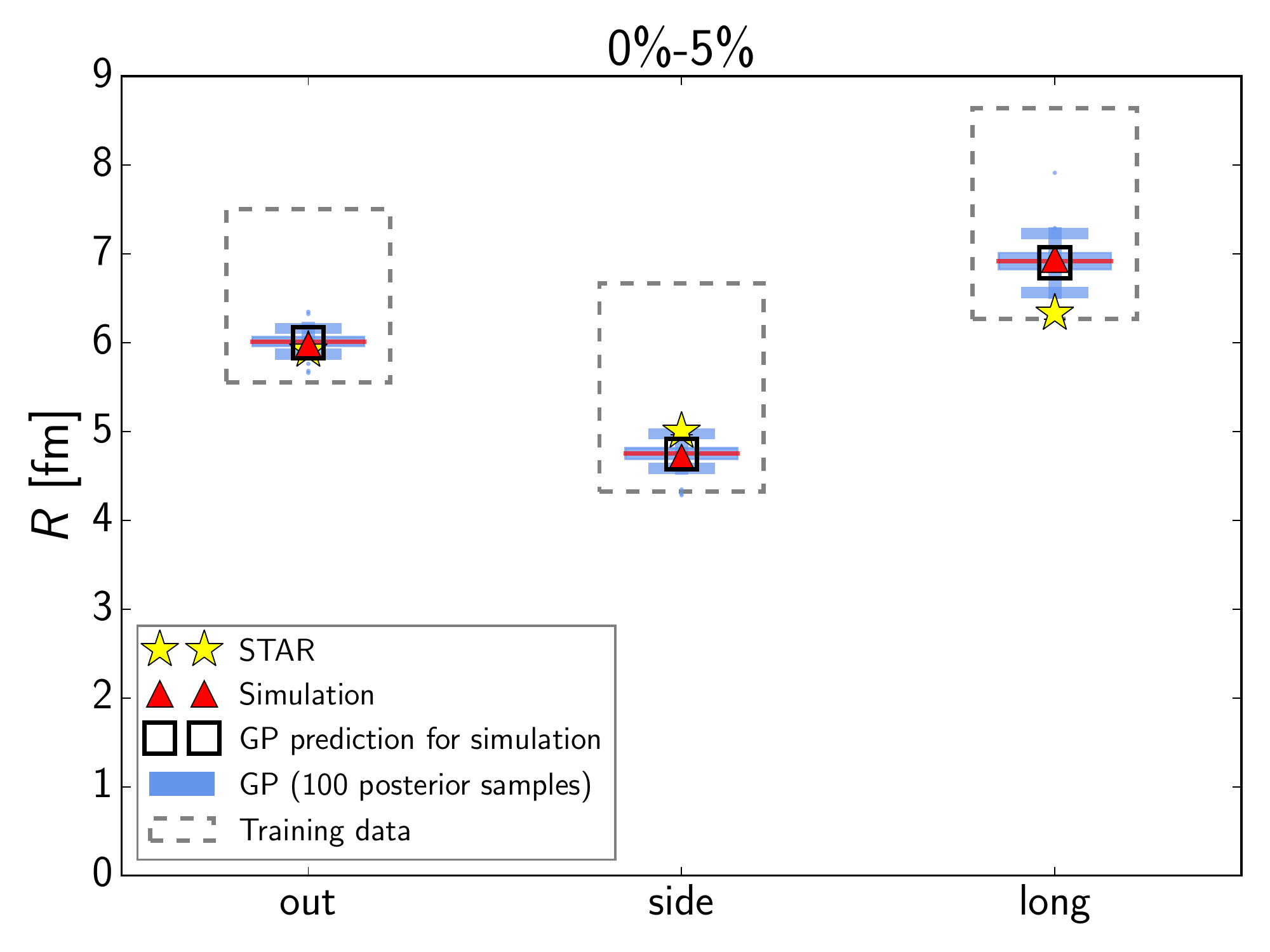}
\includegraphics[width=6cm]{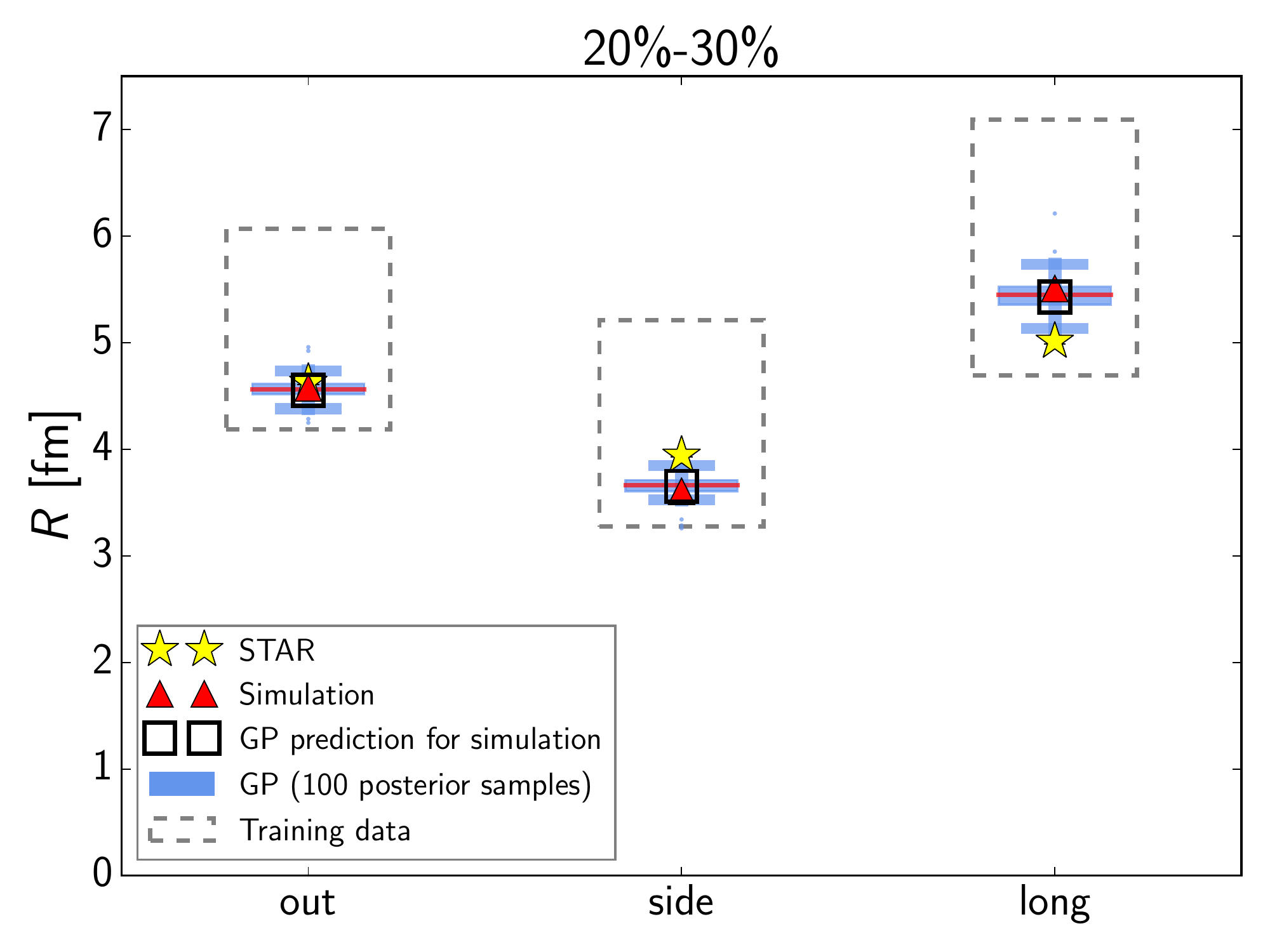}
\caption{(Color online) $R_{\text{out}}$, $R_{\text{side}}$, $R_{\text{long}}$ of charged $\pi$ \mbox{at $\langle k_T \rangle \approx 0.22$ GeV/$c$}
at (0-5)\% centrality (top) and (20-30)\% centrality (bottom) at $\sqrt{s_{NN}}=62.4$ GeV.
Experimental data from \cite{Adamczyk:2014mxp}.}
\label{fig:hbt_onebin_62}
\end{figure}

\begin{figure}[htb]
\includegraphics[width=6cm]{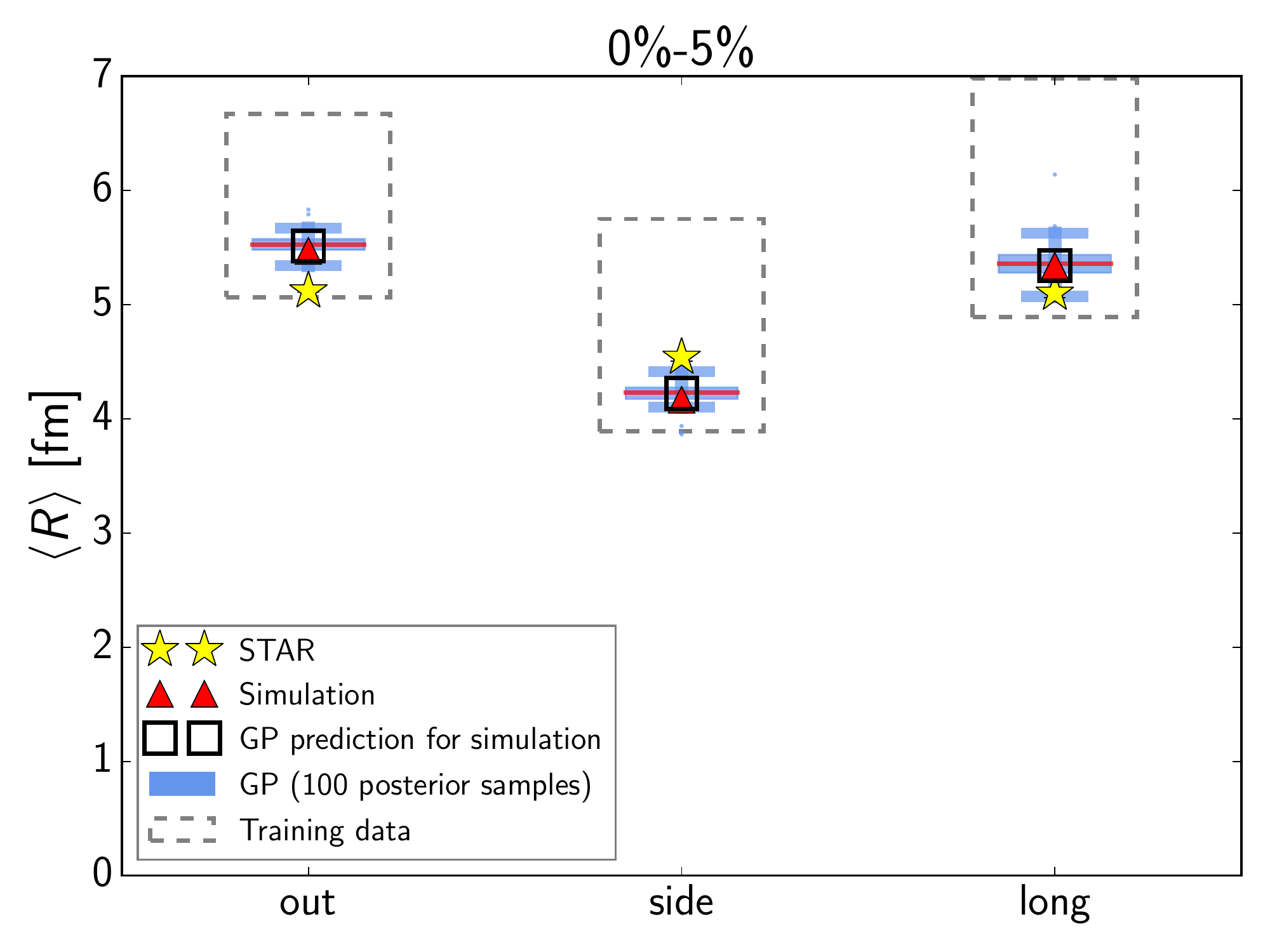}
\includegraphics[width=6cm]{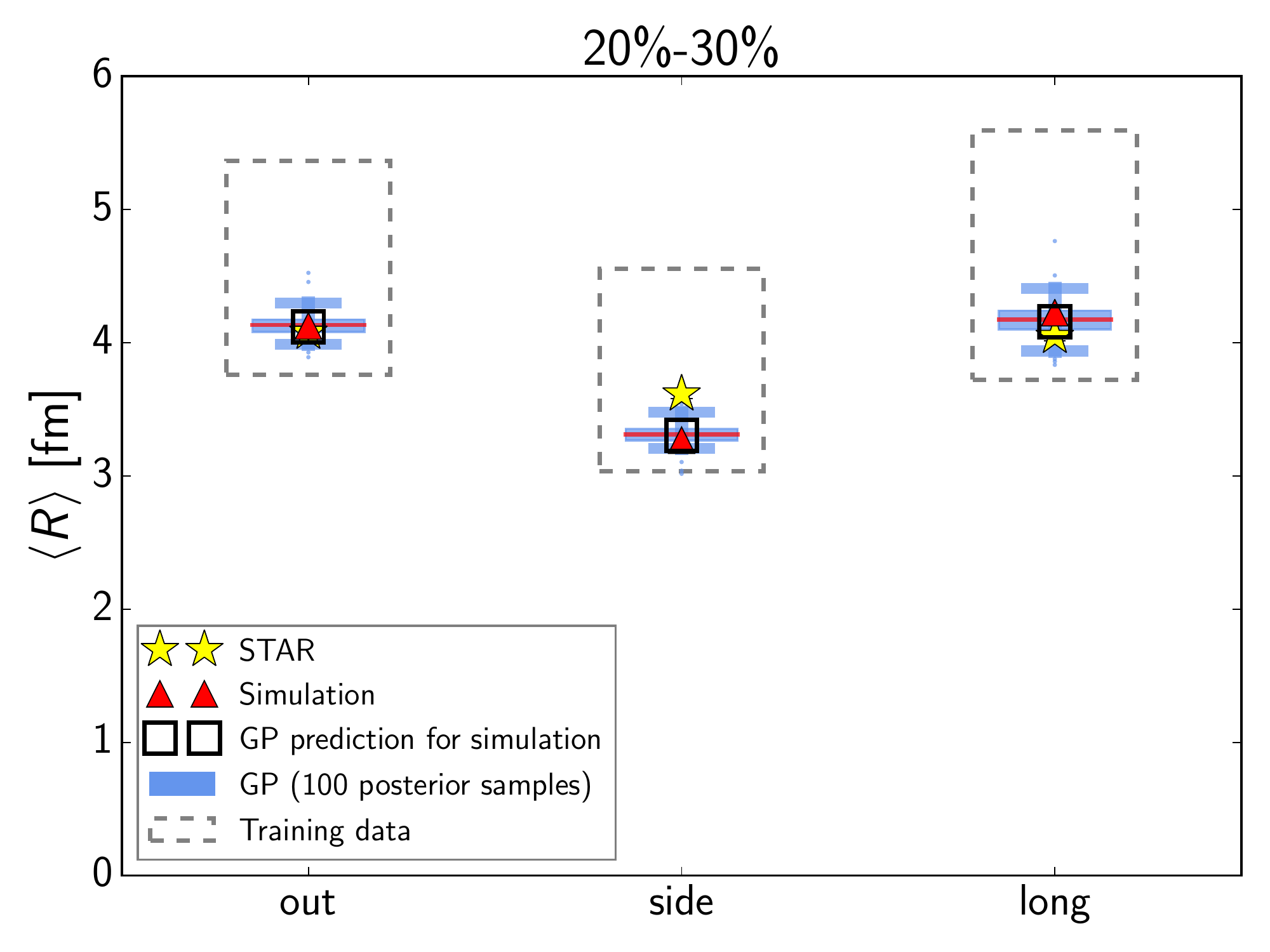}
\caption{(Color online) $R_{\text{out}}$, $R_{\text{side}}$, $R_{\text{long}}$ of charged $\pi$ averaged over 0.15 GeV/$c$ $<k_T<$ 0.6 GeV/$c$
at (0-5)\% centrality (top) and (20-30)\% centrality (bottom) at $\sqrt{s_{NN}}=62.4$ GeV.
Experimental data from \cite{Adamczyk:2014mxp}.}
\label{fig:hbt_mean_62}
\end{figure}

The HBT analysis results presented in Figs.~\ref{fig:hbt_onebin_62} and \ref{fig:hbt_mean_62} are similar to the results at lower energies,
in particular the results at $\sqrt{s_{NN}}=39$ GeV:
$R_{\text{out}}$ is described well, while $R_{\text{long}}$ values from model output are systematically too large compared to data,
while $R_{\text{side}}$ is notably smaller than the value determined from measurements.
The fact that the behavior of HBT radii at 39 GeV and 62.4 GeV is very similar,
despite the large difference in the applied value of the switching energy density,
suggests that $R_{\text{long}}$ is more sensitive to $\eta/s$ than $\epsilon_{SW}$.

\begin{figure}[htb]
\includegraphics[width=8cm]{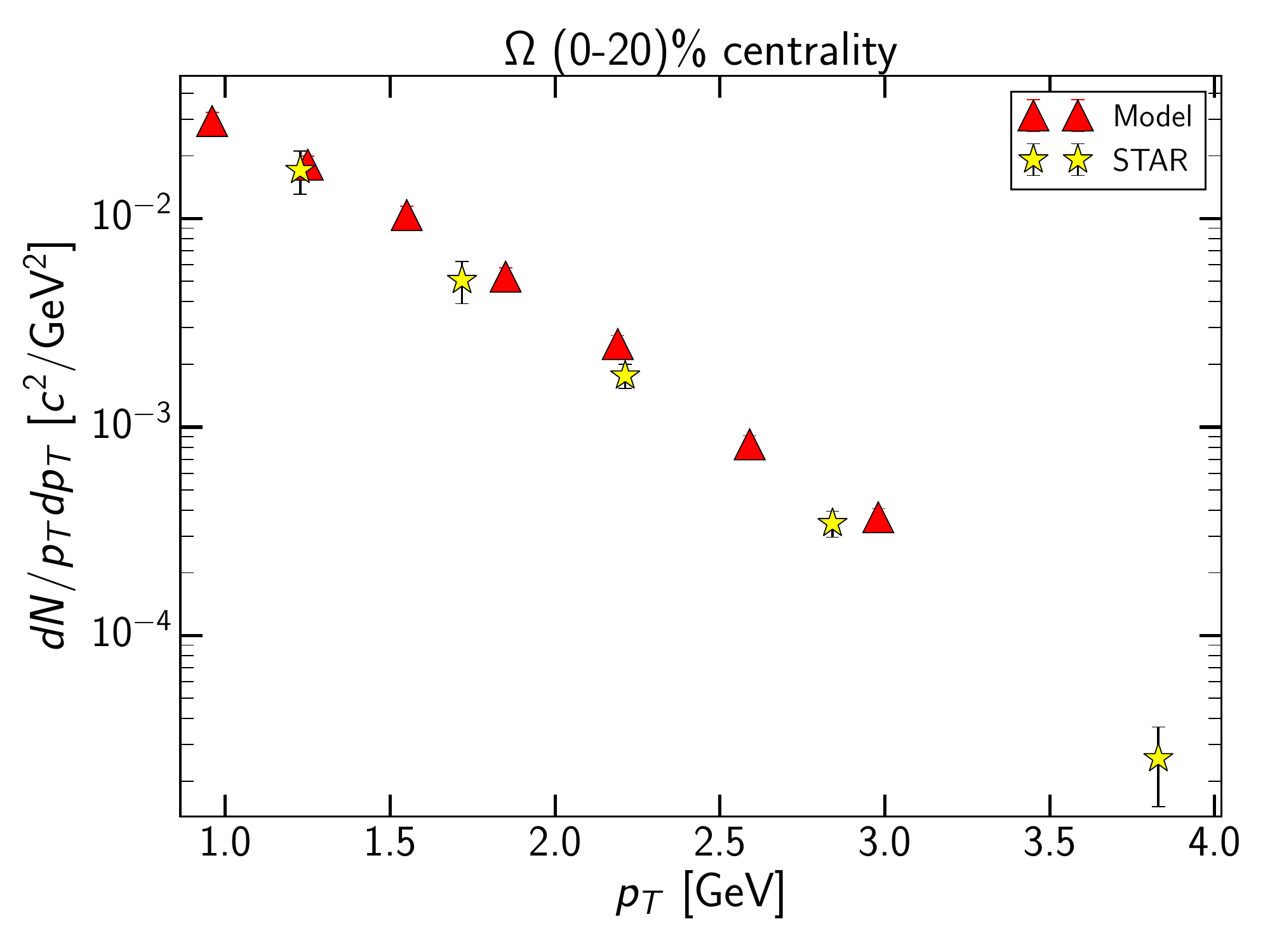}
\includegraphics[width=8cm]{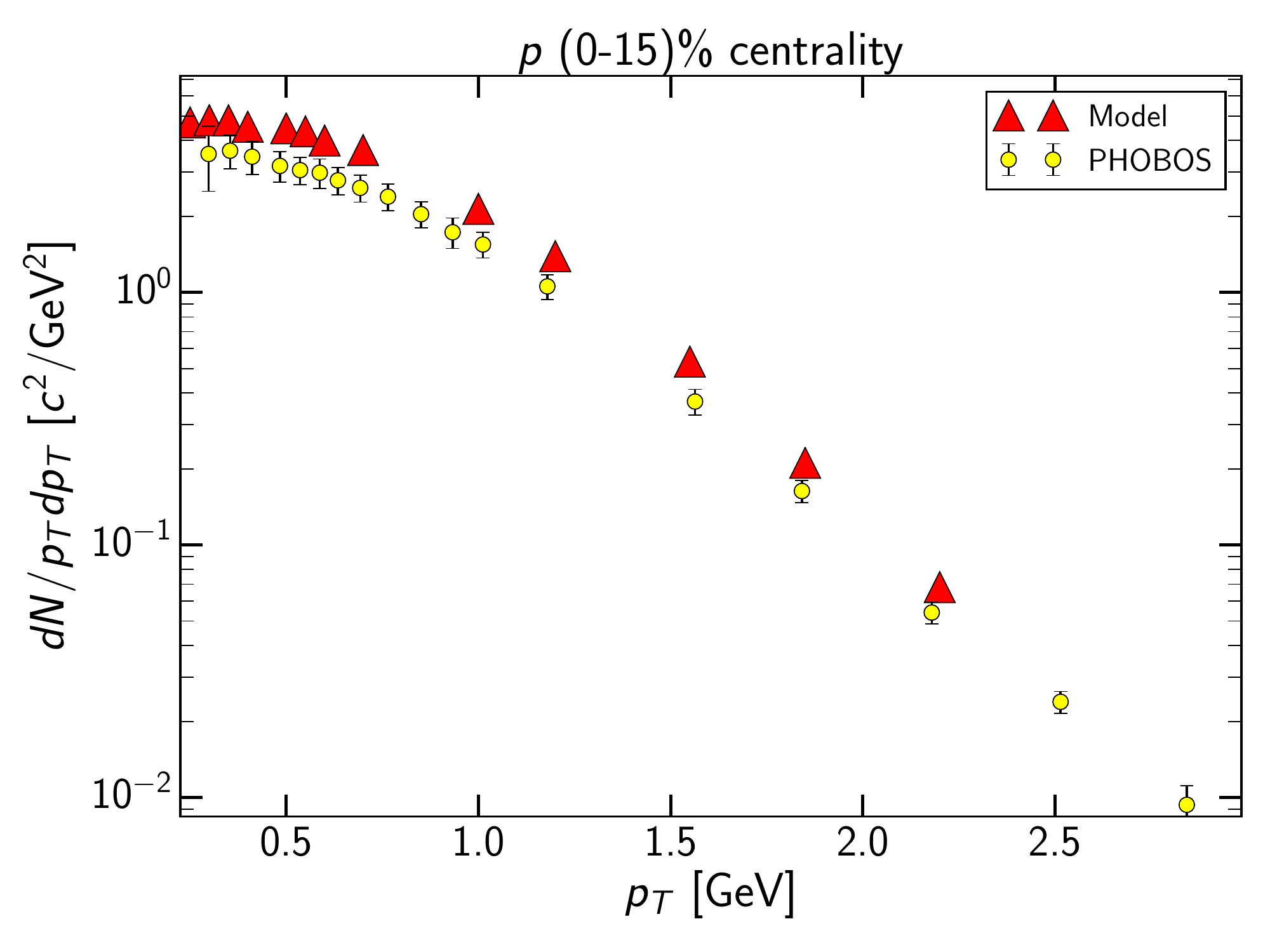}
\caption{(Color online) Transverse momentum spectrum of $\Omega$ (top) and proton (bottom) at $\sqrt{s_{NN}}=62.4$ GeV.
Note that proton data was not used in the data calibration process.
STAR data from \cite{Aggarwal:2010ig} and PHOBOS data from \cite{Back:2006tt}.}
\label{fig:dndptomega_62}
\end{figure}

Finally, we show the transverse momentum spectra of $\Omega$ and proton in Fig.~\ref{fig:dndptomega_62}.
In contrast to the results at lower collision energies,
fully $p_T$-integrated $\Omega$ multiplicity and mean transverse momentum were used in the data calibration,
leading to a good agreement with measured $p_T$ spectra;
the total multiplicity is slightly overestimated.
As we observe an excess of strange particles both at 19.6 GeV and 62.4 GeV,
while at $\sqrt{s_{NN}}=39$ GeV the kaon multiplicities are very well reproduced,
we can conclude that values of $\epsilon_{SW} \approx 0.4-0.5$ GeV/fm$^3$ would be preferred for strange particle production.
Also the proton spectrum, which was {\em not} part of the data calibration, matches quite well with the PHOBOS data,
although slight overabundance of protons is evident from the figure.

\clearpage

\subsection{Collision energy dependence of the Bayesian posteriors}

\begin{figure}[htb]
\includegraphics[width=8cm]{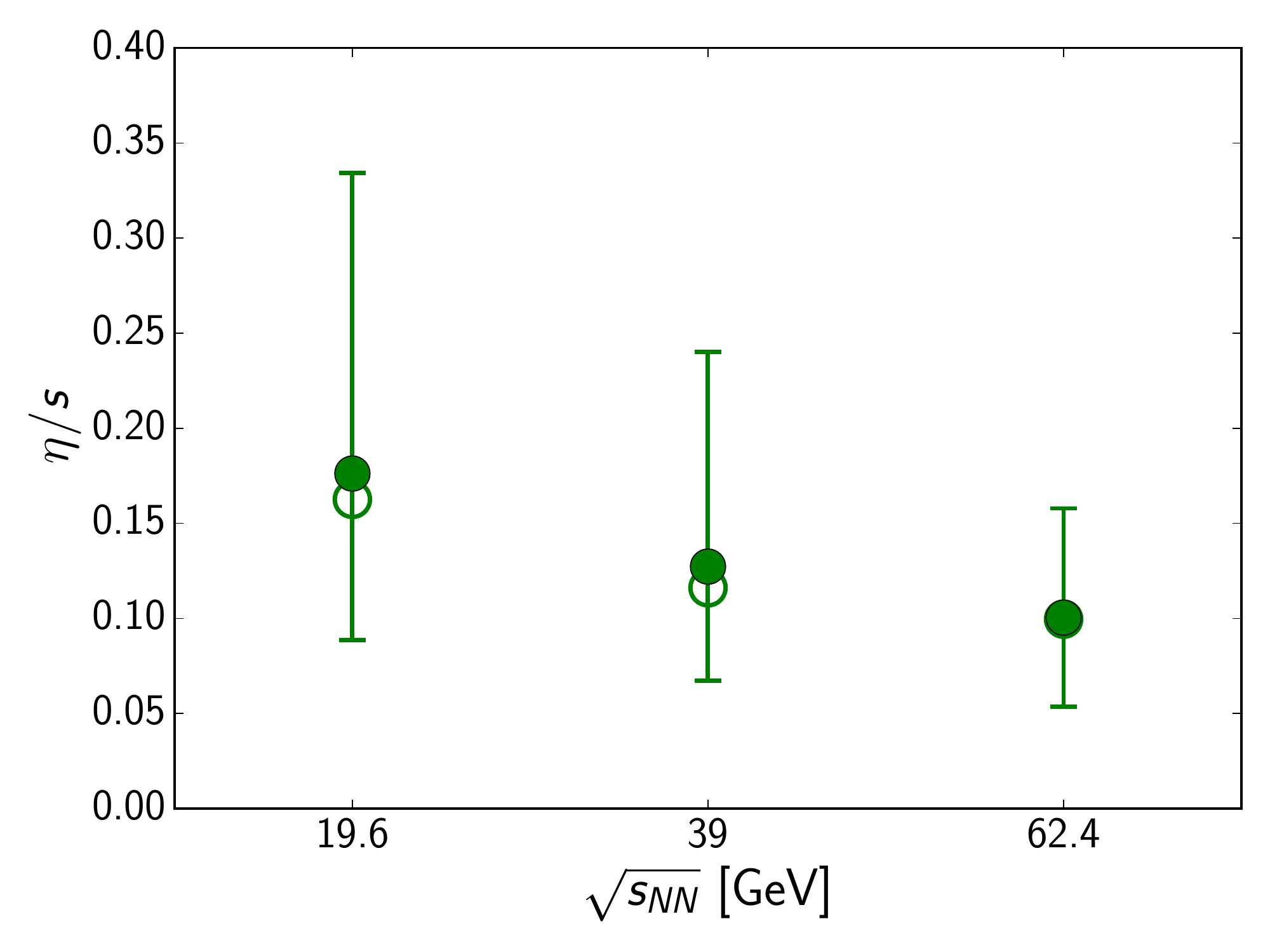}
\caption{(Color online) Collision energy dependence of the shear viscosity over entropy density ratio $\eta/s$.
Error bars represent 90\% confidence range around the median value (filled circles).
Open symbols indicate the peak position of the distribution.}
\label{fig:bes_etas}
\end{figure}

\begin{figure}[htb]
\includegraphics[width=8cm]{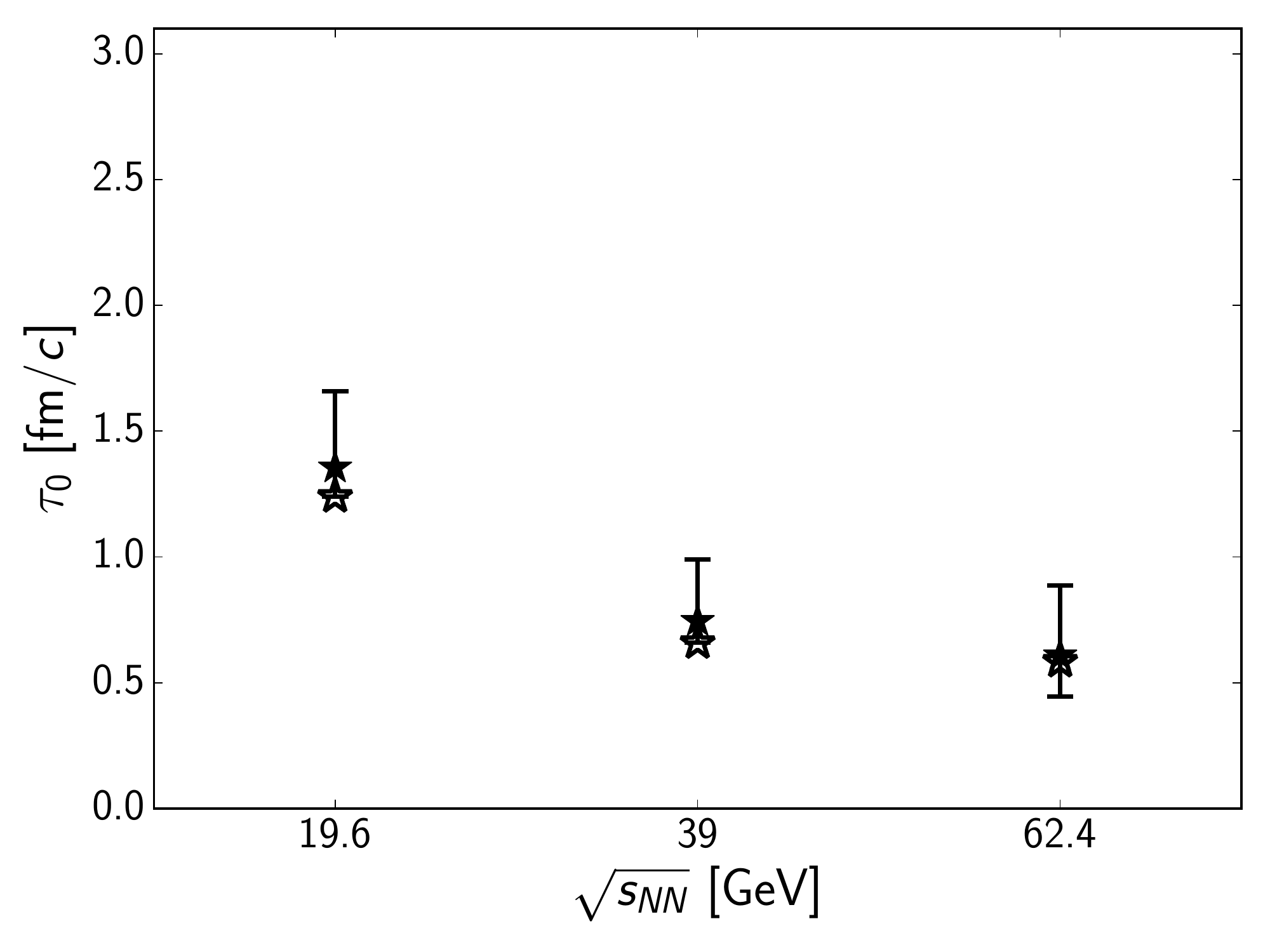}
\caption{(Color online) Collision energy dependence of the hydro starting time $\tau_0$.
Error bars represent 90\% confidence range around the median value (filled stars).
Open symbols indicate the peak position of the distribution.}
\label{fig:bes_tau0}
\end{figure}

\begin{figure}[htb]
\includegraphics[width=8cm]{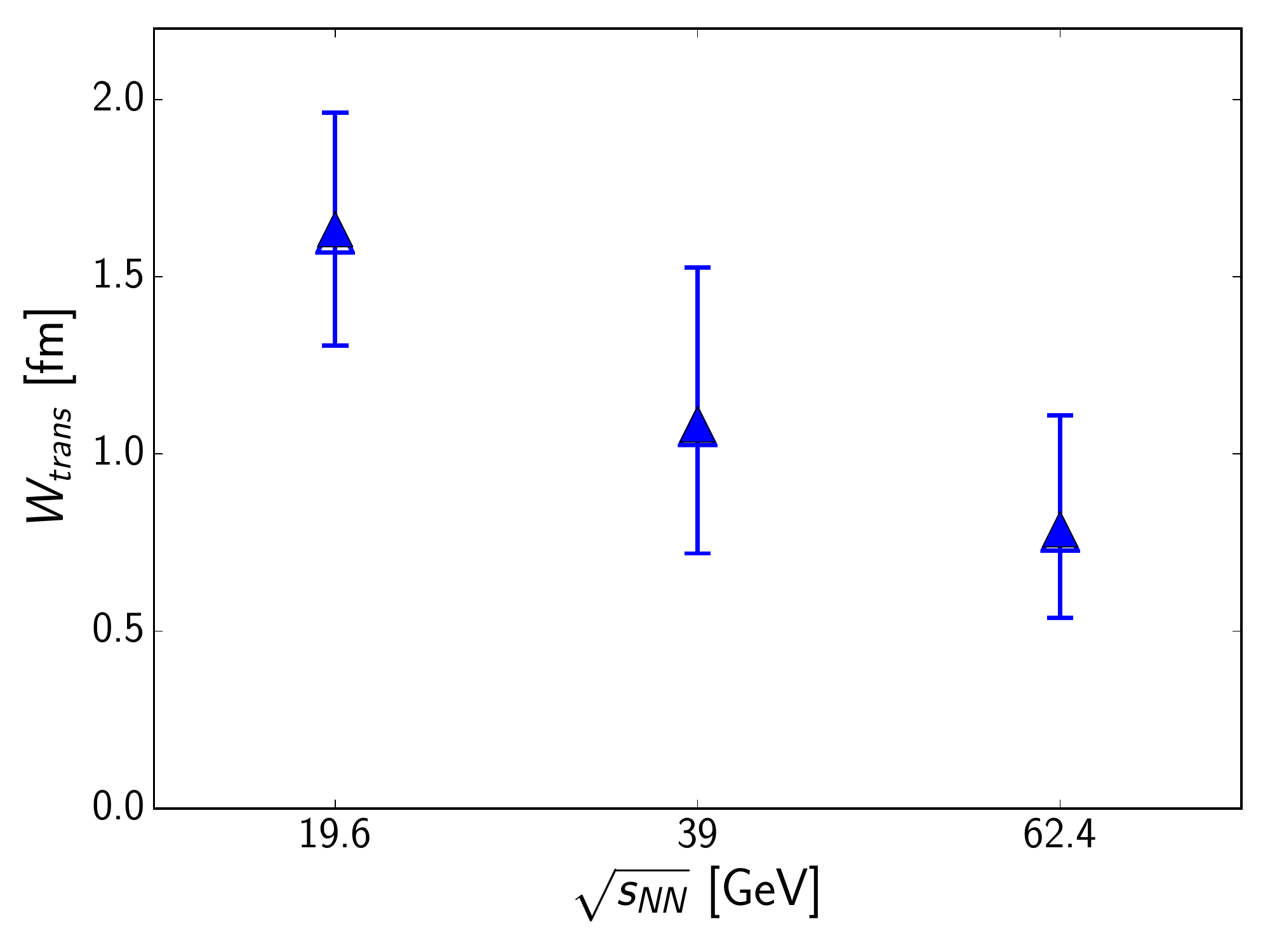}
\caption{(Color online) Collision energy dependence of the smearing factor $W_{\text{trans}}$.
Error bars represent 90\% confidence range around the median value (filled triangles).
Open symbols indicate the peak position of the distribution.}
\label{fig:bes_rtrans}
\end{figure}

\begin{figure}[htb]
\includegraphics[width=8cm]{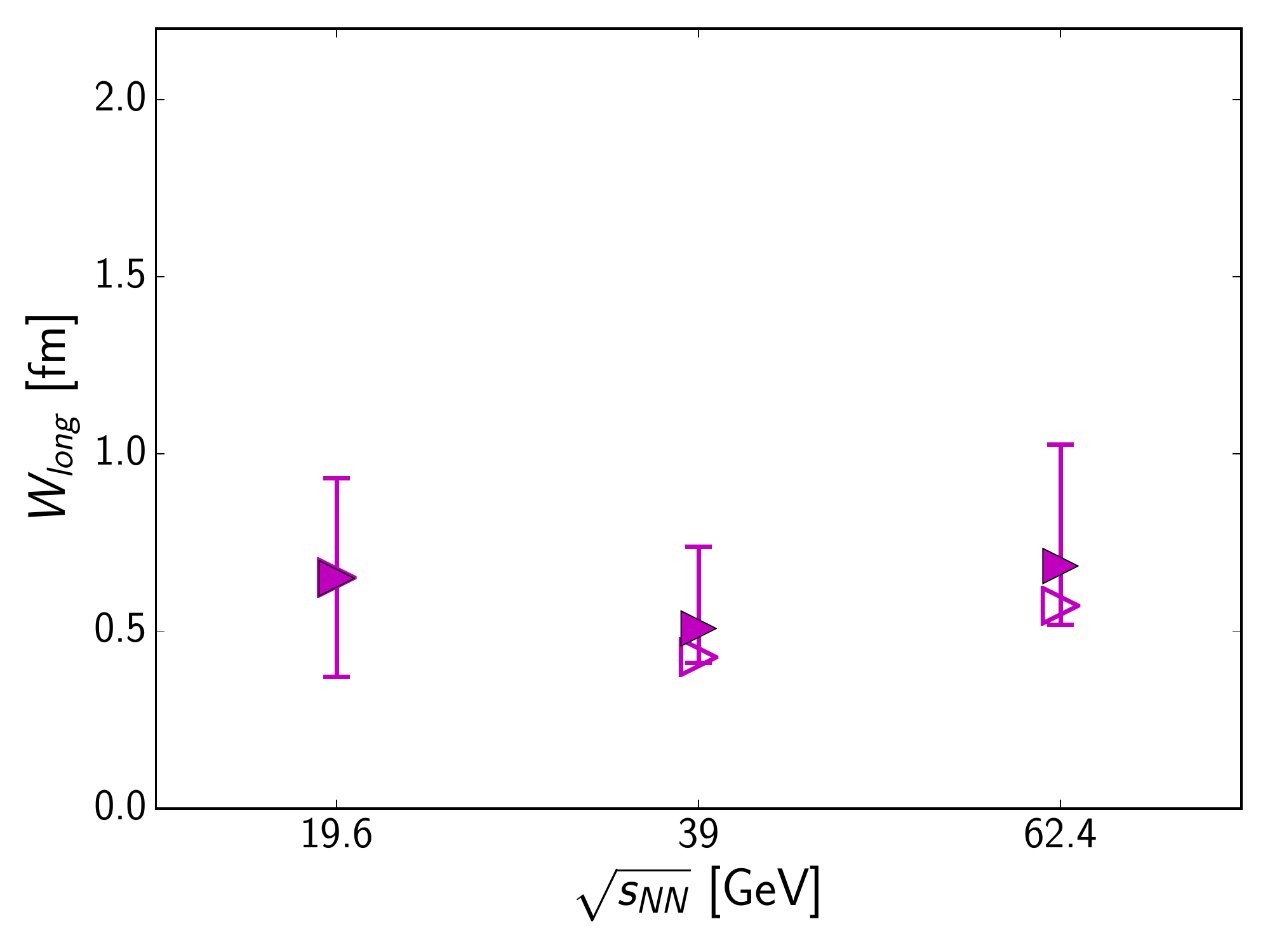}
\caption{(Color online) Collision energy dependence of the smearing factor $W_{\text{long}}$.
Error bars represent 90\% confidence range around the median value (filled triangles).
Open symbols indicate the peak position of the distribution.}
\label{fig:bes_rlong}
\end{figure}

\begin{figure}[htb]
\includegraphics[width=8cm]{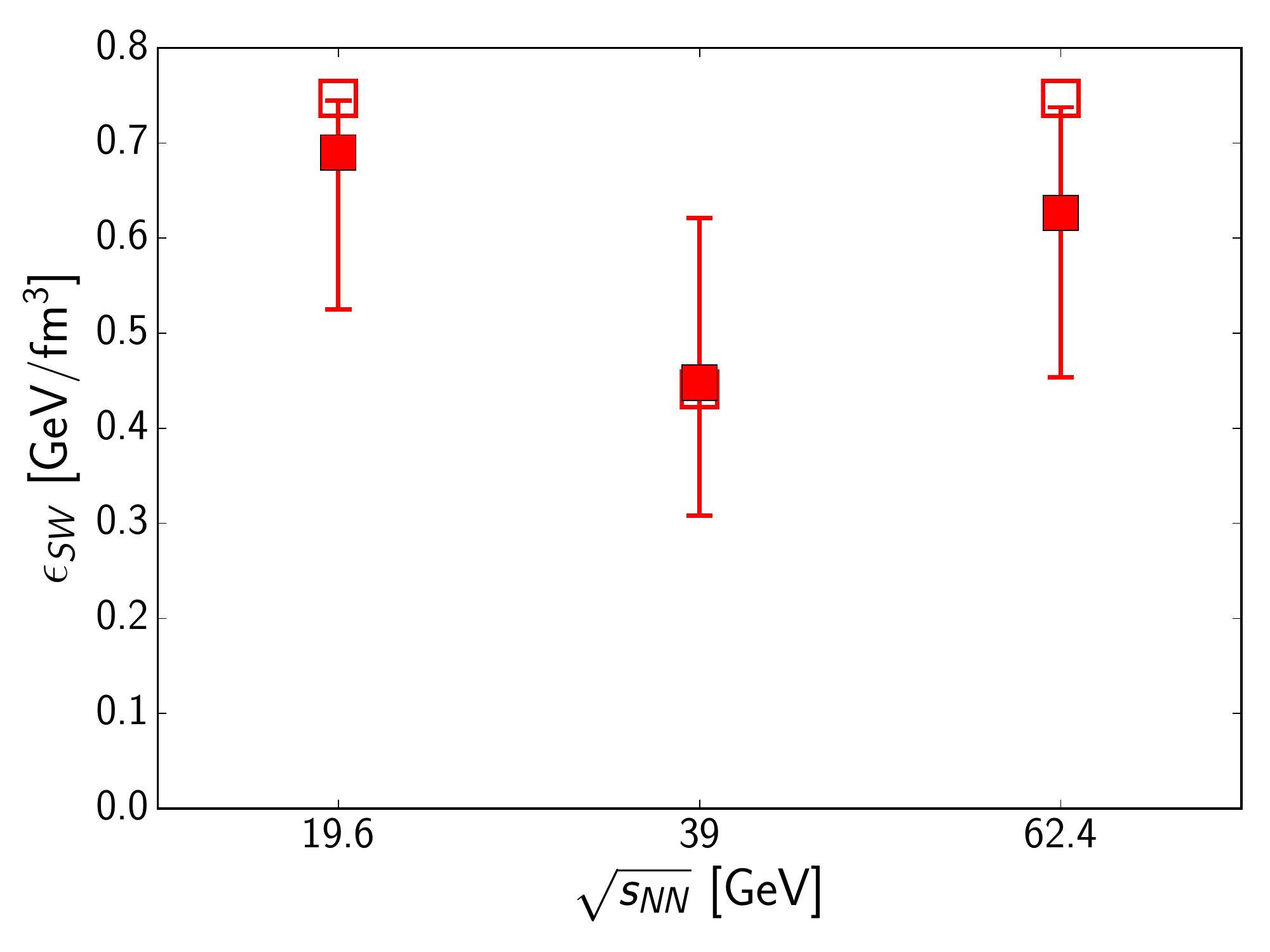}
\caption{(Color online) Collision energy dependence of the switching energy density $\epsilon_{SW}$.
Error bars represent 90\% confidence range around the median value (filled squares).
Open symbols indicate the peak position of the distribution.}
\label{fig:bes_ecrit}
\end{figure}

The collision energy dependence of the system properties is illustrated
in Figs.~\ref{fig:bes_etas}, \ref{fig:bes_tau0}, \ref{fig:bes_rtrans}, \ref{fig:bes_rlong}, and \ref{fig:bes_ecrit}.
The symbols indicate the median values, while the error bars represent the 90\% confidence range
(i.e. smallest 5\% and largest 5\% of the parameter values from the posterior distribution are excluded).

The present analysis finds $\eta/s$ to have a modest dependence on collision energy,
with median value decreasing from 0.176 at $\sqrt{s_{NN}}=19.6$ GeV to 0.100 at  $\sqrt{s_{NN}}=62.4$ GeV (see Fig.~\ref{fig:bes_etas}.
However, as the uncertainties increase towards lower value of $\sqrt{s_{NN}}$,
a possibility for a constant value of $\eta/s \approx 0.10 - 0.15$ cannot be yet excluded.

Regarding the three parameters controlling the initial state for the hydrodynamic model,
the transport-to-hydro switching time parameter $\tau_0$ is found to favor the earliest possible starting times for the hydrodynamical evolution.
and the observed collision energy dependence mainly stems from the constraint on the minimum time, Eq.~\eqref{eq:taumin}.
The transverse width of the Gaussian representation of initial particles $W_{\text{trans}}$ shows a strong collision energy dependence,
with larger values at lower collision energies,
while $W_{\text{long}}$ is almost constant with respect to collision energy,

For the hydro-to-transport switching energy density $\epsilon_{SW}$,
an early switch from hydrodynamics to hadron transport is preferred for $\sqrt{s_{NN}}=19.6$ and $62.4$ GeV,
while a lower value is found to be more probable for $\sqrt{s_{NN}}=39$ GeV.
As the kaon and $\Omega$ yields are currently overestimated at $\sqrt{s_{NN}}=19.6$ and $62.4$ GeV,
the preferred values of $\epsilon_{SW}$ may decrease if more emphasis is put on matching the observed strangeness production.

\section{Summary}
\label{sec:summary}

Using state-of-the-art statistical methods, we have determined the highest likelihood input parameters and related uncertainties
of the state-of-the-art (3+1)D viscous hydrodynamics + transport hybrid model for three collision energies $\sqrt{s_{NN}}=$ 19.6, 39, and 62.4 GeV.

An improved treatment of the uncertainty quantification yields a moderate collision energy dependence of shear viscosity over entropy density ratio.
As the uncertainties at lower energies still remain considerable,
it remains possible that the optimal value of $\eta/s$ is a constant over the investigated beam energy range,
with a value between 0.10 and 0.15.
The previously observed stronger collision energy dependence of $\eta/s$ is now partly absorbed in the transverse smearing parameter $W_{\text{trans}}$,
which has its most likely value $W_{\text{trans}} \approx 1.6$ fm at $\sqrt{s_{NN}}=19.6$ GeV
decreasing to $W_{\text{trans}} \approx 0.8$ fm at $\sqrt{s_{NN}}=62.4$ GeV.
On the other hand, the longitudinal smearing value is found to be roughly constant, $W_{\text{trans}} \approx 0.5-0.7$ fm.
As the collision energy dependence of these smearing parameters lack an obvious physical motivation,
it becomes necessary to impose more stringent theoretical constraints on the initial state for the future studies.

We find that the fluidization -- the transition from a transport description of the system to hydrodynamics -- is likely to happen early, between 0.5 - 1.0 fm.
The increase to 1.36 fm for $\sqrt{s_{NN}}=19.6$ GeV is induced by the increase in the earliest possible starting time at lower energies,
due to the longer time it takes for the two nuclei to interpenetrate each other.

The particlization energy density $\epsilon_{SW}$ (the transition from hydrodynamics back to a hadron transport afterburner)
has a notably different probability distribution at $\sqrt{s_{NN}}=39$ GeV compared to $\sqrt{s_{NN}}=19.6$ and $62.4$ GeV.
However, as there are some discrepancies on $K$ and $\Omega$ yields between model and measurements at the latter two energies
using higher values of the switching energy density,
the preferred value of $\epsilon_{SW}$ may be subject to change once more strange particle data becomes available.

While the values of the medium parameters extracted in the present analysis already provide a very good global fit on the beam energy scan data,
there remains some room for improvement on pseudorapidity distribution and HBT radii, for example.
The current model setup also ignores bulk viscosity and net-baryon diffusion;
testing the latter also requires a solution for the proton feed-down correction problem
currently hampering model-to-data comparisons on baryon observables in beam energy scan range.

\section{Acknowledgements}

This work has been supported by NSF grant no.~NSF-ACI-1550225 and by DOE grant no.~DE-FG02-05ER41367.
CPU time was provided by the Open Science Grid, supported by DOE and NSF.
We thank the Institute for Nuclear Theory at the University of Washington for its hospitality
and the Department of Energy for partial support during the completion of this work.

\appendix

\section{Experimental data}
\label{app:expdata}

The observables used in the analysis are summarized in Tables \ref{tb:expdata19}-\ref{tb:expdata62}.

Mean transverse momentum of $\Omega$ at $\sqrt{s_{NN}}=19.6$ and $39$ GeV has been computed from published $p_T$ spectrum data.
To avoid introducing additional uncertainties related to extrapolation of the spectra at low $p_T$, the mean is computed within the range of data points, $p_T = 1-5$ GeV/$c$.

$\langle R_{\text{out}} \rangle$, $\langle R_{\text{side}} \rangle$, $\langle R_{\text{long}} \rangle$ refer to simple averages over the 4 $k_T$ bins, covering the range 0.15 GeV/$c$ $<k_T<$ 0.6 GeV/$c$. 

The data in Ref.~\cite{Adamczyk:2012ku} for event plane elliptic flow $v_2\{\text{EP}\}$ at $\sqrt{s_{NN}}=62.4$ GeV is presented as a function of pseudorapidity $\eta$.
To improve our accuracy on determining $v_2$,
we have taken advantage of the fact that $v_2\{\text{EP}\}$ is practically constant between $-0.3 < \eta < 0.3$ at this energy
and compare the value of $v_2\{\text{EP}\}$ for charged particles within ${|\eta|<0.3}$ against the experimental value at $\eta=0.1$.

We have used the event plane $v_2$ also for comparisons with $v_2\{\text{corr}\}$,
which experimentally is determined with the correlation function method \cite{Magdy:2017kji}.
However, the preliminary data for $v_2\{\text{corr}\}$ at $\sqrt{s_{NN}}=19.6$ GeV and $39$ GeV suggests
that the magnitude of $v_2$ obtained by these two methods should be comparable in the investigated centrality range.

\begin{table}
\caption{Calibration data at $\sqrt{s_{NN}}=19.6$ GeV.}
\label{tb:expdata19}
\begin{tabular}{|p{4.5cm}|c|c|}
 \hline
  \centering Observable & Centrality (\%) & Ref \\
 \hline
  \centering $dN_{\text{ch}}/d\eta$ at $\eta=0.1$ & 0-6, 6-15, 15-25 & \cite{Alver:2010ck} \\
 \hline
  \centering $\langle \eta \rangle$ for $\eta>0.0$ & 0-6, 6-15, 15-25 & \cite{Alver:2010ck} \\
 \hline
  \centering $N(\pi^+)$, $N(\pi^-)$ & 0-5, 5-10, 10-20, 20-30 &  \cite{Adamczyk:2017iwn} \\
 \hline
  \centering $N(K^+)$, $N(K^-)$ & 0-5, 5-10, 10-20, 20-30 &  \cite{Adamczyk:2017iwn} \\
 \hline
  \centering $\langle p_T\rangle$ for $\pi^{+},\pi^{-},K^{+},K^{-}$ & 0-5, 5-10, 10-20, 20-30 & \cite{Adamczyk:2017iwn} \\
 \hline
  \centering Charged particle $v_2\{2\}$ \mbox{in $|\eta|<0.5$} & 5-10, 10-20, 20-30 & \cite{Adamczyk:2012ku} \\
 \hline
  \centering $dN(\Omega)/dp_T$ \mbox{at $p_T$=1.01 GeV/$c$} & 0-10 & \cite{Adamczyk:2015lvo} \\
 \hline
  \centering $\langle p_T\rangle$ for $\Omega$ \mbox{at $p_T=1.0-5.0$ GeV/$c$} & 0-10 & \cite{Adamczyk:2015lvo} \\
 \hline
  \centering $R_{\text{out}}$, $R_{\text{side}}$, $R_{\text{long}}$ of charged $\pi$ \mbox{at $\langle k_T \rangle \approx 0.22$ GeV/$c$} & 0-5, 20-30 & \cite{Adamczyk:2014mxp} \\
 \hline
  \centering $\langle R_{\text{out}} \rangle$, $\langle R_{\text{side}} \rangle$, $\langle R_{\text{long}} \rangle$ \mbox{of charged $\pi$} & 0-5, 20-30 & \cite{Adamczyk:2014mxp} \\
 \hline
\end{tabular}
\end{table}

\begin{table}
\caption{Calibration data at $\sqrt{s_{NN}}=39$ GeV.}
\label{tb:expdata39}
\begin{tabular}{|p{4.5cm}|c|c|}
 \hline
  \centering Observable & Centrality (\%) & Ref \\
 \hline
  \centering $N(\pi^+)$, $N(\pi^-)$ & 0-5, 5-10, 10-20, 20-30 &  \cite{Adamczyk:2017iwn} \\
 \hline
  \centering $N(K^+)$, $N(K^-)$ & 0-5, 5-10, 10-20, 20-30 &  \cite{Adamczyk:2017iwn} \\
 \hline
  \centering $\langle p_T\rangle$ for $\pi^{+},\pi^{-},K^{+},K^{-}$ & 0-5, 5-10, 10-20, 20-30 & \cite{Adamczyk:2017iwn} \\
 \hline
  \centering $v_2\{2\}$ in $|\eta|<0.5$ & 5-10, 10-20, 20-30 & \cite{Adamczyk:2012ku} \\
 \hline
  \centering $dN(\Omega)/dp_T$ at $p_T$=0.96 GeV/$c$ & 0-10 & \cite{Adamczyk:2015lvo} \\
 \hline
  \centering $\langle p_T\rangle$ for $\Omega$ \mbox{at $p_T=1.0-5.0$ GeV/$c$} & 0-10 & \cite{Adamczyk:2015lvo} \\
 \hline
  \centering $R_{\text{out}}$, $R_{\text{side}}$, $R_{\text{long}}$ of charged $\pi$ \mbox{at $\langle k_T \rangle \approx 0.22$ GeV/$c$} & 0-5, 20-30 & \cite{Adamczyk:2014mxp} \\
 \hline
  \centering $\langle R_{\text{out}} \rangle$, $\langle R_{\text{side}} \rangle$, $\langle R_{\text{long}} \rangle$ \mbox{of charged $\pi$} & 0-5, 20-30 & \cite{Adamczyk:2014mxp} \\
 \hline
\end{tabular}
\end{table}

\begin{table}
\caption{Calibration data at $\sqrt{s_{NN}}=62.4$ GeV.}
\label{tb:expdata62}
\begin{tabular}{|p{4.2cm}|c|c|}
 \hline
  \centering Observable & Centrality (\%) & Ref \\
 \hline
  \centering $N_{\text{ch}}$ in $|\eta|<0.5$ & 0-5, 5-10, 10-20, 20-30, 30-40 & \cite{Abelev:2008ab} \\
 \hline
  \centering $dN_{\text{ch}}/d\eta$ at $\eta=0.1$ & 0-3, 20-25 & \cite{Alver:2010ck} \\
 \hline
  \centering $\langle \eta \rangle$ for $\eta>0.0$ & 0-6, 6-15, 15-25 & \cite{Alver:2010ck} \\
 \hline
  \centering $N(K^+)/N(\pi^+)$ in $|\eta|<0.5$ & 0-5, 5-10, 10-20, 20-30, 30-40 & \cite{Abelev:2008ab} \\
 \hline
  \centering $\langle p_T\rangle$ for $\pi^{-},K^{+}$ & 0-5, 10-20, 20-30 & \cite{Abelev:2008ab} \\
 \hline
  \centering $v_2\{\text{EP}\}$ in $|\eta|<0.3$ & 10-40 & \cite{Adamczyk:2012ku} \\
 \hline
  \centering $v_2\{\text{corr}\}$ & 10-20, 20-30, 30-40 & \cite{Magdy:2017kji} \\
 \hline
  \centering $N(\Omega)$ in $|y|<0.5$ & 0-20 & \cite{Chatterjee:2015fua} \\
 \hline
  \centering $\langle p_T\rangle$ for $\Omega$ in $|y|<0.5$ & 0-20 & \cite{Aggarwal:2010ig} \\
 \hline
  \centering $R_{\text{out}}$, $R_{\text{side}}$, $R_{\text{long}}$ of charged $\pi$ \mbox{at $\langle k_T \rangle \approx 0.22$ GeV/$c$} & 0-5, 20-30 & \cite{Adamczyk:2014mxp} \\
 \hline
  \centering $\langle R_{\text{out}} \rangle$, $\langle R_{\text{side}} \rangle$, $\langle R_{\text{long}} \rangle$ \mbox{of charged $\pi$} & 0-5, 20-30 & \cite{Adamczyk:2014mxp} \\
 \hline
\end{tabular}
\end{table}

\section{Principal components}
\label{app:pcatables}

The number of principal components used is 7 for each energy.
This set of PCs explains over 99\% of the total variance in each case.
In the following we list the most important observables for the first 3 principal components for each collision energy,
as these already cover over 95\% of the total variance.

\begin{table}[htb]
\caption{Seven observables at $\sqrt{s_{NN}}=19.6$ GeV with most weight on principal component 0 (89.43\% of total variance).}
\begin{tabular}{|c|c|}
\hline
Observable & Weight\\
\hline
$N(K^-)$ (20-30)\% & 0.0634\\
\hline
$N(K^-)$ (10-20)\% & 0.0576\\
\hline
$N(\pi^+)$ (20-30)\% & 0.0568\\
\hline
$N(K^+)$ (20-30)\% & 0.0566\\
\hline
$N(\pi^-)$ (20-30)\% & 0.0548\\
\hline
$N(K^-)$ (5-10)\% & 0.0534\\
\hline
$dN_{\text{ch}}/d\eta$ at $\eta=0.1$ (15-25)\% & 0.0516\\
\hline
\end{tabular}
\end{table}

\begin{table}[htb]
\caption{Seven observables at $\sqrt{s_{NN}}=19.6$ GeV with most weight on principal component 1 (5.99\% of total variance).}
\begin{tabular}{|c|c|}
\hline
Observable & Weight\\
\hline
$N(\Omega)$ (0-10)\% & 0.2759\\
\hline
$\langle p_T\rangle (K^-)$ (20-30)\% & 0.0484\\
\hline
$\langle p_T\rangle (K^-)$ (10-20)\% & 0.0451\\
\hline
$\langle p_T\rangle (K^-)$ (5-10)\% & 0.0422\\
\hline
$v_2\{2\}$ (20-30)\% & 0.0417\\
\hline
$\langle p_T\rangle (K^-)$ (0-5)\%  & 0.0401\\
\hline
$\langle p_T\rangle (K^+)$ (20-30)\% & 0.0381\\
\hline
\end{tabular}
\end{table}

\begin{table}[htb]
\caption{Seven observables at $\sqrt{s_{NN}}=19.6$ GeV with most weight on principal component 2 (2.32\% of total variance).}
\begin{tabular}{|c|c|}
\hline
Observable & Weight\\
\hline
$v_2\{2\}$ (20-30)\%  & 0.3366\\
\hline
$v_2\{2\}$ (10-20)\%  & 0.2925\\
\hline
$v_2\{2\}$ (5-10)\%  & 0.2353\\
\hline
$N(\Omega)$ (0-10)\% & 0.0376\\
\hline
$\langle p_T\rangle (K^+)$ (0-5)\%  & 0.0064\\
\hline
$\langle p_T\rangle (K^+)$ (5-10)\% & 0.0058\\
\hline
$\langle p_T\rangle (K^-)$ (0-5)\% & 0.0056\\
\hline
\end{tabular}
\end{table}

\begin{table}[htb]
\caption{Seven observables at $\sqrt{s_{NN}}=39$ GeV with most weight on principal component 0 (84.47\% of total variance).}
\begin{tabular}{|c|c|}
\hline
Observable & Weight\\
\hline
$N(\Omega)$ (0-10)\% & 0.1311\\
\hline
$N(\pi^-)$ (20-30)\%  & 0.0659\\
\hline
$N(\pi^+)$ (20-30)\%  & 0.0643\\
\hline
$N(K^-)$ (20-30)\%  & 0.0639\\
\hline
$N(K^-)$ (10-20)\%  & 0.0572\\
\hline
$N(K^+)$ (20-30)\%  & 0.0562\\
\hline
$N(\pi^-)$ (10-20)\%  & 0.0547\\
\hline
\end{tabular}
\end{table}

\begin{table}[htb]
\caption{Seven observables at $\sqrt{s_{NN}}=39$ GeV with most weight on principal component 1 (9.37\% of total variance).}
\begin{tabular}{|c|c|}
\hline
Observable & Weight\\
\hline
$N(\Omega)$ (0-10)\% & 0.4506\\
\hline
$v_2\{2\}$ (20-30)\%  & 0.0389\\
\hline
$v_2\{2\}$ (10-20)\%  & 0.0291\\
\hline
$\langle p_T\rangle (K^-)$ (20-30)\% & 0.0282\\
\hline
$\langle p_T\rangle (K^-)$ (10-20)\%  & 0.0273\\
\hline
$\langle p_T\rangle (K^-)$ (5-10)\%  & 0.0272\\
\hline
$\langle p_T\rangle (K^-)$ (0-5)\%  & 0.0264\\
\hline
\end{tabular}
\end{table}

\begin{table}[htb]
\caption{Seven observables at $\sqrt{s_{NN}}=39$ GeV with most weight on principal component 2 (3.36\% of total variance).}
\begin{tabular}{|c|c|}
\hline
Observable & Weight\\
\hline
$N(\Omega)$ (0-10)\% & 0.3287\\
\hline
$v_2\{2\}$ (20-30)\%  & 0.1671\\
\hline
$v_2\{2\}$ (10-20)\%  & 0.1255\\
\hline
$v_2\{2\}$ (5-10)\%  & 0.1004\\
\hline
$\langle p_T\rangle (K^-)$ (20-30)\%  & 0.0157\\
\hline
$\langle p_T\rangle (K^-)$ (10-20)\%  & 0.0152\\
\hline
$\langle p_T\rangle (K^-)$ (5-10)\%  & 0.0151\\
\hline
\end{tabular}
\end{table}

\begin{table}[htb]
\caption{Seven observables at $\sqrt{s_{NN}}=62.4$ GeV with most weight on principal component 0 (82.42\% of total variance).}
\begin{tabular}{|c|c|}
\hline
Observable & Weight\\
\hline
$N(\Omega)$ (0-20)\% & 0.1511\\
\hline
$N_{\text{ch}}$ (30-40)\%  & 0.1338\\
\hline
$dN_{\text{ch}}/d\eta$ at $\eta=0.1$ (20-25)\%  & 0.1231\\
\hline
$N_{\text{ch}}$ (20-30)\%  & 0.1205\\
\hline
$N_{\text{ch}}$ (10-20)\%  & 0.1051\\
\hline
$N_{\text{ch}}$ (5-10)\%  & 0.0930\\
\hline
$N_{\text{ch}}$ (0-5)\%  & 0.0870\\
\hline
\end{tabular}
\end{table}

\begin{table}[htb]
\caption{Seven observables at $\sqrt{s_{NN}}=62.4$ GeV with most weight on principal component 1 (10.4\% of total variance).}
\begin{tabular}{|c|c|}
\hline
Observable & Weight\\
\hline
$N(\Omega)$ (0-20)\% & 0.2199\\
\hline
$v_2\{\text{EP}\}$ (30-40)\%  & 0.1323\\
\hline
$v_2\{\text{EP}\}$ (10-40)\%  & 0.1313\\
\hline
$v_2\{\text{EP}\}$ (20-30)\%  & 0.1077\\
\hline
$v_2\{\text{EP}\}$ (10-20)\%  & 0.0806\\
\hline
$\langle p_T\rangle (\Omega)$ (0-20)\%  & 0.0548\\
\hline
$\langle p_T\rangle (\pi^-)$ (20-30)\% & 0.0314\\
\hline
\end{tabular}
\end{table}

\begin{table}[htb]
\caption{Seven observables at $\sqrt{s_{NN}}=62.4$ GeV with most weight on principal component 2 (3.86\% of total variance).}
\begin{tabular}{|c|c|}
\hline
Observable & Weight\\
\hline
$N(\Omega)$ (0-20)\% & 0.2674\\
\hline
$v_2\{\text{EP}\}$ (10-40)\%  & 0.1463\\
\hline
$v_2\{\text{EP}\}$ (30-40)\%  & 0.1412\\
\hline
$v_2\{\text{EP}\}$ (20-30)\%  & 0.1123\\
\hline
$v_2\{\text{EP}\}$ (10-20)\%  & 0.0819\\
\hline
$\langle p_T\rangle (\Omega)$ 0-20\%  & 0.0243\\
\hline
$R_{\text{long}}$ (0-5)\%  & 0.0207\\
\hline
\end{tabular}
\end{table}

\end{document}